\documentclass[preprint,12pt,doublespacing]{elsarticle}
\usepackage{setspace}
\usepackage[top=2 cm, bottom=2 cm, left=2 cm, right=2 cm]{geometry}
\usepackage{graphicx}
\usepackage{amssymb,amsmath,amsbsy}
\usepackage{epsf}
\newcommand{\mb}[1]{\mbox{\boldmath$#1$}}
\usepackage{subfigure}

\usepackage{appendix}
\usepackage{natbib}

\begin{document}

\begin{frontmatter}

\title{Magnetically-actuated artificial cilia for microfluidic propulsion}


\author{S. N. Khaderi}

\address{Zernike Institute for Advanced Materials, University of Groningen, Groningen, The Netherlands.}%

\author{M. G. H. M. Baltussen}
\author{P. D. Anderson}
\address{Eindhoven University of Technology, Eindhoven, The Netherlands.}%

\author{D. Ioan}
\address{Universitatea Politehnica din Bucuresti, Bucharest, Romania.}%

\author{J. M. J. den Toonder}
\address{Eindhoven University of Technology, Eindhoven, The Netherlands.}%

\author{P. R. Onck}
\address{Zernike Institute for Advanced Materials, University of Groningen, Groningen, The Netherlands.}%

\begin{abstract}
Natural cilia are hair-like microtubule-based structures that are able to move fluid at low Reynolds number through asymmetric motion. In this paper we follow a biomimetic approach to design artificial cilia lining the inner surface of microfluidic channels with the goal to propel fluid. The artificial cilia consist of polymer films filled with magnetic nanoparticles. The asymmetric, non-reciprocating motion is generated by tuning an external magnetic field. To obtain the magnetic field and associated magnetization local to the cilia we solve the Maxwell equations, from which the magnetic torques can be deduced. To obtain the ciliary motion we solve the dynamic equations of motion which are then fully coupled to the fluid dynamic equations that describe fluid flow around the cilia.  By doing so we show that by properly tuning the applied magnetic field, asymmetric ciliary motion can be generated that is able to propel fluid in a microchannel.
The results are presented in terms of three dimensionless parameters that fully delineate the asymmetry and cycle time as a function of the relative contribution of elastic, inertial, magnetic and viscous fluid forces. 

\end{abstract}


\end{frontmatter}


      \section{Introduction}

In the biomedical field there is an increasing need for miniaturized lab-on-a-chip analysis systems that are able to analyze small quantities of biological samples such as biofluids (e.g. blood, saliva, urine) \cite{0960-1317-14-6-R01, whitesides_nature, chang_nat_mat}. These so-called biosensors are microfabricated total-analysis-systems that typically consist of a system of microscopic channels, connecting microchambers (the labs) where dedicated tests are carried out. Classical means for fluid-propulsion do no longer suffice at these small length scales, which has led to a search for new methods dedicated for fluid propulsion at the micron-scale, such as micropumps \cite{0960-1317-14-6-R01}, syringe pumps \cite{SchillingE.A._ac015640e, JeonN.L._la000600b} or by exploiting electro-magnetic actuation, as in electro-osmotic \cite{shulin, lingxin} and magnetohydrodynamic devices \cite{asuncion,  alexandra}. The use of electric fields, however, in transporting biological fluids (which usually have high conductivity) may induce heating, bubble formation and pH gradients from electrochemical reactions \cite{b408382m, wu:234103, henrik_2004_conference}. In this work, we explore a new way to manipulate fluids in microfluidic systems, inspired by nature, through the magnetic actuation of artifical cilia.

\begin{figure}[ht]\centering
      \includegraphics[scale=0.5]{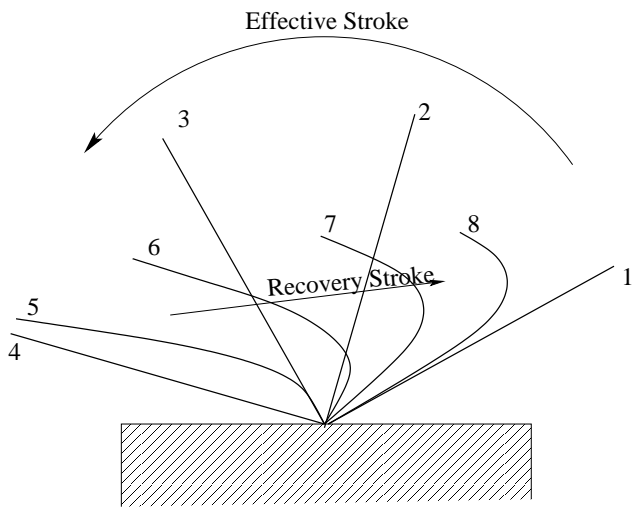}
      \label{fig:cilia_motion}
       \caption{Asymmetric motion of a  cilium. Instances 1-4 show the effective stroke and 5-8 show the   recovery stroke. The asymmetry in the motion of cilia during forward and return stroke propels the fluid effectively in one direction.}
\end{figure}

Akin to size-effects in the mechanical properties of solid materials, the dependence of size in fluid dynamics is captured by the Reynolds number. At small length scales and low Reynolds number, gravity does not play a role and fluid dynamics is usually dominated by viscosity rather than inertia. An important consequence of this is that fluids can only be propelled by motions that are initiated by actuators whose movement is cyclic but asymmetric in time. Nature has solved this problem by means of hair-like structures, called cilia, whose beating pattern is non-reciprocating and consists of an effective and a recovery stroke, as shown in  Fig.~\ref{fig:cilia_motion}. In this work we design artificial cilia that can be actuated by an external magnetic field. The artificial cilia are thin films consisting of a polymer matrix filled with magnetic nanoparticles. Depending on the nature of the particles, the film can be super-paramagnetic or ferromagnetic with a remanent magnetization. The applied magnetic field is uniform but its magnitude and direction can be manipulated in time to get the desired asymmetric motion.
The Lagrangian model for the cilia is based on a dynamic finite-element representation, accounting for elastic, inertia and drag forces in a non-linear geometry setting. Simultaneously, Maxwell’s equations are solved at each configuration, yielding the local magnetic field and magnetic induction from which the magnetic forces can be obtained. This magneto-mechanical model allows studying the overall deformation of the cilia actuated by an external magnetic field as a function of the mechanical and magnetic material parameters of the film’s microstructure. The Lagrangian solid model is also coupled to an Eulerian formulation of the fluid, which enables to generate fluid flow through the magnetically-induced film motion. The goal of this paper is understand the multiphysics interplay between magnetostatics, solid mechanics and fluid dynamics and to explore ways to exploit this interplay to design non-reciprocating motion in artifical cilia. 

The paper is organized as follows.  In section 2 we discuss the magnetomechanical model consisting of a finite element discretization of the dynamic principle of virtual work that is explicitly coupled to a discretized form of an integral formulation of Maxwell’s equations. In section 3.1 and 3.2 we study several fundamental magnetic loading situations in which the interaction of elastic, inertial, magnetic and viscous drag forces is summarized in a deformation mechanism map that features three distinct regimes of operation. In section 3.3 we identify four different configurations that show asymmetric motion, followed by a parametric study in section 4 to explore the efficiency of ciliary motion as a function of the three governing dimensionless numbers. Finally, we couple the magnetostatic model a fluid-dynamics model and show that fluid can be propelled and that a linear relation exists between the asymmetry of the ciliary motion and the amount of fluid propelled per cycle. Section 5 contains the conclusions.
%
%
%

\section{Method}\label{sec:magnetomechanical_model}
The proposed actuators are polymers films filled with magnetic particles.  Depending on the magnetic nature of the particles, the film can either be super-paramagnetic (SPM) or permanently magnetic. The thickness of the film is much smaller than its length, which allows us to use  Euler-Bernoulli kinematics. In the following two Sections we will discuss the numerical methods  used to solve the equations of motion for the dynamic deformation of the film and  Maxwell's  equations for the magnetostatics problem.

\subsection{Equations of motion}\label{sec:eom}
As a starting point for the Euler-Bernoulli beam element formulation we use the principle of virtual work \cite{malvern}. In this weak form of the equations of motion, the virtual work of the external forces is equal to the internal work and is given by
\begin{equation}
 \delta W_\text{int}^t = \delta W_\text{ext}^t,
\end{equation}
with
\begin{equation}
 \delta W_{int}^t = \int_V\left( \sigma \delta\epsilon + \rho(\ddot{u}\delta u+\ddot{v}\delta v)\right)dV,
\end{equation}
where $u$ and $v$ are the axial and transverse displacements along the beam length (axial coordinate $x$) and $\rho$ is the density of the film. Furthermore,  $\sigma$ is the axial stress and $\epsilon$ is the corresponding strain, given by
\begin{eqnarray}
 \epsilon&=&\frac{\partial u}{\partial x}+\frac{1}{2}\left(\frac{\partial v}{\partial x} \right)^2 -y\frac{\partial^2 v}{\partial x^2}\nonumber \equiv \bar{\epsilon}-y\chi\nonumber.
\end{eqnarray}
Its first variation is thus
\begin{eqnarray}
\delta \epsilon &=& \frac{\partial\delta u}{\partial x}+\frac{\partial v}{\partial x}\frac{\partial \delta v}{\partial x}-y\frac{\partial ^2\delta v}{\partial x^2} \equiv \delta\bar{\epsilon}-y\delta\chi.
\end{eqnarray}
 By substituting the strains and defining
$\int \sigma dA=P$ and $-\int \sigma y dA=M$ ($A$ is the area of the cross section), the internal virtual work at time $t$ can be written as the sum of an elastic and an inertial part
\begin{eqnarray}
	\delta W_\text{int}^t=\int \left( P^t\delta\bar{\epsilon}^t+M^t\delta\chi^t+ \rho A(\ddot{u^t}\delta u^t+\ddot{v^t}\delta v^t)\right) dx.
\end{eqnarray}
The internal virtual work at time $t+\Delta t$ is written as
\begin{eqnarray}\label{eqn:internal_virtual_work}
	\delta W_\text{int}^{t+\Delta t}=\int \left( P^{t+\Delta t}\delta\bar{\epsilon}^{t+\Delta t}+M^{t+\Delta t}\delta\chi^{t+\Delta t}+ \rho A(\ddot{u}^{t+\Delta t}\delta u^{t+\Delta t}+\ddot{v}^{t+\Delta t}\delta v^{t+\Delta t})\right) dx.
\end{eqnarray}
The corresponding external virtual work is
\begin{equation}\begin{split}
     \delta W_\text{ext}^{t+\Delta t}=&\int \left(f_x^{t+\Delta t} \delta u^{t+\Delta t}+f_y^{t+\Delta t} \delta v^{t+\Delta t}+ N_z^{t+\Delta t} \frac{\partial \delta v^{t+\Delta t}}{\partial x}  \right)Adx \\
     &+ \int \left(t_x^{t+\Delta t}\delta u^{t+\Delta t}+t_y^{t+\Delta t}\delta v^{t+\Delta t}\right)b dx,
\end{split}\end{equation}
where $f_x$ and $f_y$  are the magnetic body forces in axial and transverse directions, $N_z$ is the magnetic body couple in the out-of-plane direction, $t_x$ and $t_y$ are the surface tractions due to the fluid drag and $b$ is the out-of-plane thickness of the film.
We now expand the elastic part of the internal work linearly in time by substituting  $(Q^{t+\Delta t}=Q^{t}+\Delta Q)$ for the field parameters in Eqn.~\ref{eqn:internal_virtual_work} which gives
\begin{eqnarray}
    \delta W_\text{int}^{t + \Delta t}&=&\int \left[ \left( P^t\delta\bar{\epsilon}^t+M^t\delta\chi^t \right)+ \left(\Delta P\delta\bar{\epsilon}^t+\Delta M\delta\chi^t \right)+P^t\Delta\delta\bar{\epsilon} +  \rho A(\ddot{u}^{t+\Delta t}\delta u^{t}+\ddot{v}^{t+\Delta t}\delta v^{t}) \right]dx,
\end{eqnarray}
in which terms of order higher than one are neglected.
The axial and transverse displacements are linearly and cubically interpolated in terms of the nodal degrees of freedom, the displacements  and rotations,
\begin{eqnarray}
	u=\mb{N_u} \mb{p}, \ \ \ \ \ \ 
	v=\mb{N_v} \mb{p}, 
\end{eqnarray}
where
\[\mb{p}=\{u_1\  v_1\  \phi_1l_0\  u_2\  v_2\  \phi_2l_0 \}^T,\] $\mb{N_u}$ and $\mb{N_v}$ being the standard interpolation matrices, given in Appendix \ref{int_func}.
By using the following notation
\begin{eqnarray}
\frac{\partial u}{\partial x}=\mb{B_u}\mb{p},\ \ \frac{\partial v}{\partial x}=\mb{B_v}\mb{p},\ \ \frac{\partial^2 u}{\partial x^2}=\mb{C_v}\mb{p},
\end{eqnarray}
and the   constitutive relations
\begin{eqnarray}
\Delta P=EA\Delta\bar{\epsilon},\ \ \   \ \Delta M=EI\Delta\chi,
\end{eqnarray}
with $E$ being the elastic modulus and $I$ being the second moment of area defined as $ I=\frac{1}{12}bh^3 $,  the internal virtual work can be written as
\begin{equation}\begin{split}
\delta W_\text{int}^{t+\Delta t}=&\delta \mb{p}^T\int\left[P^t\left(\mb{B_u}^T+\mb{B_v}^T\mb{B_vp}\right)+M^t\mb{C_v}^T  \right]dx\\&+\delta\mb{p}^T\int\left[ EA\left(\mb{B_u}^T+\mb{B_v}^T\mb{B_vp}\right)\left(\mb{B_u}+\mb{B_v}\mb{B_vp}\right) \right]dx \Delta\mb{p}\\
&+\delta \mb{p}^T\int EI \mb{C_v}^T\mb{C_v} dx  \Delta\mb{p}+\delta\mb{p}^T\int P^t \mb{B_v}^T\mb{B_v} dx \Delta\mb{p}\\
&+\delta \mb{p}^T\int\rho A(\mb{N_u}^T\mb{N_u}+\mb{N_v}^T\mb{N_v})dx\ddot{\mb{p}}^{t+\Delta t}.
\end{split}\end{equation}
By choosing the domain of integration to be the current configuration (i.e. using an updated Lagrangian framework), the total displacements are zero, $\mb{p}=0$, and we get
\begin{equation}\begin{split}
\delta W_\text{int}^{t+\Delta t}
=\delta\mb{p}^T\mb{f}_\text{int}^t+\delta\mb{p}^T\mb{K}\Delta \mb{p} +\delta\mb{p}^T\mb{M}\ddot{\mb{p}}^{t+\Delta t},
\end{split}\end{equation}
where
\[\mb{f}_\text{int}^t=\int\left[P^t\mb{B_u}^T+M^t\mb{C_v}^T  \right]dx\] is the nodal internal force vector,
\[  \mb{K}=\int EA\mb{B_u}^T\mb{B_u} dx 
+\int EI \mb{C_v}^T\mb{C_v} dx  +\int P^t \mb{B_v}^T\mb{B_v} dx \]
 is the stiffness matrix, the first two terms of which represent the material stiffness and the third term  represents the geometric stiffness, and 
\[ \mb{M}=\int\rho A(\mb{N_u}^T\mb{N_u}+\mb{N_v}^T\mb{N_v})dx \] is the mass matrix.
The external virtual work is
\begin{equation}\begin{split}\label{eqn:ext_v_work}
	\delta W_\text{ext}^{t+\Delta t}=&\delta\mb{p}^T\int\left[(f_x^{t+\Delta t}\mb{N_u}^T+f_y^{t+\Delta t}\mb{N_v}^T+N_z^{t+\Delta t}\mb{B_v}^T) A+b\left( t_x^{t+\Delta t} \mb{N_u}^T+t_y^{t+\Delta t} \mb{N_v}^T\right)\right]dx,\\
	=&\delta\mb{p}^T\mb{f}_\text{ext}^{t+\Delta t}.
\end{split}
\end{equation}
By equating the internal and external virtual work and noting that the resulting equation holds for arbitrary $\delta \mb{p}$ we get

\begin{equation}\label{eq:after_linearization}
 \mb{K}\Delta\mb{p}+\mb{M}\ddot{\mb{p}}^{\text{t}+\Delta \text{t}}=\mb{f}^\text{t+$\Delta $t}_\text{ext}-\mb{f}^\text{t}_\text{int}. \end{equation}

The motion of the film with time is obtained by solving Eqn.~\ref{eq:after_linearization} with appropriate initial and boundary conditions. Newmark's method is employed to integrate Eqn.~\ref{eq:after_linearization} in time, viz,
\begin{subequations}
\begin{eqnarray}\label{eq:newmark_disp}
\mb{p}^\text{t+$\Delta$t}&=&\mb{p}^\text{t}+\dot{\mb{p}}^\text{t}\Delta t+\frac{1}{2}\Delta t^2 \left[(1-2\beta)\ddot{\mb{p}}^\text{t}+2\beta\ddot{\mb{p}}^{\text{t}+\Delta\text{t}} \right],\\ \label{eq:newmark_velo}
\dot{\mb{p}}^{\text{t}+\Delta\text{t}}&=&\dot{\mb{p}}^\text{t}+\Delta t\left[(1-\gamma)\ddot{\mb{p}}^\text{t}+ \gamma\ddot{\mb{p}}^{\text{t}+\Delta \text{t}} \right],
\end{eqnarray}
\end{subequations}
where $\gamma$ and $\beta$ are integration parameters. By using Eqn.~\ref{eq:newmark_disp}, Eqn.~\ref{eq:after_linearization}  can be written as
\begin{equation}\label{eqn:disc_eqn_to_solve}
         (\Delta t^2\beta\mb{K}+\mb{M})\ddot{\mb{p}}^\text{t+$\Delta$t}=\mb{f}^\text{t+$\Delta $t}_\text{ext}-\mb{f}^\text{t}_\text{int}-\Delta t\mb{K}\left[ \dot{\mb{p}}^\text{t} + \frac{1}{2}\Delta t(1-2\beta)\ddot{\mb{p}}^\text{t} \right],
\end{equation}
the solution of which will give the nodal accelerations, which are integrated using Eqn.~\ref{eq:newmark_disp} and \ref{eq:newmark_velo} to get the nodal displacements and velocities. 

The drag forces of the fluid on the film are accounted for through the surface tractions in Eqn.~\ref{eqn:ext_v_work}. The tractions are proportional to the velocity in the low Reynolds number regime, with the normal and tangential coefficients of proportionality (the drag coefficients) denoted by $C_x$ and $C_y$. The tractions are calculated from the velocity at time $t$. The magnetic  forces come in through the body forces and couples in Eqn.~\ref{eqn:ext_v_work}. These are calculated by solving Maxwell's equations on the current configuration, i.e. the configuration at time $t$ (see Section \ref{sec:magnetostatics}).

\subsection{Magnetostatics}\label{sec:magnetostatics}
Maxwell's equations for the magnetostatic problem with no currents are
\begin{eqnarray}
     \mb{\nabla}\cdot\mb{B}&=&0\label{eqn:divbzero}\\
     \mb{\nabla}\times\mb{H}&=&0,
\end{eqnarray}
with    the constitutive relation
\begin{eqnarray}\label{eqn:mag_cons_relation}
      \mb{B}&=&\mu_0(\mb{M}+\mb{H}),
\end{eqnarray}
where $\mb{B}$ is the magnetic flux density (or magnetic induction), $\mb{H}$ is the magnetic field, $\mb{M}$ is the magnetization which includes the remnant magnetization,   and $\mu_0$ is the permeability of vacuum. Substituting Eqn.~\ref{eqn:mag_cons_relation} into Eqn.~\ref{eqn:divbzero} yields
\begin{eqnarray}\label{eqn:divhdivm}
        \mb{\nabla} \cdot\mb{H} &=&-\mb{\nabla} \cdot\mb{M}.
\end{eqnarray}
As $\mb{\nabla}\times\mb{H}=0$, a scalar potential $\phi$ exists, such that $\mb{H}=-\mb{\nabla}\phi$. Substituting this in Eqn.~\ref{eqn:divhdivm} yields a Poisson equation for $\phi$, $\nabla^2 \phi=-\mb{\nabla}\cdot\mb{M}$. By taking into consideration the effect of discontinuity in the medium, the general solution of the Poisson equation can be found \cite{jackson}, resulting in  
\begin{eqnarray}\label{eq:solnforh}
      \mb{H}(\mb{x})=-\frac{1}{4\pi}\mb{\nabla}\oint\frac{ \mb{n}'\cdot\mb{M}(\mb{x'})  }{|\mb{x}-\mb{x'}|} dS' + \frac{1}{4\pi}\mb{\nabla} \int \frac{  \mb{\nabla}'\cdot \mb{M}(\mb{x'})  }{|\mb{x}-\mb{x'}|}  dV'.
\end{eqnarray}
where $\mb{n}'$ is the outward normal to the surface of  $V$. By assuming that  {\it the  magnetization is  uniform inside the volume}, so $\mb{\nabla}'\cdot \mb{M}=0$, the volume integral vanishes, and the field is only due  to the jump of magnetization across the surface, as reflected by the surface integral in Eqn.~\ref{eq:solnforh}.

We now discretize the film  into a chain of rectangular segments.  On the surface of each segment there is a jump of magnetization. $\mb{M}$ is uniform inside the segment and zero outside the segment. The magnetic field in local coordinates  (denoted by $\ \hat{}$) due to the four surfaces of a segment  can now be calculated at any position $(\hat{x}, \hat{y})$  by evaluating the surface integral in Eqn.~\ref{eq:solnforh}, resulting in
\begin{equation}\label{eqn:local_field}\begin{split}
\hat{H}_x&=\frac{\hat{M}_x  \left(-\tan^{-1}\left[\frac{-\frac{h}{2}-\hat{y}}{-\frac{l}{2}+\hat{x}}\right]+\tan^{-1}\left[\frac{\frac{h}{2}-\hat{y}}{-\frac{l}{2}+\hat{x}}\right]\right)}{2
\pi }+\frac{-\hat{M}_x \left(-\tan^{-1}\left[\frac{-\frac{h}{2}-\hat{y}}{\frac{ l}{2}+\hat{x}}\right]+\tan^{-1}\left[\frac{\frac{h}{2}-\hat{y}}{\frac{ l}{2}+\hat{x}}\right]\right)}{2
\pi }+\\&
\frac{\hat{M}_y \left(\frac{1}{2} \ln\left[\left(-\frac{l}{2}-\hat{x}\right)^2+\left(-\frac{h}{2}+\hat{y}\right)^2\right]-\frac{1}{2} \ln\left[\left(\frac{l}{2}-\hat{x}\right)^2+\left(-\frac{h}{2}+\hat{y}\right)^2\right]\right)}{2
\pi }+\\&\frac{-\hat{M}_y \left(\frac{1}{2} \ln\left[\left(-\frac{l}{2}-\hat{x}\right)^2+\left(\frac{h}{2}+\hat{y}\right)^2\right]-\frac{1}{2} \ln\left[\left(\frac{l}{2}-\hat{x}\right)^2+\left(\frac{h}{2}+\hat{y}\right)^2\right]\right)}{2
\pi }\\
\hat{H}_y&=\frac{\hat{M}_y \left(-\tan^{-1}\left[\frac{-\frac{l}{2}-\hat{x}}{-\frac{h}{2}+\hat{y}}\right]+\tan^{-1}\left[\frac{\frac{l}{2}-\hat{x}}{-\frac{h}{2}+\hat{y}}\right]\right)}{2
\pi }+\frac{-\hat{M}_y \left(-\tan^{-1}\left[\frac{-\frac{l}{2}-\hat{x}}{\frac{h}{2}+\hat{y}}\right]+\tan^{-1}\left[\frac{\frac{l}{2}-\hat{x}}{\frac{h}{2}+y}\right]\right)}{2
\pi }+\\&
\frac{\hat{M}_x \left(\frac{1}{2} \ln\left[\left(-\frac{l}{2}+\hat{x}\right)^2+\left(-\frac{h}{2}-\hat{y}\right)^2\right]-\frac{1}{2} \ln\left[\left(-\frac{l}{2}+\hat{x}\right)^2+\left(\frac{h}{2}-\hat{y}\right)^2\right]\right)}{2
\pi }+\\&\frac{-\hat{M}_x \left(\frac{1}{2} \ln\left[(\frac{ l}{2}+\hat{x})^2+\left(-\frac{h}{2}-\hat{y}\right)^2\right]-\frac{1}{2} \ln\left[(\frac{ l}{2}+\hat{x})^2+\left(\frac{h}{2}-\hat{y}\right)^2\right]\right)}{2
\pi },\end{split}
\end{equation}
where $\hat{M}_x$, $\hat{M}_y$ are  magnetizations in the tangential (or length) and normal (or thickness) directions,  
 $h$ is the thickness  and  $l$ is  the length of the segment. Here, $\hat{x}$ and $\hat{y}$ are the local coordinates having their origin in the center of the segment.  On rearranging, the magnetic field due to segment $i$ can be written as
\begin{eqnarray}
   \hat{\mb{H}}_i &=\mb{G}_i\ \hat{\mb{M}}_i.
\end{eqnarray}
where $\hat{\mb{M}}=[\begin{array}{c c} \hat{M}_x &\hat{M}_y  \end{array}]^T$, $\hat{\mb{H}}=[\begin{array}{c c} \hat{H}_x &\hat{H}_y  \end{array}]^T$ and $\mb{G}_i$  can be obtained from Eqn.~\ref{eqn:local_field}.
The field due to segment $i$ with respect to the global coordinates is

\begin{eqnarray}
     \mb{H}_i=\mb{R}_i\   \hat{\mb{H}}_i,
\end{eqnarray}
where $\mb{R}_i$ is
\begin{equation}
      \mb{R}_i=\left[ \begin{array}{c c }  \cos\theta_i &-\sin\theta_i\\  \sin\theta_i &\cos\theta_i    \end{array}\right],
\end{equation}
with $\theta _i$ the orientation of the segment $i$ with respect to the global coordinates.
The field at any element $j$ because of the magnetization of all the segments throughout the film is
\begin{eqnarray}\label{eqn:self_field}
        \mb{H}_j&=&\mb{H}_0+\sum_{i=1}^{N}\mb{R}_i\mb{G}_{ij}\hat{\mb{M}}_i,
\end{eqnarray}
where $\mb{G}_{ij}$ properly accounts for the relative positioning of segments $i$ and $j$, $\mb{H}_0$ is the externally applied magnetic field, far away from the film, and $N$ is the total number of segments.  By rotating $\mb{H}_j$ back to the local coordinates we get

\begin{eqnarray}\label{eq:total_field}
        \hat{\mb{H}}_j&=&\mb{R}^T_j\mb{H}_0+\sum_{i=1}^{N}\mb{R}^T_j\mb{R}_i\mb{G}_{ij}\hat{\mb{M}}_i.
\end{eqnarray}
For the situation of a permanently magnetized film with magnetization $\hat{\mb{M}}_i, i=1, \dots , N$, Eqn.~\ref{eq:total_field} gives the magnetic field in all the segments.

However, in case of a super-paramagnetic film, the magnetization $\hat{\mb{M}}_j$ is not known a-priori, but depends on the local magnetic field through
\begin{eqnarray}\label{eqn:implicit_eqn_in_mag}
\hat{\mb{M}}_j
     &=&\hat{\mb{\chi}} \ \hat{\mb{H}}_j\nonumber\\
  &=&\hat{\mb{\chi}}\mb{R}^T_j\mb{H}_0+\sum_{i=1}^{N}\hat{\mb{\chi}}\mb{R}^T_j\mb{R}_i\mb{G}_{ij}\hat{\mb{M}}_i,
\end{eqnarray}
with \[\mb{\chi}= \left[\begin{array}{c c} \hat{\chi}_x &0 \\ 0& \hat{\chi}_y \end{array}\right].\]

There are $N$ similar pairs of equations. In total these are $2 \times N$ equations for the $2 \times N$ unknown magnetizations. This set of equations is solved to get the magnetization with respect to the local coordinate frame. Once the magnetization is known the field can be found from Eqn.~\ref{eq:total_field}. Then the magnetic flux density can be found by using Eqn.~\ref{eqn:mag_cons_relation}. The magnetic couple per unit volume  is given by $\mb{N}=\mb{M}\times \mb{B}$,  which can be obtained in local coordinates through $N_z=\hat{M}_x\hat{B}_y-\hat{M}_y\hat{B}_x$. Magnetic body forces due to field gradients will be neglected in the present study.

\section{Results}
      \subsection{Fundamental loading situations}\label{sec:fundamental_loading_situations}
In this Section we study the behavior of the film under simple loading conditions. We will consider a horizontal  film that is clamped at the left end and free elsewhere. Three cases are considered: (i) a permanently magnetic film with a field applied in the transverse (vertical) direction, (ii) a permanently magnetic film subjected to a rotating magnetic field and (iii) a super-paramagnetic (SPM) film subjected to a rotating magnetic field.  

The following reference parameters are considered: Elastic modulus $E = 1$ MPa, drag coefficients $C_x = 10\  \text{Ns/m}^3$, $C_y$ = 20 $\text{Ns/m}^3$, remnant magnetization for permanently magnetic films $M = 15$ kA/m, susceptibility for the SPM film $\hat{\chi}_x$ =4.6,  $\hat{\chi}_y$ =0.8, reference time during which the field is applied $t_\text{ref}=1\  \text{ms} $, length of the film $L$ = 100 $\mu \text{m} $,  thickness $h$ = 2 $\mu \text{m} $ and the magnitude of the applied field at $t=t_\text{ref}$ is $B_{0\text{max}}=\mu_0 H_{0\text{max}}=12.5$ mT. In the simulations the values of  $\gamma$ and $\beta$ are $1.0$ and $0.5$, respectively.
\newline

\par
\begin{figure}[htbp]\centering
     \subfigure[ ] {\label{fig:mag_vertical_field_disp_with_time_with_markers}\includegraphics[width=80 mm]{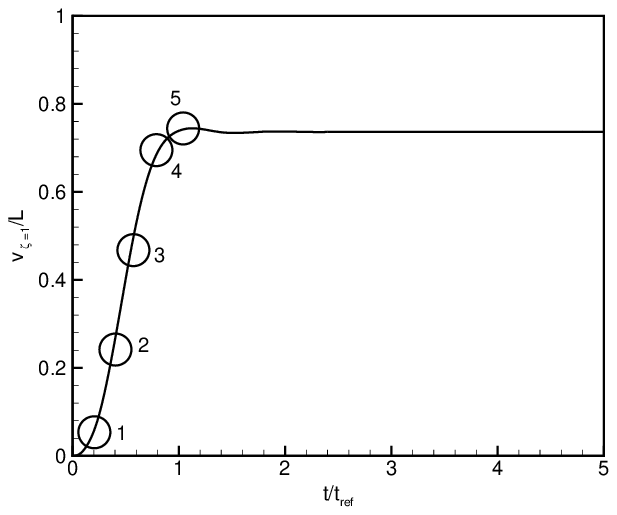}}
     \subfigure[ ] {\label{fig:mag_vertical_field_deformed_geometry}\includegraphics[width=80 mm]{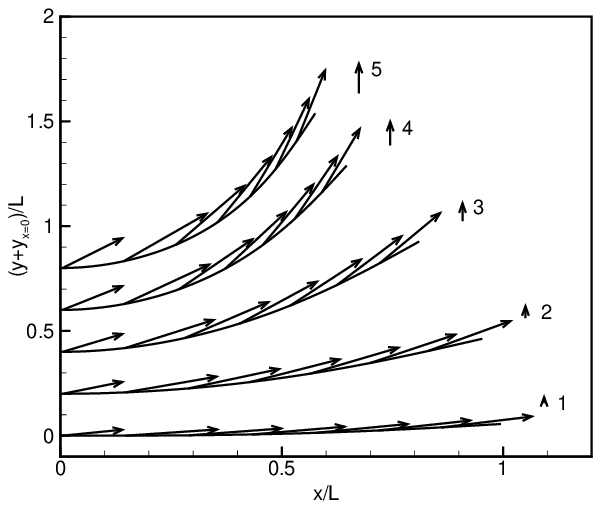}}
     \subfigure[ ] {\label{fig:mag_vertical_field_bx}\includegraphics[width=80 mm]{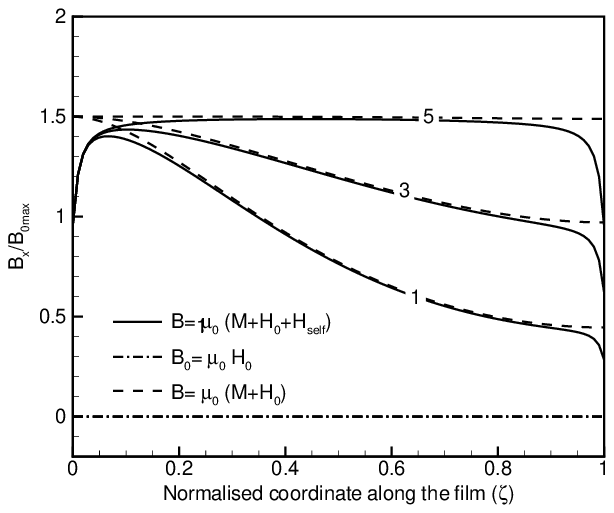}}
     \subfigure[ ]{\label{fig:mag_vertical_field_by}\includegraphics[width=80 mm]{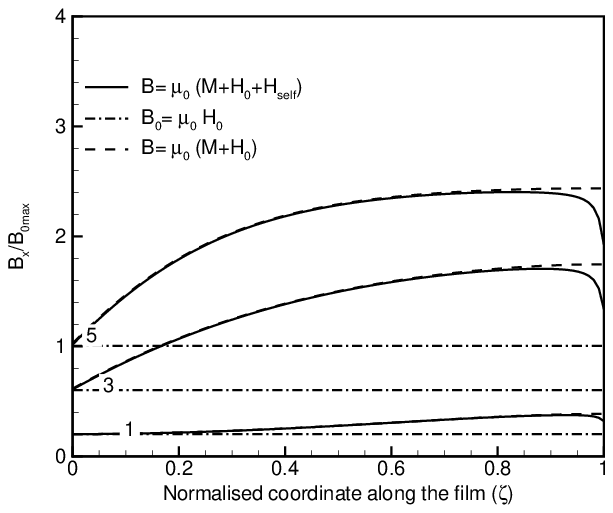}}
     \subfigure[ ] {\label{fig:mag_vertical_field_couple}\includegraphics[width=80 mm]{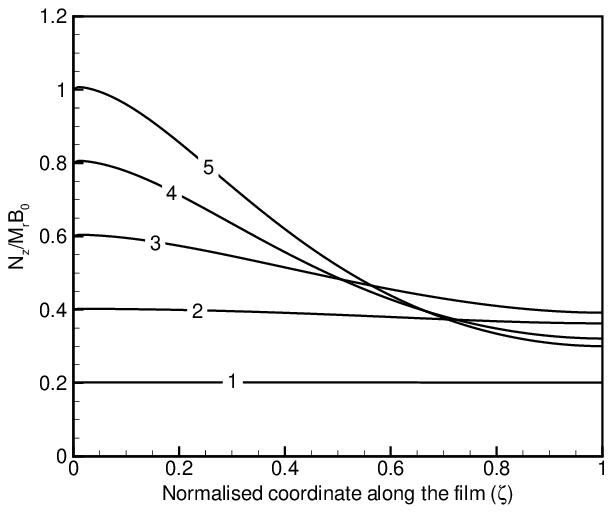}}
     \caption{A horizontal permanently magnetic film subjected to an increasing field in transverse (i.e. vertical) direction. (a) The displacement of the free end in time. (b) Deformed geometry: The  arrows on  the film  show the direction and magnitude of $\mb{B}$ and the separate arrows at the right show the applied field $\mb{B}_0$. (c) Contributions to $B_x$ from various sources. (d) Contributions to $B_y$ from various sources. (e) Variation of body couple along the film at different instances of time. }
\label{fig:permanently_magnetic_film_with_transverse_field}
\end{figure}

 \noindent \textbf{Permanently magnetic film in transverse and rotating field.}
A magnetic field is applied in the vertical ($y$) direction and is increased linearly from zero to $B_{0\text{max}}$ in $t=t_\text{ref}$. Then the applied magnetic field is maintained at this value. 
Fig.~ \ref{fig:mag_vertical_field_disp_with_time_with_markers} shows the vertical displacement of the free end of the film as a function of time. After time $t=\ t_\text{ref}$ the applied field is kept constant, but due to the presence of the fluid and inertia in the film, it reaches a steady state at time $t=1.5 \ t_\text{ref}$.  In Fig.~\ref{fig:mag_vertical_field_deformed_geometry} the entire film is depicted showing the film deformation in time at five instances identified in Fig.~\ref{fig:mag_vertical_field_disp_with_time_with_markers}. Next to the film, at the right, the arrow  denotes  the applied magnetic field $\mb{B}_0$ at five instances. In addition, the magnetic induction $\mb{B}$ is plotted by means of arrows along the film. Fig.~\ref{fig:mag_vertical_field_bx} shows the variation  of the horizontal component $B_x$ (with respect to the global coordinate axes)  along the film. Since no external field is applied in the $x$ direction, the primary cause of $B_x$ is the permanent magnetization in the film, $\mb{M}$. For a permanently magnetic film, we can write the magnetic field as the sum of the applied field $\mb{H}_0$ and the field generated by the magnetization in the rest of the film (the 'self-field'), $\mb{H}=\mb{H}_0+\mb{H}_\text{self}$, see Eqn.~\ref{eqn:self_field}. It can be clearly seen that at instant 1 when the film is still horizontal, that it is this self-field that tends to decrease the magnetic induction near the ends of the film, effectively de-magnetizing the film. When the film deforms, the horizontal component of the field, $B_x$, decreases near the free end while $B_y$ increases (see Fig.~\ref{fig:mag_vertical_field_by}). This is due to the rotation of $\mb{M}$
 caused by the deformation of the film, except near the ends where the self-field is operative. The $B_y$ near the fixed end of the film is nearly equal to the applied field (Fig.~\ref{fig:mag_vertical_field_by}). This is a consequence of the magnetostatic boundary condition, which dictates that the normal component of $\mb{B}$ will be continuous across a boundary.  The  variation of the magnetic body couple is shown in Fig.~\ref{fig:mag_vertical_field_couple}. As the field is applied in the transverse direction, in the initial stages the couple is almost uniform throughout the film. As the film deforms, the free end  gets more aligned with the applied field and the couple intensity near the free end is considerably reduced. 

\begin{figure}[htbp]\centering
     \subfigure[ ] {\label{fig:mag_rotating_field_disp_with_time_with_markers}\includegraphics[width=80 mm]{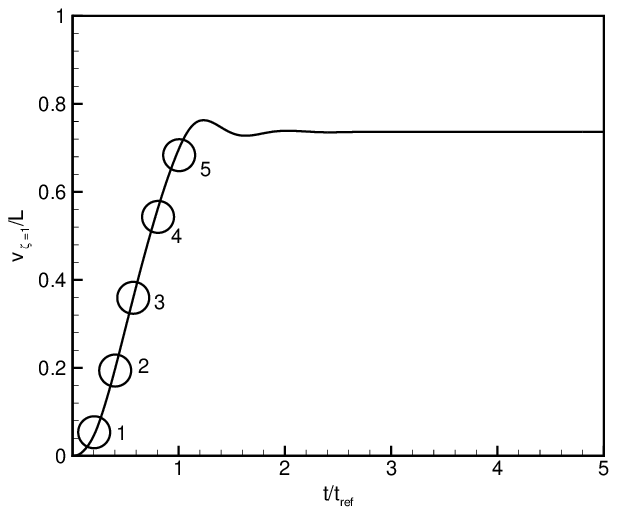}}
     \subfigure[ ] {\label{fig:mag_rotating_field_deformed_geometry}\includegraphics[width=80 mm]{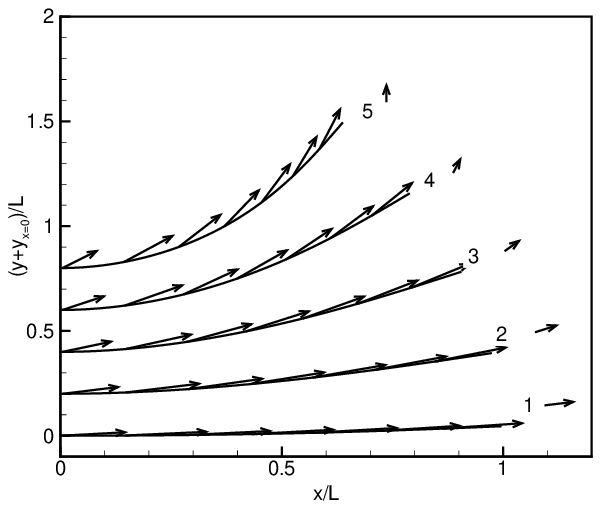}}
     \subfigure[ ] {\label{fig:mag_rotating_field_bx}\includegraphics[width=80 mm]{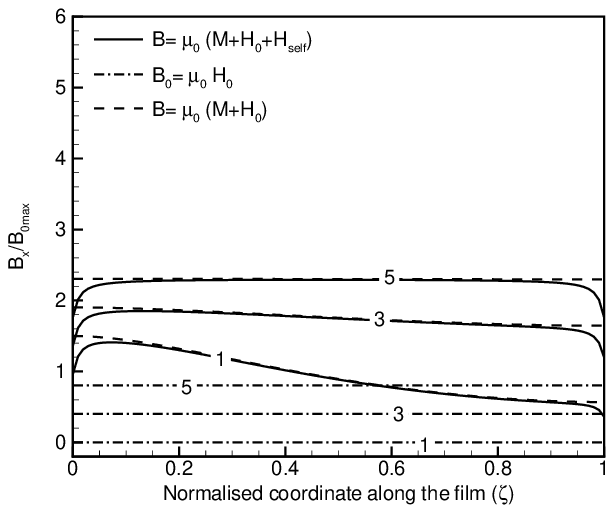}}
     \subfigure[ ]{\label{fig:mag_rotating_field_by}\includegraphics[width=80 mm]{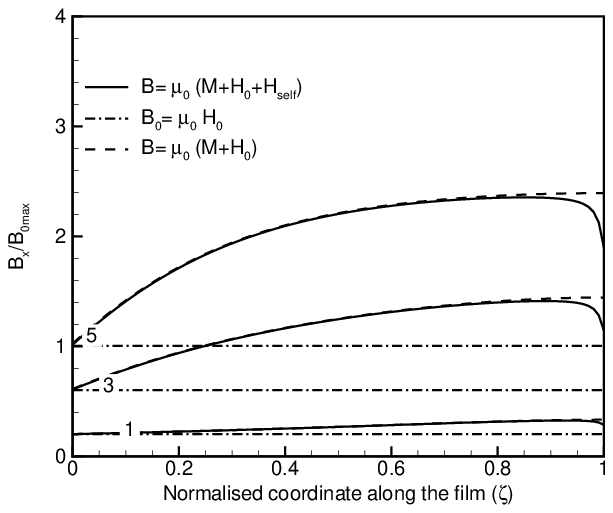}}
     \subfigure[ ] {\label{fig:mag_rotating_field_couple}\includegraphics[width=80 mm]{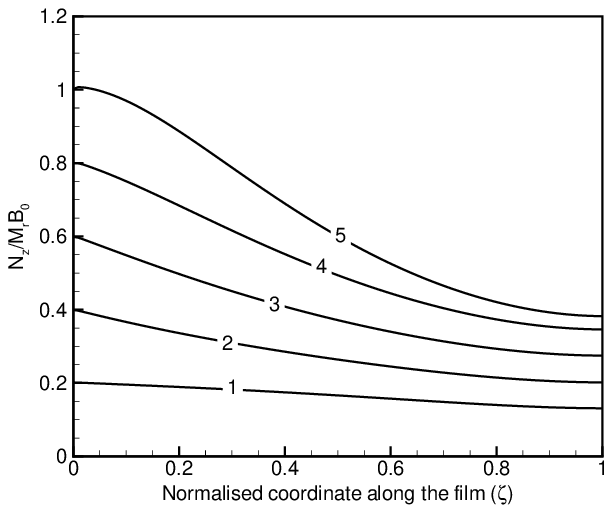}}
     \subfigure[]{\label{fig:mag_rotating_field_disp_with_time_with_vary_tref}\includegraphics[width=80 mm]{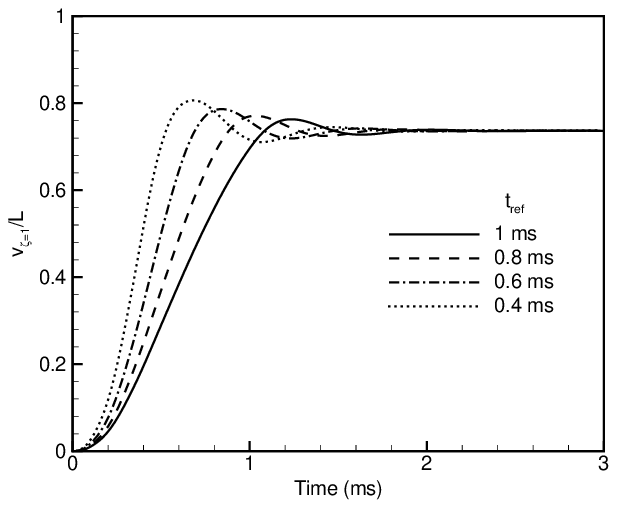}}
\caption{A horizontal permanently magnetic film subjected to a rotating field. (a) The displacement of the free end in time. (b) Deformed geometry: The  arrows on  the film  show the direction and magnitude of $\mb{B}$ and the separate arrows show the applied field. (c) Contributions to $B_x$ from various sources. (d) Contributions to $B_y$ from various sources. (e) Variation of body couple along the film at different instances of time. (f) Displacement of the free end vs time for various loading rates  $t_\text{ref}$.}
	\label{fig:permanently_magnetic_film_with_rotating_field}
\end{figure}

Next, we study the same film but now subjected to a rotating field. The initial field is in the $x$ direction and has magnitude $B_{0\text{max}}$. Then  we linearly increase the field in the $y$ direction and decrease it simultaneously in the $x$ direction. By doing so,  the magnetic field vector is 'rotated' from the $x$ to the $y$ axis. Of course, its magnitude does not remain constant during rotation. When the field has rotated, its orientation and magnitude is maintained at a constant value.  From Figs.~\ref{fig:mag_rotating_field_disp_with_time_with_markers} and \ref{fig:mag_rotating_field_deformed_geometry} it can be seen that the deformation is very similar to the previous case (cf. Figs.~\ref{fig:mag_vertical_field_disp_with_time_with_markers} and \ref{fig:mag_vertical_field_deformed_geometry}).  At time $t_\text{ref}$ (corresponding to instant 5 in Fig.~\ref {fig:mag_rotating_field_disp_with_time_with_markers}), the magnetic forces are larger than the elastic forces, hence the film continues to deform for a short time, even when the applied magnetic field is kept constant. 
 The  contributions to $\mb{B}$ from various sources is shown in Figs.~\ref{fig:mag_rotating_field_bx} and \ref{fig:mag_rotating_field_by}.   Unlike the previous case, the couple intensity always increases at  the free end (Fig.~\ref{fig:mag_rotating_field_couple}). The effect of the rate of rotation of the magnetic field is shown in Fig.~\ref{fig:mag_rotating_field_disp_with_time_with_vary_tref}. We decrease the reference time of the field rotation leading to an increased rotation rate. As the rate of rotation  is increased, the tip reaches its maximum deflection in less time, but its maximum value remains the same  irrespective of the rate of rotation.
\\
%

\par
\begin{figure}[htbp]\centering
     \subfigure[] {\label{fig:spm_rotating_field_disp_with_time_with_markers}\includegraphics[width=80 mm]{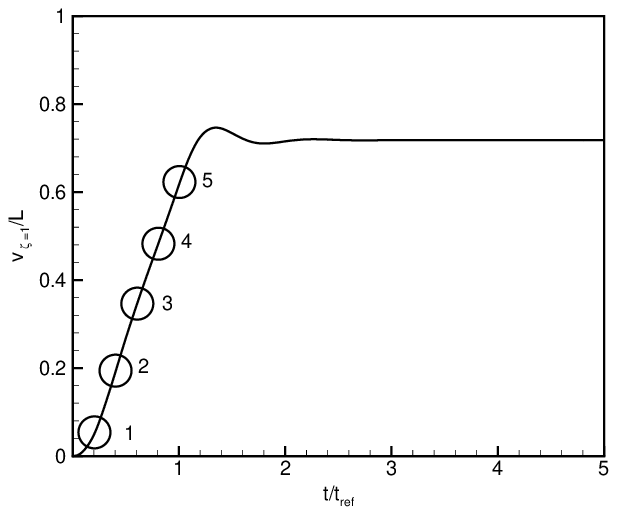}}
     \subfigure[] {\label{fig:spm_rotating_field_deformed_geometry}\includegraphics[width=80 mm]{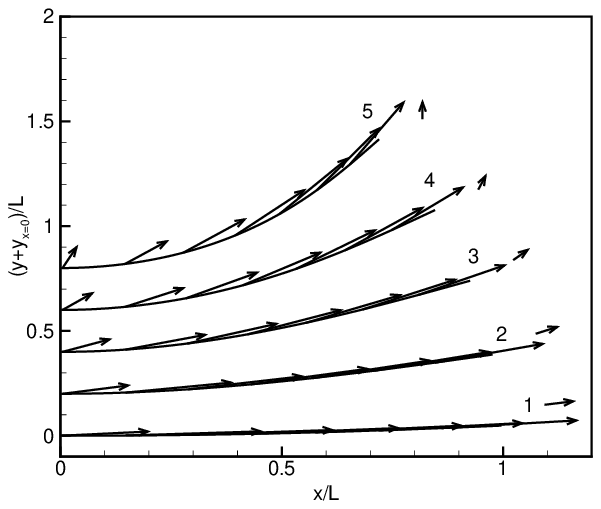}}
     \subfigure[] {\label{fig:spm_rotating_field_bx}\includegraphics[width=80 mm]{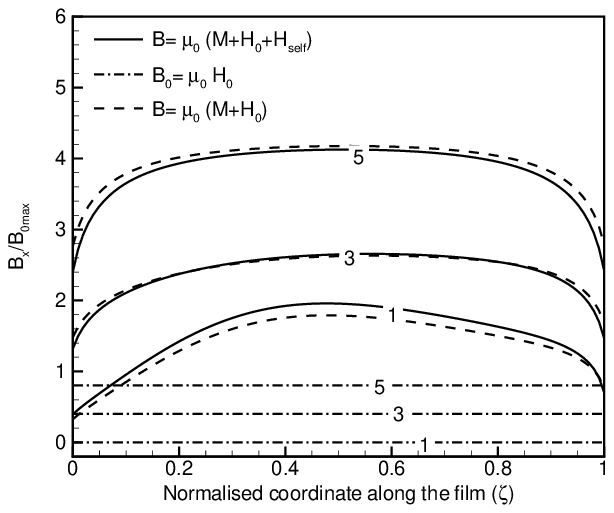}}
     \subfigure[]{\label{fig:spm_rotating_field_by}\includegraphics[width=80 mm]{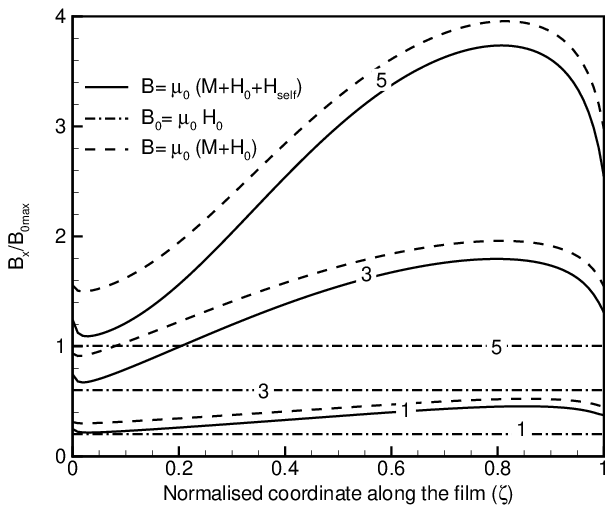}}
     \subfigure[] {\label{fig:spm_rotating_field_couple}\includegraphics[width=80 mm]{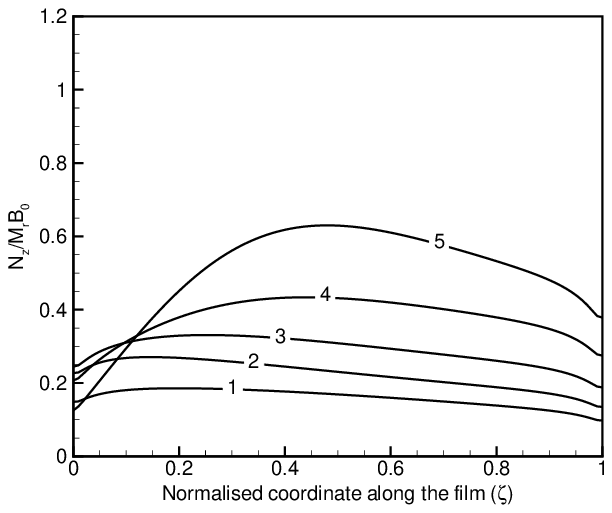}}
     \caption{A horizontal super-paramagnetic film subjected to a rotating field in transverse direction. (a) The displacement of the free end in time. (b) Deformed geometry: The  arrows on  the film  show the direction and magnitude of $\mb{B}$ and the separate arrows show the applied field. (c) Contributions to $B_x$ from various sources. (d) Contributions to $B_y$ from various sources. (e) Variation of body couple along the film at different instances of time. To normalize the couple distribution in Fig.~\ref{fig:spm_rotating_field_couple}, the magnetization is used of the permanently magnetic film (see Fig.~\ref{fig:mag_rotating_field_couple}).}
	\label{fig:spm_film_with_rotating_field}
\end{figure}

 \noindent\textbf{Super-paramagnetic film in a rotating magnetic field.}
We study the behavior of a SPM film subjected to the same rotating applied field as for the permanently magnetic  film.  The direction and magnitude  of magnetization in a super-paramagnetic film is not fixed, but changes with the applied field and the geometry of the film, given the specific magnetic susceptibility of the film (here we have $\hat{\chi}_x=4.6$ and $\hat{\chi}_y=0.8$). 
The variation of the tip displacement, the geometry, field (both applied and total) and  induced couple for a   super-paramagnetic film are shown in Fig.~\ref{fig:spm_film_with_rotating_field}.  Figs.~\ref{fig:spm_rotating_field_disp_with_time_with_markers} and \ref{fig:spm_rotating_field_deformed_geometry} show that for the given susceptibility the deformation of the film is similar to that of the permanently magnetic case.  As the film deforms,  the $x$ component of the applied field decreases and the $y$ component  increases, and accordingly, so do the components of the magnetization and the total field (Figs.~\ref{fig:spm_rotating_field_bx} and \ref{fig:spm_rotating_field_by}).  It can be observed that the field gradients are lower when compared to the permanently magnetic films (see Figs.~\ref{fig:mag_vertical_field_bx}, \ref{fig:mag_vertical_field_by}, \ref{fig:mag_rotating_field_bx} and \ref{fig:mag_rotating_field_by}). This is due to the fact that the magnetization is not a uniform constant throughout the film as in the permanently magnetic case.  As the film rotates, unlike the permanently magnetic film, the $B_y$ near the fixed end of the film is not equal to that of the applied field. This is due to the fact that field outside the film is enhanced by the normal magnetization of the film. 
For a SPM film, the magnetization is obtained by solving an implicit equation to obtain $\mb{M}$ (Eqn.~\ref{eqn:implicit_eqn_in_mag}), for which it is essential to account for the self-field, generated by the magnetization in the film (see Eqn.~\ref{eq:total_field}). Thus, the difference between the applied field $\mb{B}_0$ and the magnetic induction $\mb{B}$ in  Figs.~\ref{fig:spm_rotating_field_bx} and \ref{fig:spm_rotating_field_by} is entirely due to the segment-to-segment interaction.
 Unlike the permanently magnetic film, in the super-paramagnetic film the induced magnetic couple is not uniform, even in the initial stages of deformation. As the applied field is rotated, the couple induced at the fixed end tends to increase initially, but it gets reduced as the applied field becomes vertical. The couple distribution, for a given instant of time, along the film does not monotonically decrease towards the free end as in the permanently magnetic film, but attains a maximum along the film. Despite this different torque distribution, the deformations are identical (see Figs.~\ref{fig:mag_rotating_field_deformed_geometry} and \ref {fig:spm_rotating_field_deformed_geometry}).

\begin{figure}[htbp]\centering
     \subfigure[] {\label{fig:spm_rotating_field_with_vary_chit}\includegraphics[width=80mm]{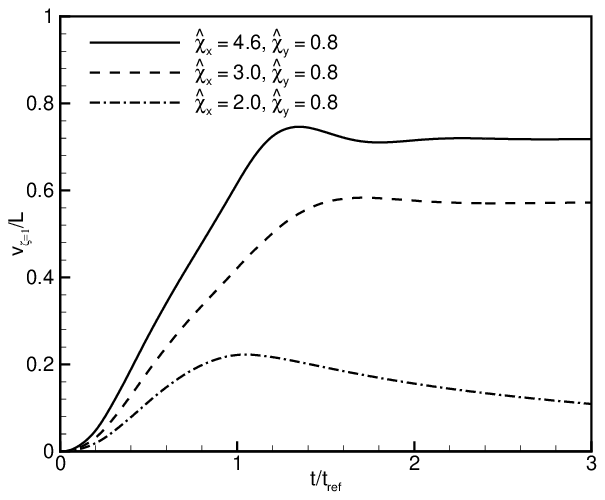}}
\subfigure[] {\label{fig:spm_rotating_field_with_same_diff}\includegraphics[width=80mm]{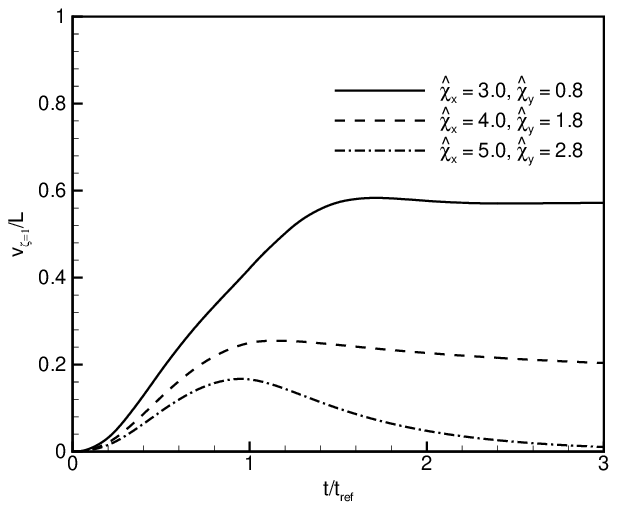}}
\subfigure[] {\label{fig:spm_rotating_field_disp_with_time_with_vary_tref}\includegraphics[width=80mm]{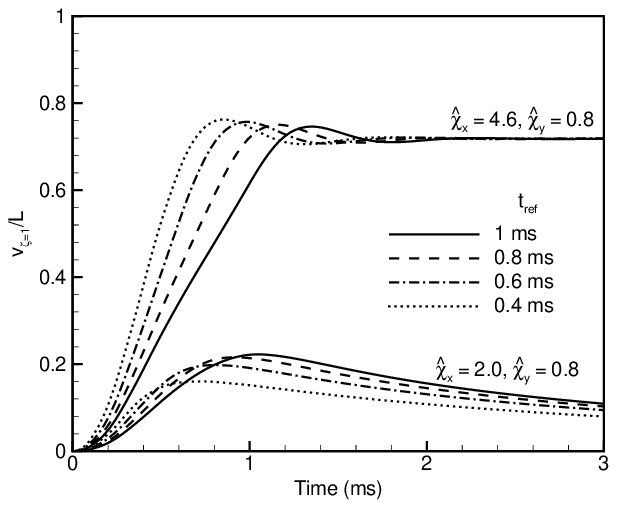}}

%

\caption{A horizontal super-paramagnetic film subjected to a rotating field:  (a) Displacement of the free end vs time for various $\hat{\chi}_x \ ( \hat{\chi}_y=0.8)$. (b) Displacement of the free end vs time for various $\hat{\chi}_x$  and $ \hat{\chi}_y$, with $\hat{\chi}_x-\hat{\chi}_y=2.2$. (c)  Displacement of the free end vs time for various $t_\text{ref}$.}
\label{fig:spm_rotating_field_varying_things}
\end{figure}

For the SPM film to deform, the material susceptibility tensor $\mb{\chi}$ needs to be anisotropic.
The induced magnetic couple depends on the difference between  $\hat{\chi}_x$ and  $\hat{\chi}_y$ and on the magnetic field, which itself depends on the magnitude of the susceptibility.
The effect of the change in $\hat{\chi}_x$, with constant $\hat{\chi}_y$, on the deflection of the free end is shown in Fig.~\ref{fig:spm_rotating_field_with_vary_chit}.  As $\hat{\chi}_x$ decreases, the induced magnetization decreases, and so does the magnetic couple, resulting in less deflection of the film. Also, for small values of $\hat{\chi}_x$ ($\le 2.0$), even when the applied magnetic field is kept constant after $t_\text{ref}$, the magnetic forces due to the induced couples are  smaller than the elastic forces. Hence, the film returns back to its initial position.
The deflection of the free end with varying $\hat{\chi}_y$ and $\hat{\chi}_x$ but keeping their difference constant is shown in Fig.~\ref{fig:spm_rotating_field_with_same_diff}. As the magnitude of the susceptibility increases, the deflection of the  film decreases. Hence, for a SPM film both the absolute value of the susceptibility tensor as well as its anisotropy determine the deformation of the film.

The effect of the  rate of rotation of the magnetic field  is shown in Fig.~\ref{fig:spm_rotating_field_disp_with_time_with_vary_tref}. We reduce the time in which the applied field is rotated from horizontal to vertical ($t_\text{ref}$) from $1$ ms to $0.4$ ms. As a result the rotation rate of the applied magnetic field is increased. This study is done for two different cases of anisotropy: $\hat{\chi}_x=4.6, \ \hat{\chi}_y=0.8$ and $\hat{\chi}_x=2.0, \ \hat{\chi}_y=0.8$. When  $\hat{\chi}_x=4.6$ and $ \ \hat{\chi}_y=0.8$, in the steady state, the deflection of the free end attains a value independent of the rate of rotation of the applied magnetic field. The only effect of the increase of the rate of rotation, the  free end attains its maximum displacement in less time. When  $\hat{\chi}_x=2.0$ and $ \ \hat{\chi}_y=0.8$, however,  the maximum deflection of the free end decreases when the rotation rate is increased. This behaviour is typical for the SPM film. The magnetic torque distribution depends sensitively on the orientation of the film with respect to the field direction. If the field is rotated too fast, the film lags behind due to the opposing fluid drag. A field oriented perpendicular to an almost undeformed film, will not induce a pronounced torque and the film returns back to its undeformed configuration by elastic recovery. This is the case when the susceptibility is unfavourable as for ($\hat{\chi}_x,\ \hat{\chi}_y$) = ($2.0,\ 0.8$) in Figs.~\ref{fig:spm_rotating_field_with_vary_chit} and \ref{fig:spm_rotating_field_disp_with_time_with_vary_tref} and for ($\hat{\chi}_x,\ \hat{\chi}_y$) = ($5.0,\ 2.8$) in Fig. \ref{fig:spm_rotating_field_with_same_diff}.

  \subsection{Dimensional analysis}
%
\newcommand{\dimanafundwidth}{70 mm}

To identify the dimensionless parameters that govern the behavior of the system, we start from the virtual work equation (see Section \ref{sec:eom}), neglecting the axial deformations:
\begin{eqnarray}
      \int EI\frac{\partial^2 v}{\partial x^2}\frac{\partial^2\delta v}{\partial x^2}dx+
      \int \rho A\frac{\partial^2 v}{\partial t^2}\delta v dx=
      \int N_z\frac{\partial \delta v}{\partial x}A dx
      -\int C_y\frac{\partial v}{\partial t}\delta v b dx\nonumber,
\end{eqnarray}
where the first term represents the virtual elastic work done by the internal moments, the second term represents the virtual work done by the inertial forces, the third term represents the virtual work done by the magnetic couple and the last term represents the work done by the fluid drag forces. This is valid for any segment $dx$, and hence,
\begin{equation}
    EI\frac{\partial ^2 v}{\partial x^2}\frac{\partial^2\delta v}{\partial x^2}+
   \rho A\frac{\partial ^2 v}{\partial t^2}\delta v=
    N_z\frac{\partial \delta v}{\partial x}A 
    - C_y\frac{\partial v}{\partial t}\delta v b \nonumber.
\end{equation}
We introduce the dimensionless variables $V$, $T$ and $X$, such that $v=VL$, $x=XL$ and $t=T\tau$, where $L$ is a characteristic length (taken to be the length of the film) and $\tau$ is a characteristic time. Substitution yields
\begin{equation}
       \frac{ E b h^3}{12L^2}\frac{\partial^2 V}{\partial X^2}\frac{\partial^2\delta V}{\partial X^2}+
       \frac{\rho b h L^2}{\tau^2}\delta V \frac{\partial^2V}{\partial T^2}=
       -\frac{C_y L^2 b }{\tau h} \frac{\partial V}{\partial T}\delta V+
       N_z h b\frac{\partial \delta V}{\partial X},
\end{equation}
from which  the elastic  ($\frac{Ebh^3}{12L^2}$), the  inertial  ($\frac{\rho b h L^2}{\tau^2}$), the viscous  ($\frac{C_y L^2 b }{\tau h}$) and the magnetic ($N_z h b$) terms can be easily identified.By normalising with the elastic term, we get

\begin{equation}
     \frac{1}{12}\left(\frac{\partial^2V}{\partial X^2}\frac{\partial^2\delta V}{\partial X^2}\right)+
     I_n\left(\frac{\partial^2V}{\partial T^2}\delta V\right)=
     M_n\left(\frac{\partial \delta V}{\partial X}\right)
     -F_n\left(\frac{\partial V}{\partial T}\delta V \right).
\end{equation}

With the three governing dimensionless numbers being defined as, $I_n=\frac{\rho L^4}{Eh^2\tau^2}$  the ratio of inertial to  elastic force, $M_n=\frac{N_zL^2}{Eh^2} $ the ratio of magnetic to elastic force and $F_n=\frac{C_yL^4}{E\tau h^3}$ the ratio of fluid to elastic force.

Before proceeding, we identify the origin  of the magnetic couple $N_z$ for the two magnetic material systems under consideration. For permanently magnetic materials, having a remanent magnetization along the axial direction of the film:$(\hat{M}_x, \hat{M}_y )=(\hat{M}_x, 0 )$,
\begin{equation}
    N_z=\hat{M}_x\hat{B}_y-\hat{M}_y\hat{B}_x=\hat{M}_x\hat{B}_y=\hat{M}_xB_0\hat{f}(\theta),
\end{equation}
where $\theta$ is the film orientation and $B_0$ is the amplitude of the applied magnetic field. For a SPM film,
\begin{equation}
\begin{split}
     N_z=&\hat{M}_x\hat{B}_y-\hat{M}_y\hat{B}_x=\mu_0\hat{H}_x\hat{H}_y(\hat{\chi}_x-\hat{\chi}_y)\\
                       =&\mu_0H_0\hat{f}_x(\hat{\chi}_x, \hat{\chi}_x)\hat{g}_x(\theta) H_0\hat{f}_y(\hat{\chi}_x, \hat{\chi}_y)\hat{g}_y(\theta)(\hat{\chi}_x-\hat{\chi}_y)\\
                      =&\frac{B_0^2}{\mu_0}\hat{h}(\hat{\chi}_x, \hat{\chi}_x, \theta),
\end{split}
\end{equation}
in which $\hat{h}$ groups the dimensionless dependence of the torque on the susceptibilities $\hat{\chi}_x$, $\hat{\chi}_y$, the film orientation $\theta$. It is to be noted that the body couple for the SPM film is proportional to the square of the applied field while in the permanently magnetic case, it is linearly proportional. As a result, the magneto-elastic number becomes $M_n=\frac{\hat{M}_xB_0L^2}{Eh^2}$ for the permanently magnetic case and $M_n=\frac{B_0^2L^2}{\mu_0Eh^2}$ for the SPM case.

\begin{figure}
     \centering
     \subfigure[]{\includegraphics[width =\dimanafundwidth]{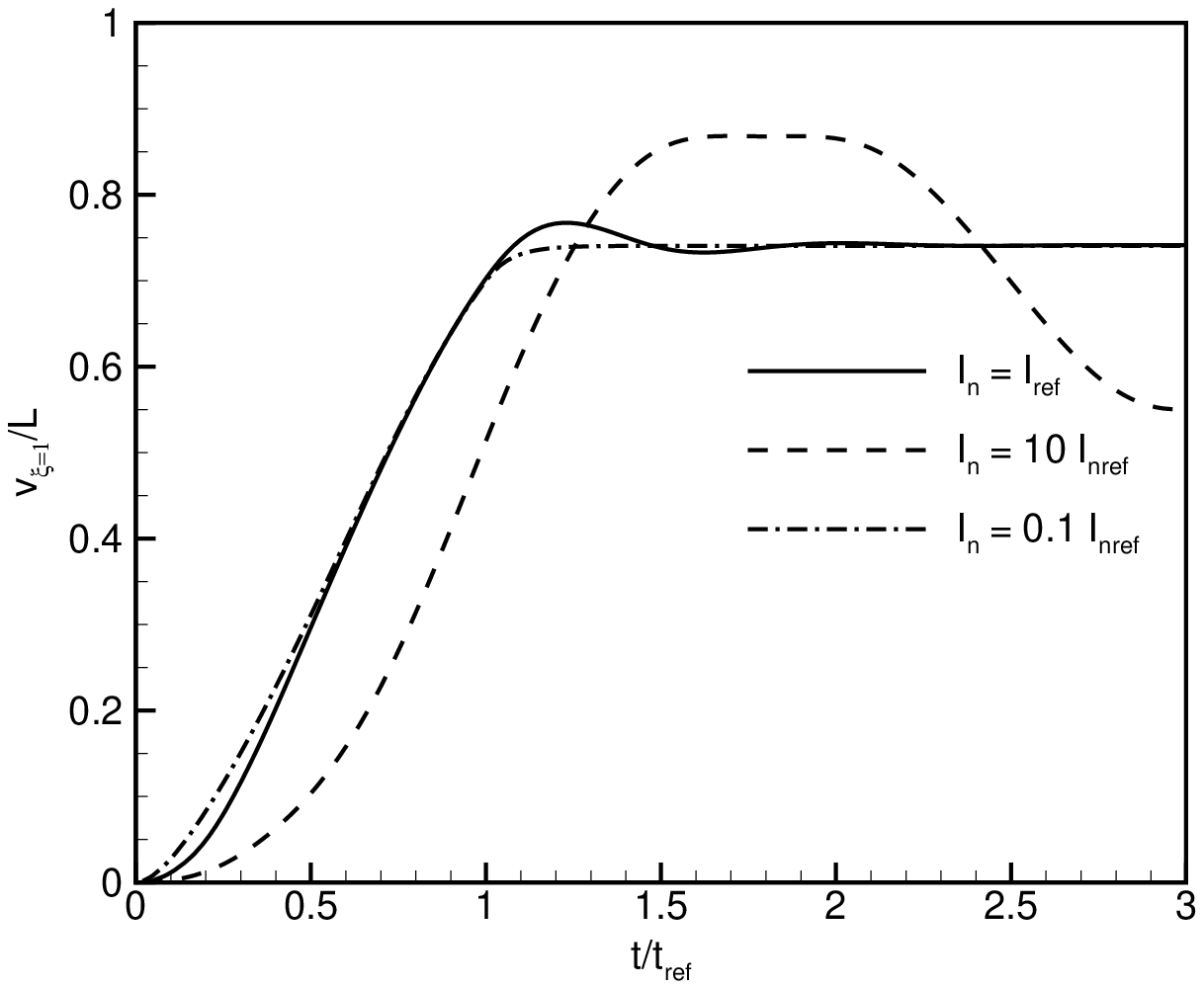}\label{fig:mag_nond_m}}
     \subfigure[]{\includegraphics[width =\dimanafundwidth]{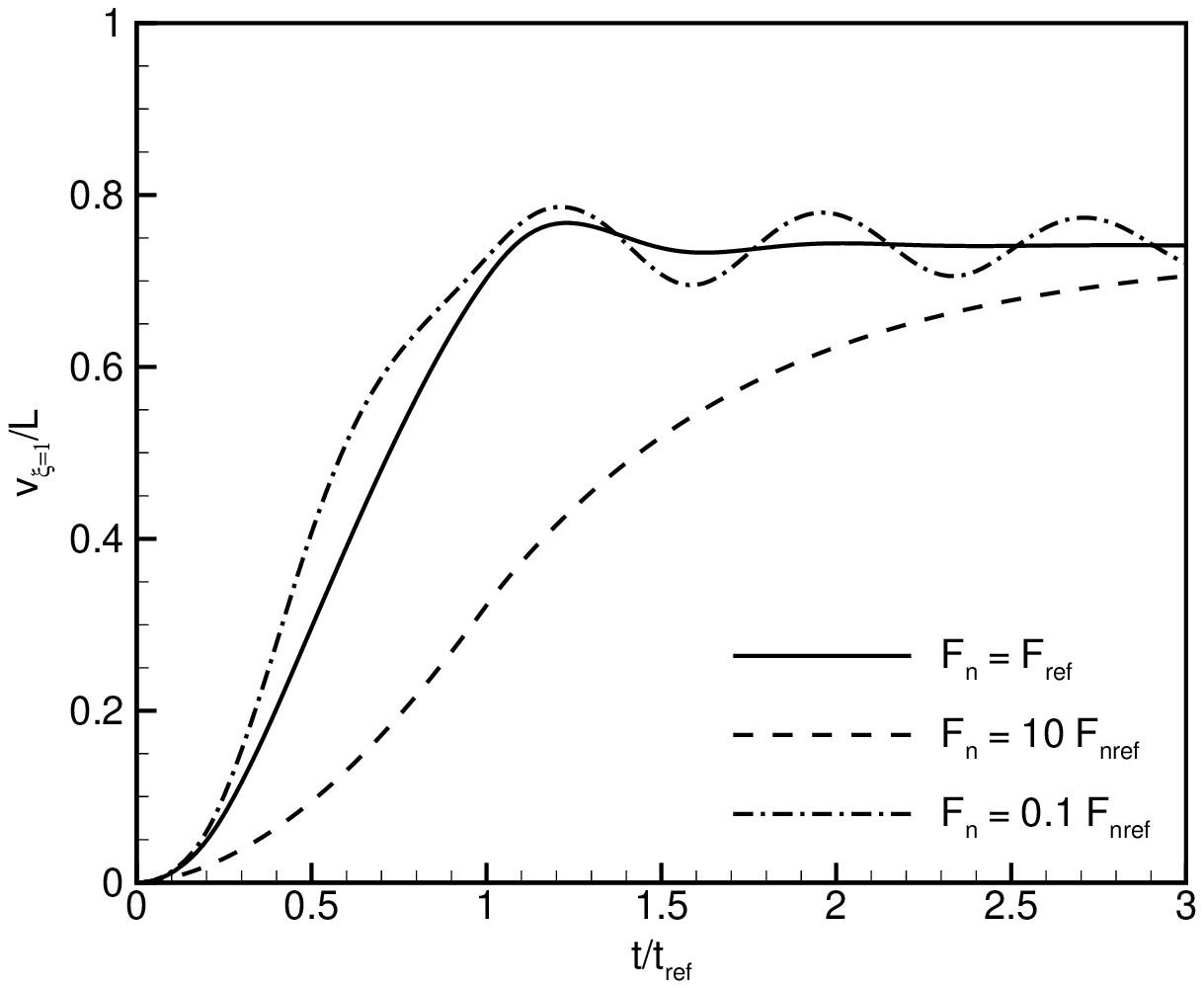}\label{fig:mag_nond_f}}
     \subfigure[]{\includegraphics[width =\dimanafundwidth]{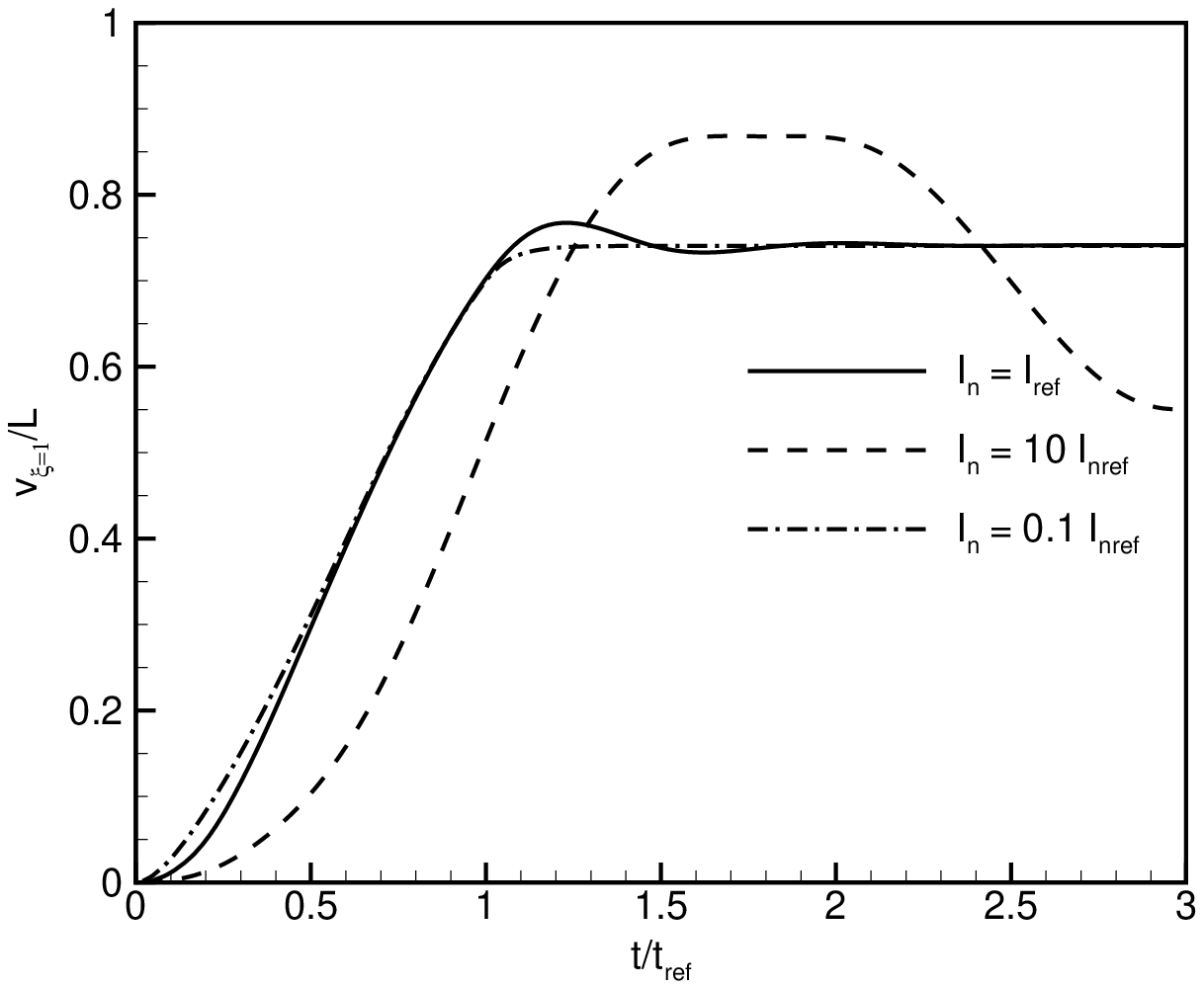}\label{fig:mag_nond_r}}
     \caption{Transverse tip displacement of a permanently magnetic film with variation of non-dimensional numbers. (a) $M_n$, (b) $F_n$, (c) $I_n$,}
     \label{fig:mag_nond}
\end{figure}

Fig.~\ref{fig:mag_nond} shows the normalised tip displacement as a function of time for permanently magnetic  film subject to a rotating field, as studied in Section \ref{sec:fundamental_loading_situations}.

The transverse displacement is plotted with varying each of the non-dimensional number keeping others constant. The reference cases ($M_n=M_{n\text{ref}}$, $I_n=I_{n\text{ref}}$ and $F_n=F_{n\text{ref}}$) are the same as those shown in the fundamental loading situations.

\begin{figure}
     \centering
     \subfigure[]{\includegraphics[width =\dimanafundwidth]{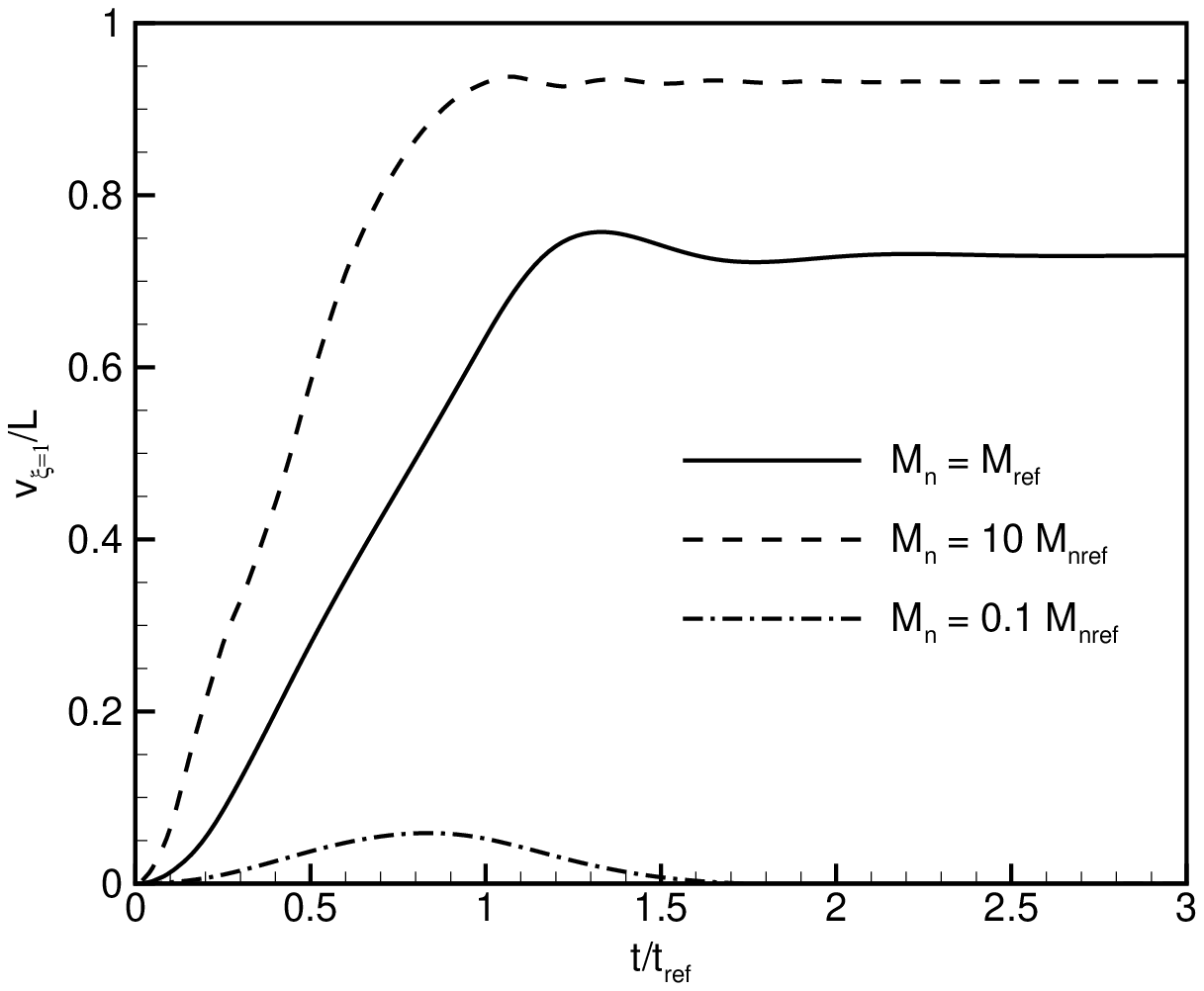}\label{fig:spm_nond_m}}
     \subfigure[]{\includegraphics[width =\dimanafundwidth]{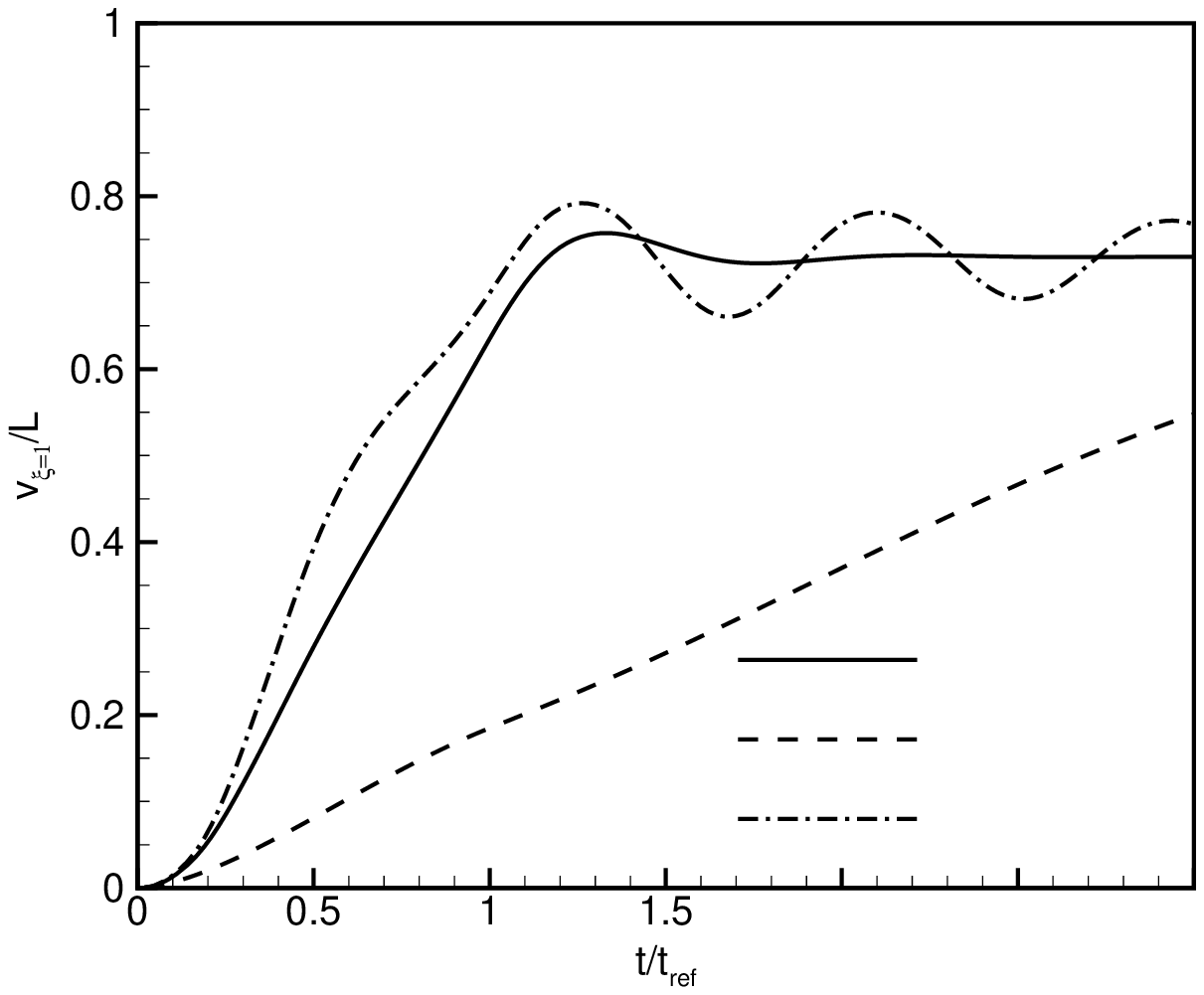}\label{fig:spm_nond_f}}
     \subfigure[]{\includegraphics[width =\dimanafundwidth]{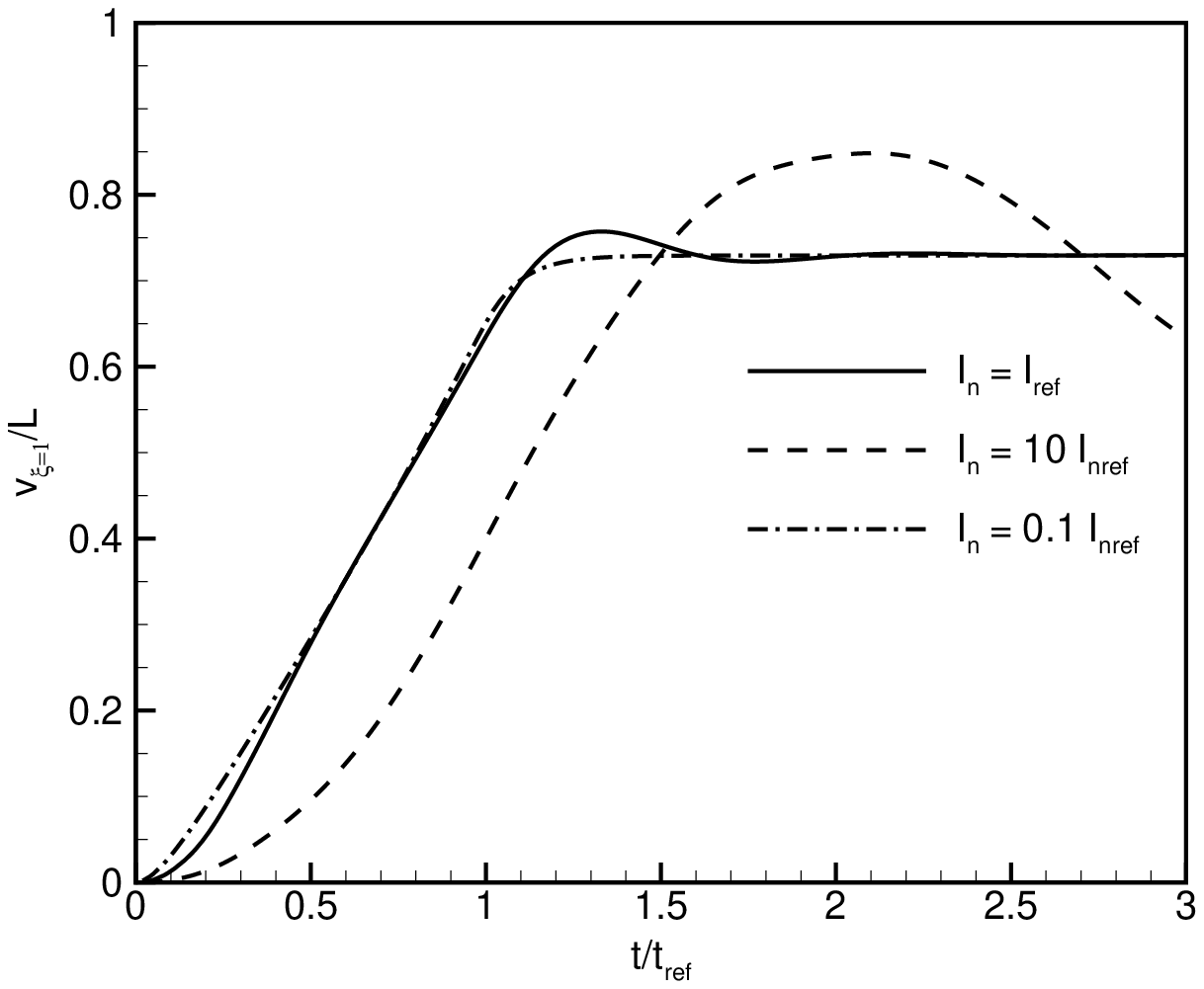}\label{fig:spm_nond_r}}
     \caption{Transverse tip displacement of a super-paramagnetic film with variation of non-dimensional numbers. (a) $M_n$, (b) $F_n$, (c) $I_n$,}
     \label{fig:spm_nond}
\end{figure}

\begin{figure}
     \centering
     \includegraphics[width =\dimanafundwidth]{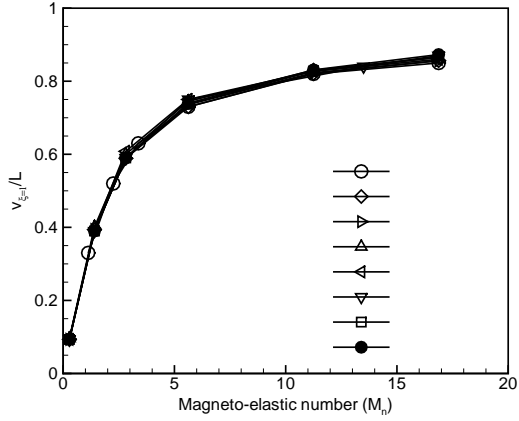}
     \caption{Variation of transverse displacement of a permanently magnetic film  with magneto-elastic number for different systems  identified by the symbols shown in Fig.~\ref{fig:contour_amp_1mn}.}
     \label{fig:funda_disp_magno_vary_in_fn}
\end{figure}

It is to be noted from Figs.~\ref{fig:mag_nond}, \ref{fig:funda_disp_magno_vary_in_fn} and \ref{fig:spm_nond}, that the steady state of the film deformation is principally governed by  the ratio of magnetic to elastic forces. With increase of the magneto-elastic number the film deformation increases (Figs.~\ref{fig:mag_nond_m}, \ref{fig:spm_nond_m}). In the case of SPM, for low magneto-elastic number the elastic forces are much larger than the magnetic forces. Hence  the film fails to deform after a certain extent and returns to its initial position.

Changing the fluid  and inertia numbers changes the dynamic properties of the system ( damping and time period). Increasing the fluid number changes the system from under-damped to an over-damped state. Increasing the inertia number changes the time period of oscillation by increasing its period and increases the amplitude of oscillation. When $I_n$ is small ($I=0.1I_\text{ref}$), the film shows a near quasi-static deformation behavior.

The increase of the inertia number makes the film heavier, hence the film moves slowly, showing an increase in its time period. As the film has higher inertia number, the inertia forces will be high, due to which the film undergoes large amplitudes oscillation about the steady state. Another way to look at it is, increase of the inertia number decreases the elastic forces ($I_n=$Inertia force/elastic force) resisting the deformation of the film, thus making the film floppy, hence its amplitude of oscillation about the steady state is large. The increase of fluid number increases the drag force, hence the deformation of the film in the transient stage is less, thus taking a long time for the film to reach the steady state. These two facts (larger inertia number leads to large deformation and large fluid number leads to less deformation) will be of help in understanding the behavior of the asymmetric configurations.

\begin{figure}
     \centering
     \subfigure{\includegraphics[width =\dimanafundwidth]{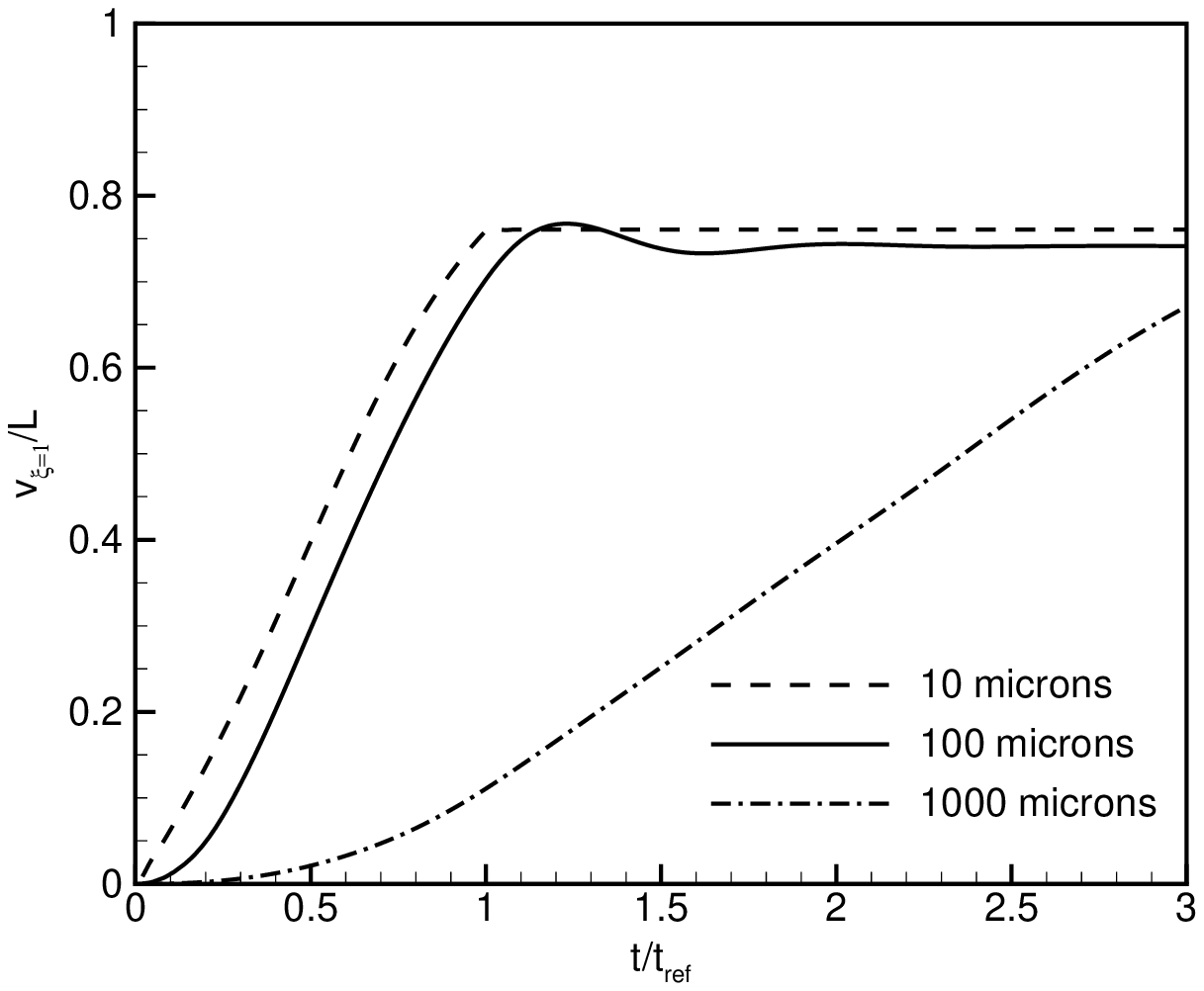}\label{fig:same_ratio_change_magnitude_lh_mag}}
     \subfigure{\includegraphics[width =\dimanafundwidth]{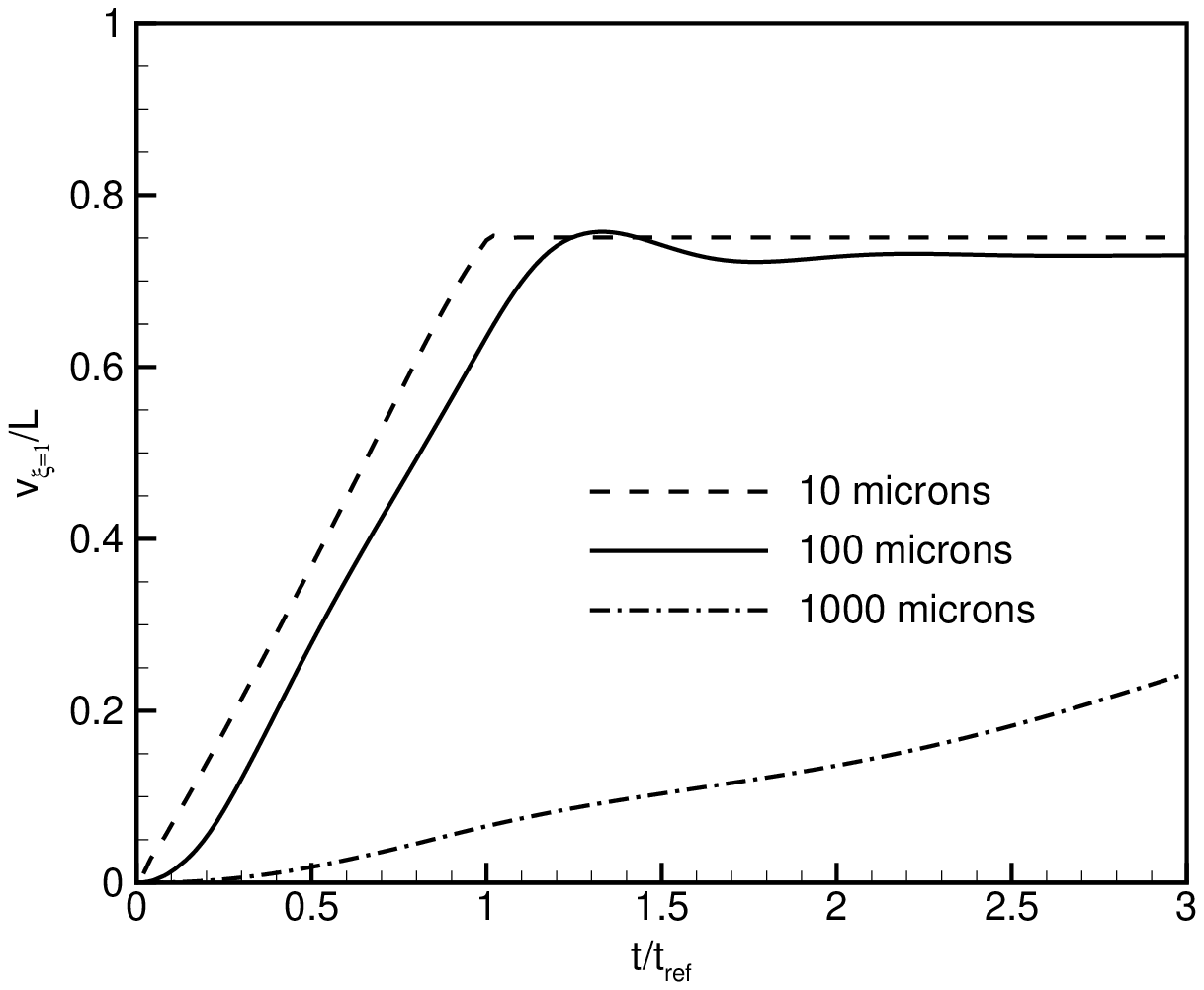}\label{fig:same_ratio_change_magnitude_lh_spm}}
     \caption{Transverse tip displacement of film, with same $L/h$ ratio while their magnitudes are varied. (a) Permanently magnetic film, (b) Super-paramagnetic film.}
     \label{fig:same_ratio_change_magnitude_lh}
\end{figure}

As can be seen in Figs.~ \ref{fig:mag_nond}, \ref{fig:spm_nond} and \ref{fig:same_ratio_change_magnitude_lh}, three modes of deformation can be identified. The under-damped motion of the film, where the film oscillates around the steady state before reaching it,  the over-damped motion, where the film takes a long time to reach the steady state and the quasi-static motion of the film where the dynamic effects are not seen. It is seen from Figs.~\ref{fig:mag_nond_m} and  \ref{fig:spm_nond_m} the amplitude of film deformation in the steady state is governed by the magneto-elastic number ($M_n$). The inertia number and the fluid number determine the way in which the steady state is reached (under-damped, over-damped or quasi-static).  

The motion of the film can be quantified with the steady state deflection, the time to reach the steady state and the amplitude of oscillation. The latter two are independent of the steady state deflection and can be used to quantify the modes of deformation according to Table \ref{tab:def_mode_table}.

\begin{figure}
     \centering
     \subfigure[]{\includegraphics[width =\dimanafundwidth ]{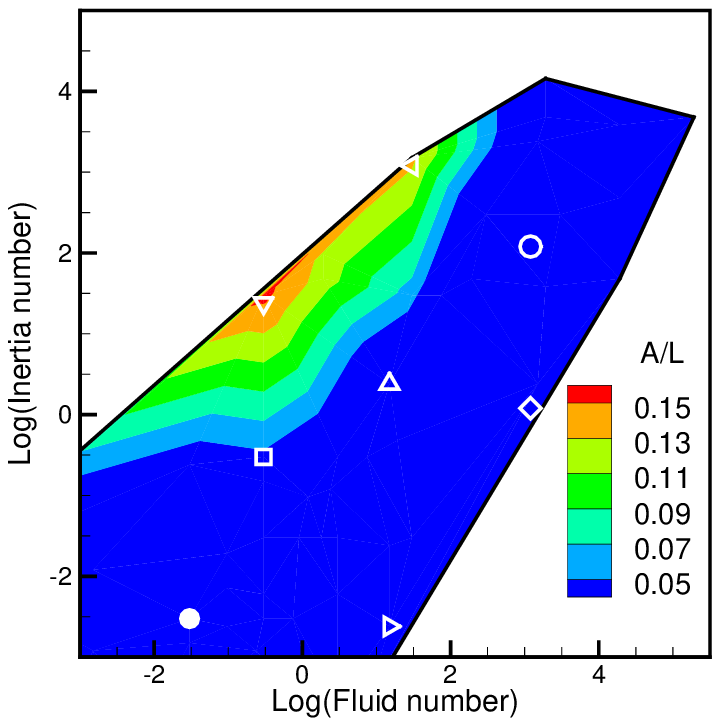}\label{fig:contour_amp_1mn}}
     \subfigure[]{\includegraphics[width =\dimanafundwidth ]{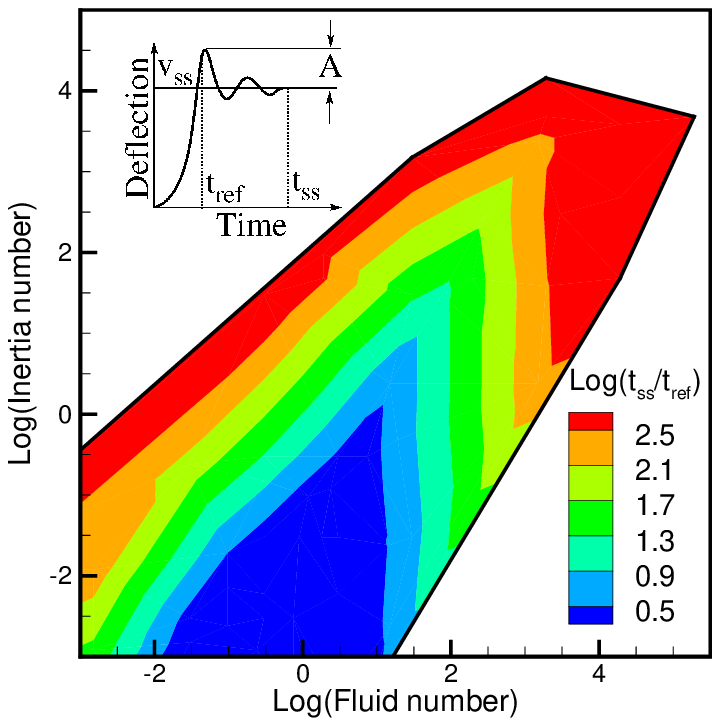}\label{fig:contour_tss_1mn}}
     \caption{$M_n$=0.4688. \subref{fig:contour_amp_1mn} Contours of amplitude, \subref{fig:contour_tss_1mn} Contours of amplitude and time to reach steady state.   }
     \label{fig:contour_amp_tss}
\end{figure}

We now extend the range of fluid and inertia number so as to include all possible materials, fluids and geometries. This is done for the case of permanently magnetic materials. The actuator properties chosen cover the length of the film from $100$ nm to $1$ mm, film densities from $1000$ to $8000$ kg$/$m$^3$, elastic moduli from 1 MPa to 200 GPa and fluid viscosities from one tenth to ten times that of water. As a result the fluid and and inertia number are scanned by 8 orders of magnitude, for a given magneto-elastic number. For a given magneto-elastic number the time to reach the steady state ($t_{ss}$) and the amplitude of oscillation are plotted against varying inertia and fluid number in Fig. \ref{fig:contour_amp_tss}.

From the contour plots of the amplitude of oscillation, the region in which the film shows oscillatory behavior before it reaches a steady state can be identified. In this region, the plots of the time to reach steady state  show that the film reaches the steady state after the field has stopped rotating. This under-damed  behavior occurs  for large inertia number  and low fluid number.

From the contour plots of the time to reach the steady state, a region can be identified when the film reaches the steady state as quickly as possible with no oscillation. Absence of oscillation implies that the inertia effects are negligible and because the film reaches the steady state as quickly as possible, the time lag due to viscous effects are absent and hence the viscous forces are negligible. The film deforms because of the competition between elastic and magnetic forces and the dynamic effects are negligible. Such a deformation behavior is named as "quasi-static". Quasi-static behavior is observed when the inertia and fluid numbers are small. 

From the contour plots of amplitude of oscillation and the time to reach the steady state, another region can be identified in which the film shows no oscillatory behavior, but reaches the steady state after the field has stopped rotating. As the film exhibits no oscillations, the inertia effects are negligible in this region. The fluid number in this region leads to large viscous forces, which make the film to deform in an over-damped manner.

The transitions between the zones happens as follows. If the system is initially in the over-damped region, reducing the fluid number leads to the following consequences: At large inertia numbers,  the system will start to show oscillations about the steady state and enter the under-damped region. At low inertia numbers, the  $t_\text{ss}$ gets reduced and at  a certain fluid number the film reaches steady state at $t_\text{ref}$. Further reduction of the fluid number does not change the $t_\text{ss}$ and brings the system in to the quasi-static region.

Decrease of the inertia number also leads to two kind of transitions: At large fluid numbers, it reduces the amplitude of oscillation, and, as the viscous effects are large for large fluid numbers, enables the system to enter the over-damped region, where no oscillations are observed. At low fluid numbers, the decrease of inertia number decreases the amplitude of oscillation, and, as the viscous effects are small, the system is shifted to the quasi-static region, where the dynamic effects are negligible.	

\begin{table}\centering
\caption{Deformation modes quantified}
\begin{tabular}{|c|c|c|}
        \hline
         Deformation mode  & time to reach steady state $t_\text{ss}$ & amplitude \\
        \hline
          under-damped & $>t_\text{ref}$ & $>0$ \\
          over-damped & $>t_\text{ref}$ & $0$ \\
          quasi-static & $=t_\text{ref}$ & $0$ \\
\hline
\end{tabular}\\
\label{tab:def_mode_table}
\end{table}

\begin{figure}
     \centering
      \includegraphics[width =\dimanafundwidth]{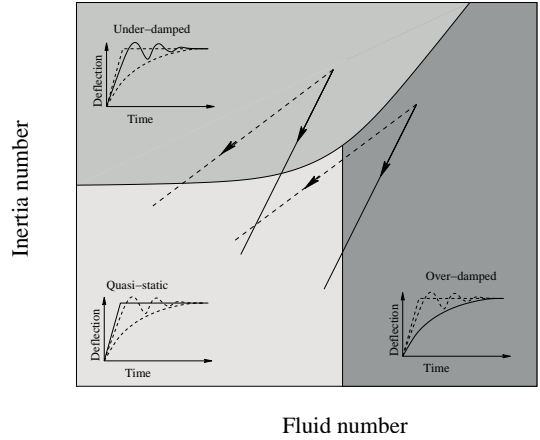}
     \caption{Schematic representation of the deformation behavior of the film at different inertia and fluid numbers. Systems along the solid lines have constant shape and the systems along the dashed lines have constant size.}
     \label{fig:deformation_modes_schematic}
\end{figure}

The deformation modes can be nicely summarized in terms of the dimensionless numbers as shown in the schematic picture shown in Fig.~\ref{fig:deformation_modes_schematic}. The transition from over-damped to quasi-static region is obtained from the plots of steady state time (Fig.~\ref{fig:contour_amp_tss}). The transition from the under-damped to the other two regions is obtained from the plots  of amplitude of oscillation. For a given system, this diagram can quantify its  the behavior at any magneto-elasic number.

The effect of size is investigated now. Keeping the ratio $L/h$ fixed, their magnitudes are varied and the corresponding displacements are plotted for permanently magnetic and SPM case (Fig.~\ref{fig:same_ratio_change_magnitude_lh}). With reference to the non-dimensional numbers, keeping the ratio $L/h$ fixed and varying their size does not change the magneto-elastic number, but has noticeable effect on the fluid number and the inertial number. When the size of the film is small, it shows a near quasi-static behavior, $L=10\mu\text{m}$ in Fig.~\ref{fig:same_ratio_change_magnitude_lh}. Which enables us to infer that, as the size of the film decreases the inertia  of the film can be neglected. This is similar to the fluid dynamic counterpart, where at small length scales the inertia and viscous effects are negligible.

Referring to the schematic diagram in Fig.~\ref{fig:deformation_modes_schematic}, the solid lines on the plot shows the effect of size (for a given modulus, density, fluid, characteristic time) on the actuator system. The solid lines are the lines of constant shape and as we decrease the size (move in the direction of the arrow), the fluid and inertia number are reduced. Thus the system moves towards the quasi-static region when the size of the actuator is reduced.  It can be shown that these lines of constant shape have a slope of 2.

The dashed lines on the schematic diagram represent the lines of constant length when the aspect ratio of the film is varied (for a given modulus, density, fluid, characteristic time). As  the aspect ratio of the film is decreased (move in the direction of the arrow) it becomes more stiff and the elastic forces in the film increase, thus reducing the inertia and fluid number. Thus, taking the film to the quasi-static region. It can be shown that these lines of constant size have a slope of $2/3$.


   \subsection{Asymmetric motion}\label{sec:configuration}
As mentioned before, the fluid propulsion in micro-channels takes place at low Reynolds number. The actuating member, therefore, should move in an asymmetric manner to  effectively propel fluid. In this Section we discuss several configurations that are able to do so. In all the configurations the applied magnetic field is uniform in space, but its magnitude and direction are varied in time. By tuning the applied field, the initial geometry of the film and its magnetic nature (permanently magnetic or super-paramagnetic), we have identified four configurations that mimic ciliary motion.  For the results presented, the thickness of the film is taken to be 2 $\mu$m and the elastic modulus to be $1$ MPa. The drag coefficients are calibrated against computational fluid dynamic (CFD) simulations, details of which will be discussed in   Section \ref{sec:drag_calib}.\\

\par
\begin{figure}[ht]\centering
     \subfigure[$t=0.2$ms]{\includegraphics[scale=0.18]{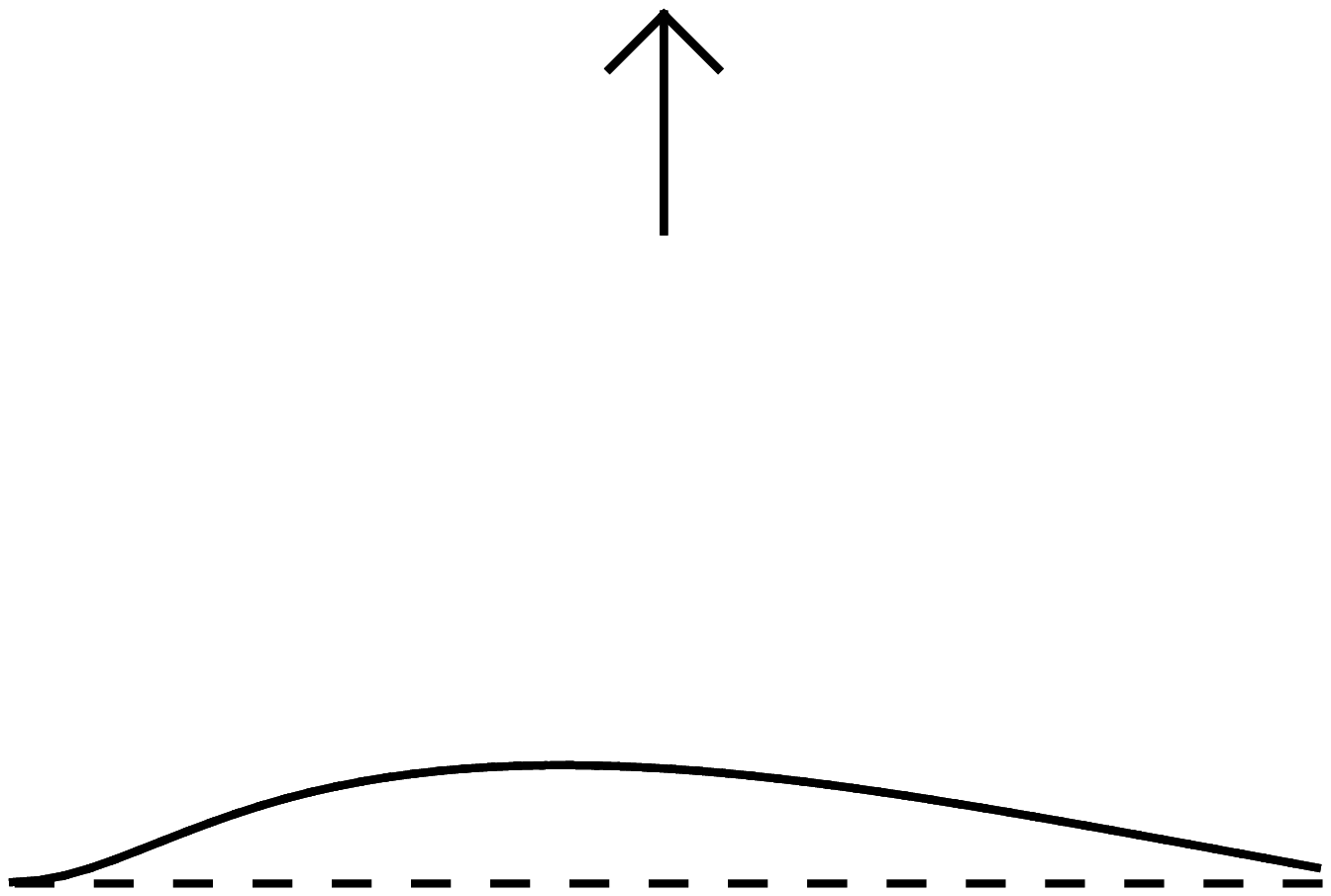}}
     \subfigure[$t=0.4$ms]{\includegraphics[scale=0.18]{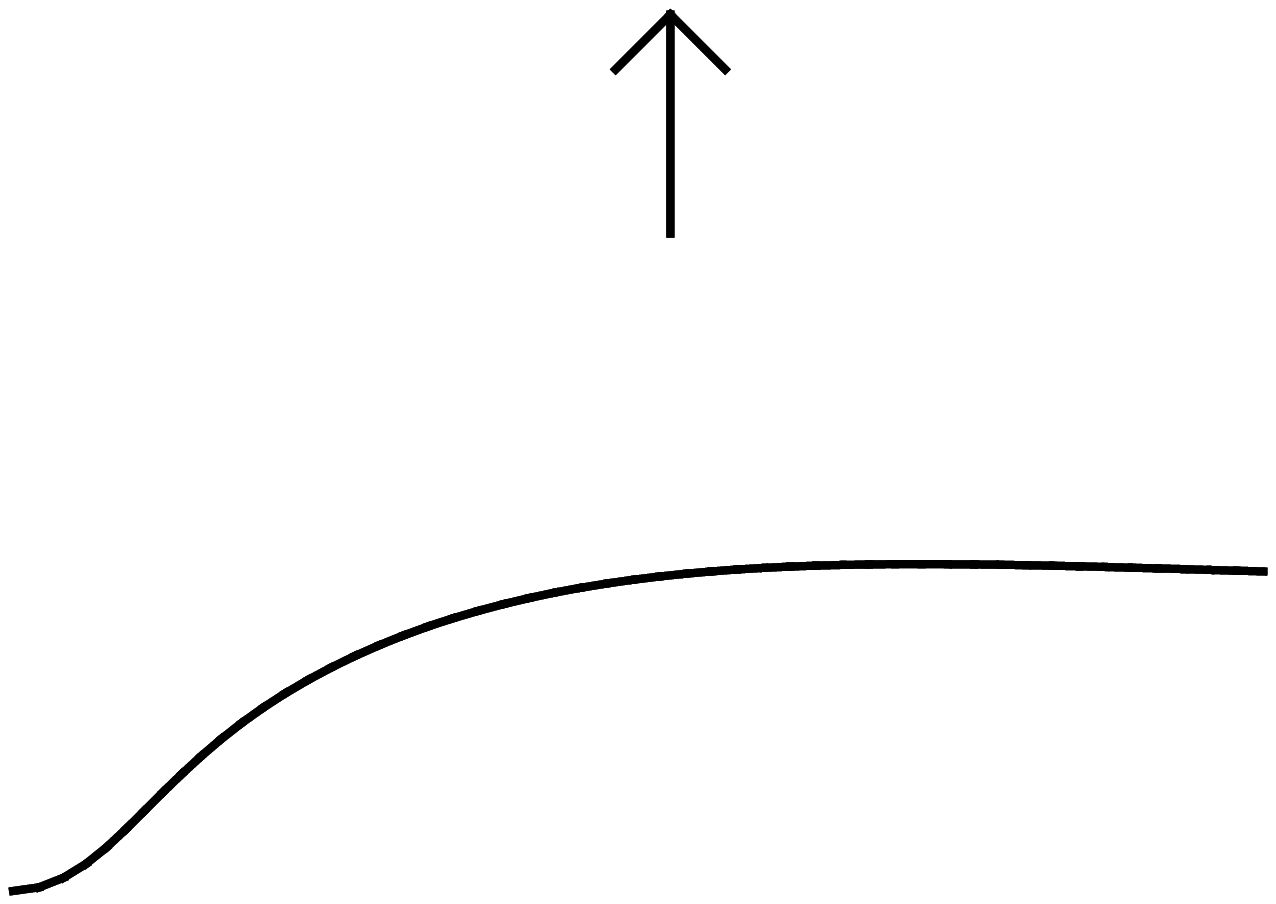}}
     \subfigure[$t=1.0$ms]{\includegraphics[scale=0.18]{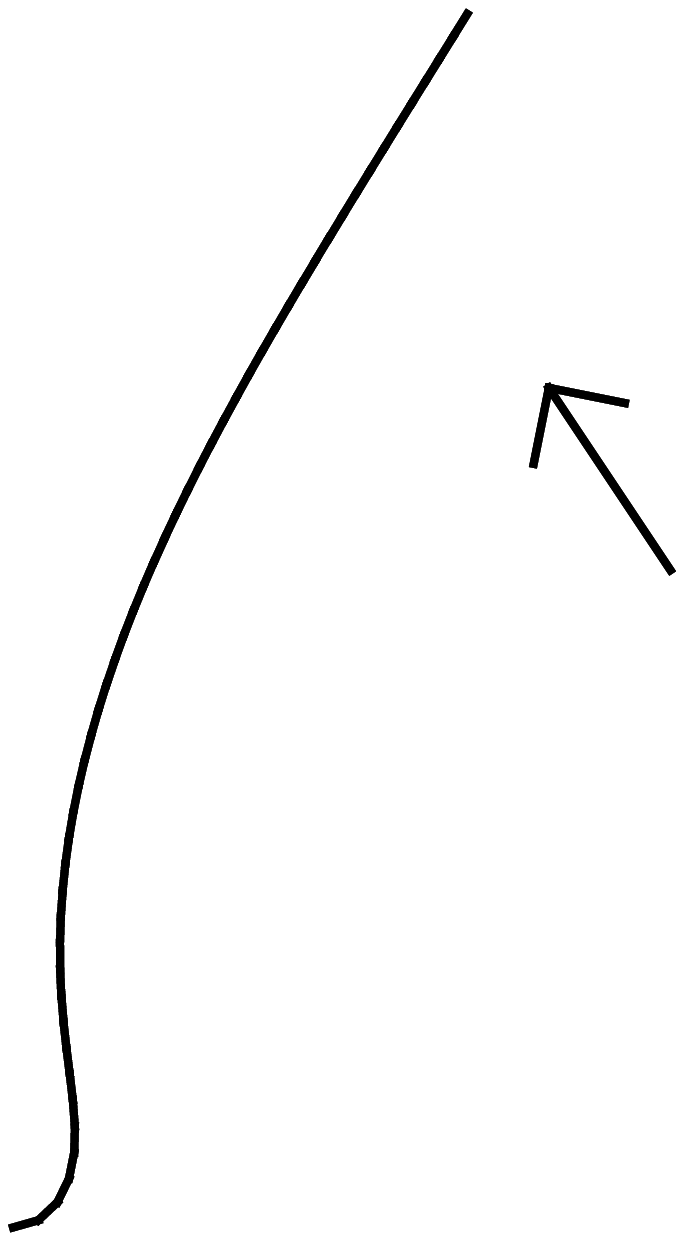}}
     \subfigure[$t=1.9$ms]{\includegraphics[scale=0.18]{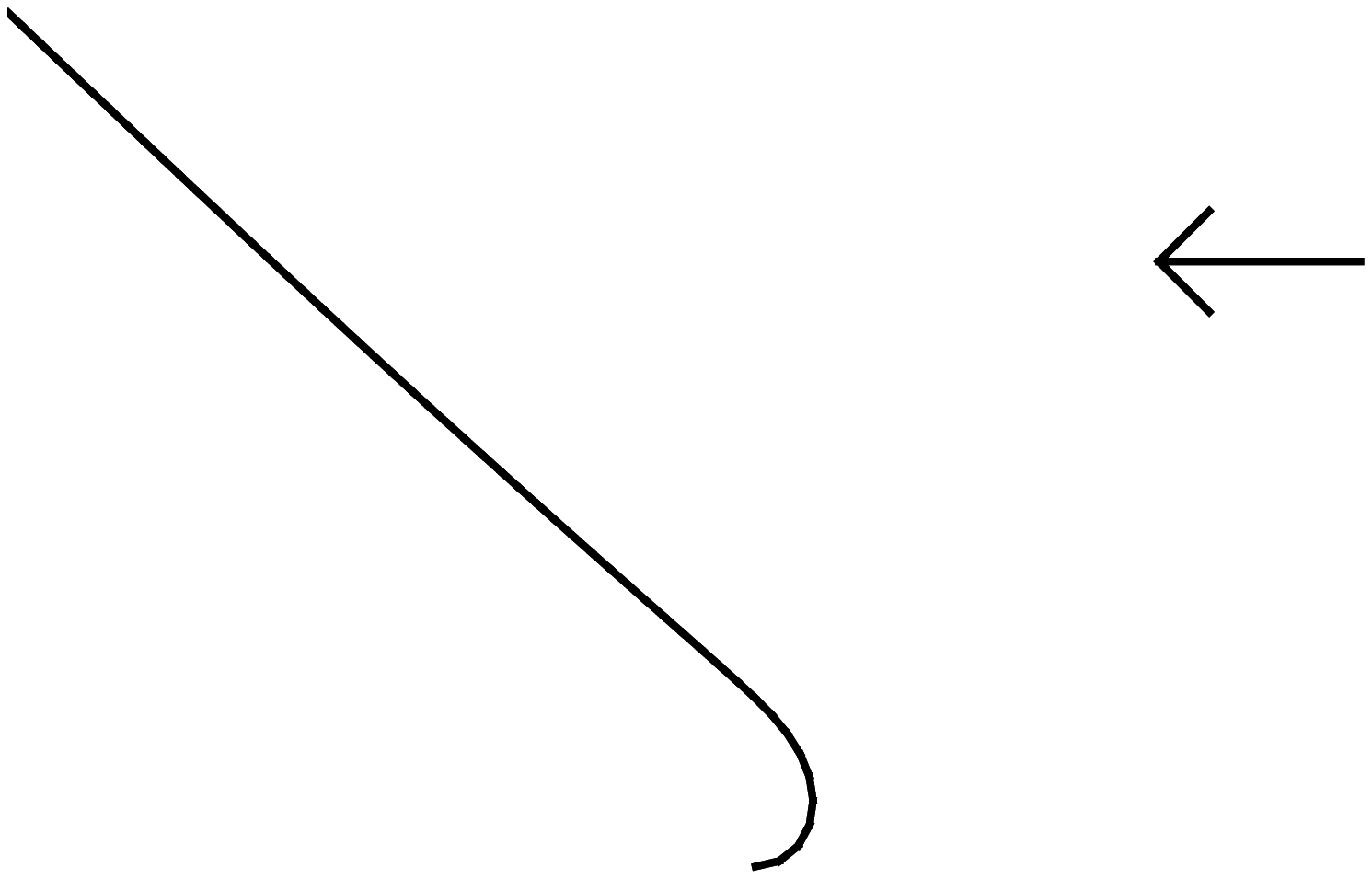}}
     \subfigure[$t=2.1$ms]{\label{fig:case0_switchoff}\includegraphics[scale=0.18]{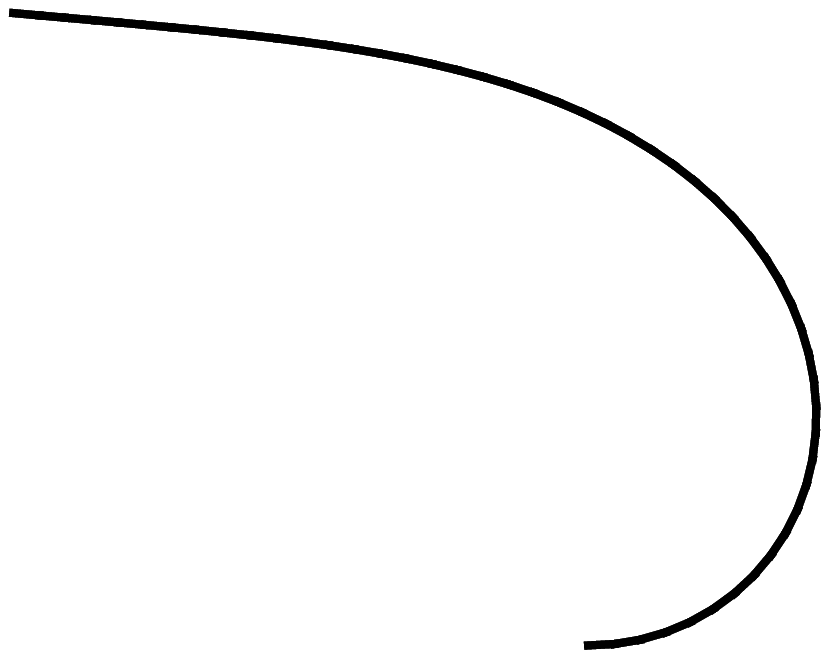}}
     \subfigure[$t=3.0$ms]{\includegraphics[scale=0.18]{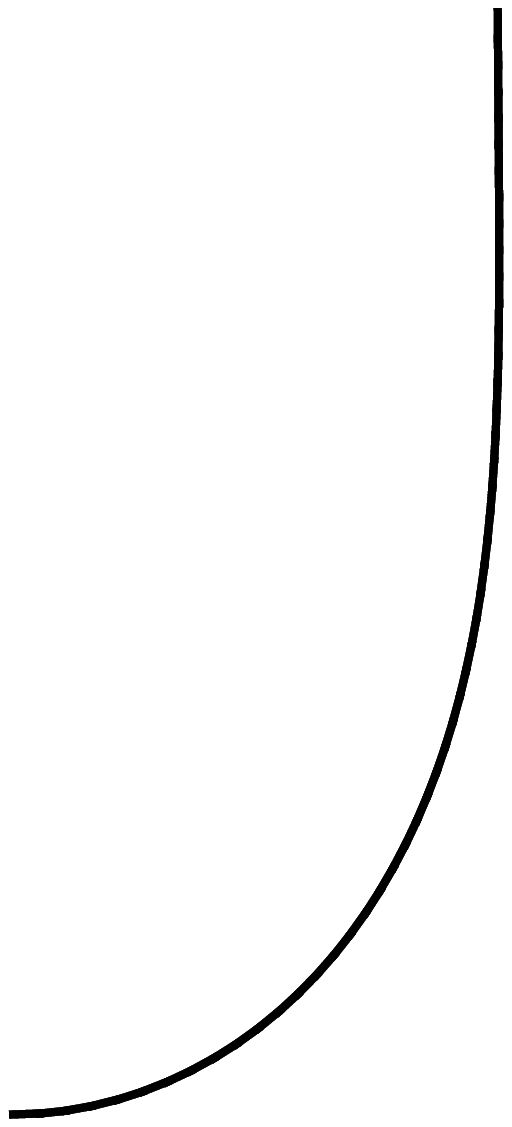}}
     \subfigure[$t=3.5$ms]{\includegraphics[scale=0.18]{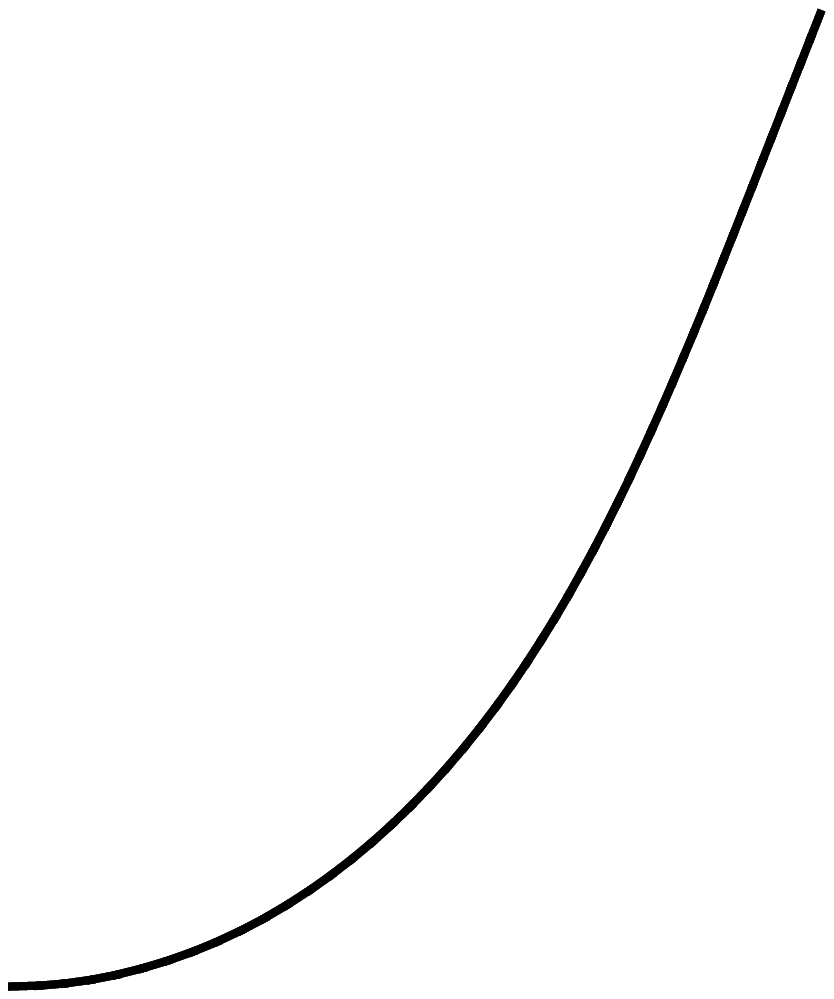}}
     \subfigure[$t=6.5$ms]{\includegraphics[scale=0.18]{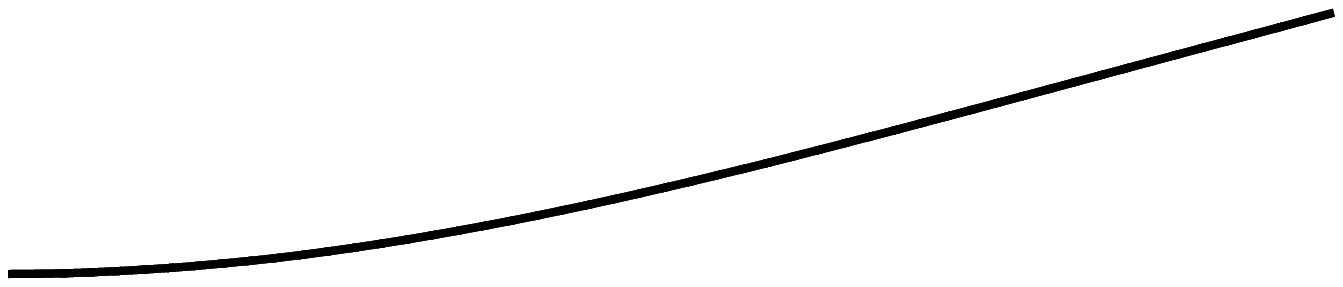}}
     \subfigure[Trajectory of the free end]{\includegraphics[scale=0.27]{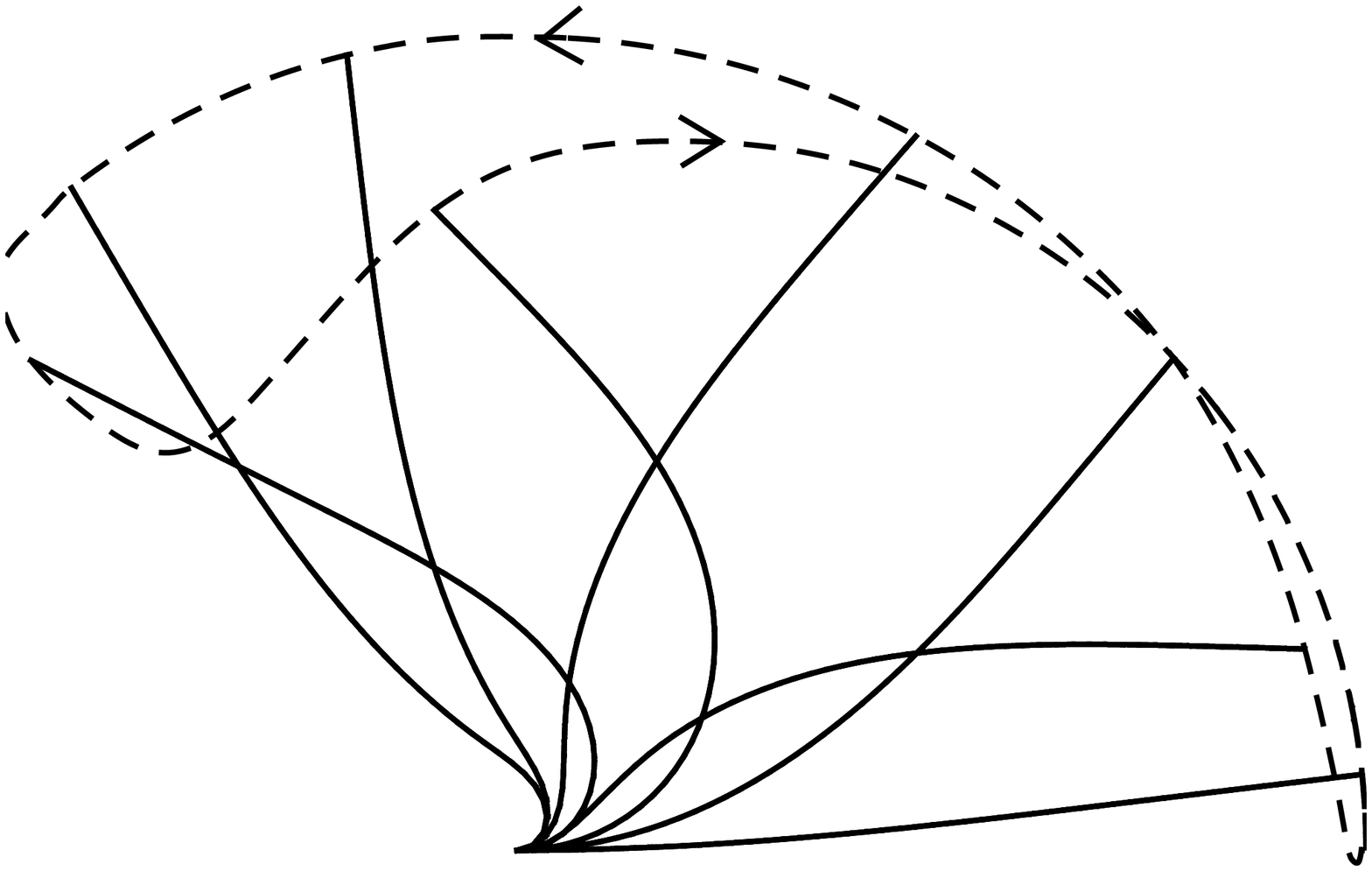}}

     \caption{Film with cracks with only a part ($20\%$) magnetized.  The dashed line shows the initial position of the film.  The arrow shows the direction of the applied field.}
     \label{fig:film_with_cracks}
\end{figure}

\noindent
 \textbf{1. Partly magnetic film with cracks.} The natural cilium is found to have a varying stiffness in the effective and recovery stroke \cite{gray_1922}. To use this concept we need to have the film to possess a large bending stiffness in the effective stroke while pushing  the fluid and to possess a low stiffness during the recovery stroke. This can be achieved by introducing cracks in one side of the film, while only a part of the film is magnetic. The film is straight initially, is attached  at the left end and has cracks of size 0.75 $\mu$m at the bottom. By magnetizing only a part of the film, the film is expected to behave like a flexible oar (as also mentioned by Purcell \cite{purcell}). Only $20\%$ of the film, the end near to the fixed part, is magnetic. The assumed remnant magnetization is $15$ kA/m, with the magnetization vector pointing from the fixed end to the free end. The drag coefficients used are $C_x = 30$ $\text{Ns/m}^3$ and $C_y = 60$ $\text{Ns/m}^3$.   The applied magnetic field is increased linearly to $145$ mT in the $y$ direction in $0.6 $ ms, then rotated by $90^\circ$ in the next $1.2$ ms and finally reduced to zero in the next $0.2$ ms. These system properties lead to a magneto-elastic number of $5.43$ or $1.08\ (0.2\times 5.43)$, a fluid number of $0.416$ and an inertia number of $0.012$.  The movement of the film under the action of the applied magnetic field is shown in Fig.~\ref{fig:film_with_cracks}.  When the external magnetic field is applied, the magnetic couples act on the magnetized portion of the film in a counter-clockwise manner, thus rotating the film about the fixed end. Now the drag forces are acting on the top part of the film, which close the cracks, making the film stiff.

When the applied field is switched off (Fig.~\ref{fig:case0_switchoff}), the film will recover elastically and  the drag forces act on the bottom part of the film which open the cracks making the film compliant. Such an interaction of magnetic couples, elastic forces and drag forces results in an asymmetric motion, as is clear from Fig.~\ref{fig:film_with_cracks}.\\

\par\noindent
\textbf{2. Buckling of a straight magnetic film.} A straight horizontal magnetic film with a slight perturbation  is used to get the desired asymmetric motion.  The film is assumed to  have a uniform magnetization with the magnetization vector pointing along the film length, from the fixed end at the left to the free end at the right. The remnant magnetization of the film is taken  to be $15$ kA/m. The length of the film is 100 $\mu$m. The drag coefficients used are $C_x = 30$ $\text{Ns/m}^3$ and $C_y = 60$ $\text{Ns/m}^3$. The  external field is applied as follows: A field of $30$ mT is applied in the negative $x$ direction from $t=0$  to $t=1$ ms and the field is reduced to zero in the next $0.2$ ms. These system properties lead to a magneto-elastic number of $1.125$, a fluid number of $0.75$ and an inertia number of $0.04$. Initially, the magnetization and the applied field are parallel, but with opposite sign, so that the magnetic couple is zero. However, with any perturbation of the film, the equilibrium state becomes unstable and the film will buckle away from the straight configuration. By assuming  a uniform magnetization in the film and neglecting drag forces, the critical field can be calculated (see Appendix \ref{sec:film_buckling}).
\begin{figure}[ht]\centering
     \subfigure[$t=0.6$ms]{\label{case1_0p6}\includegraphics[scale=0.18]{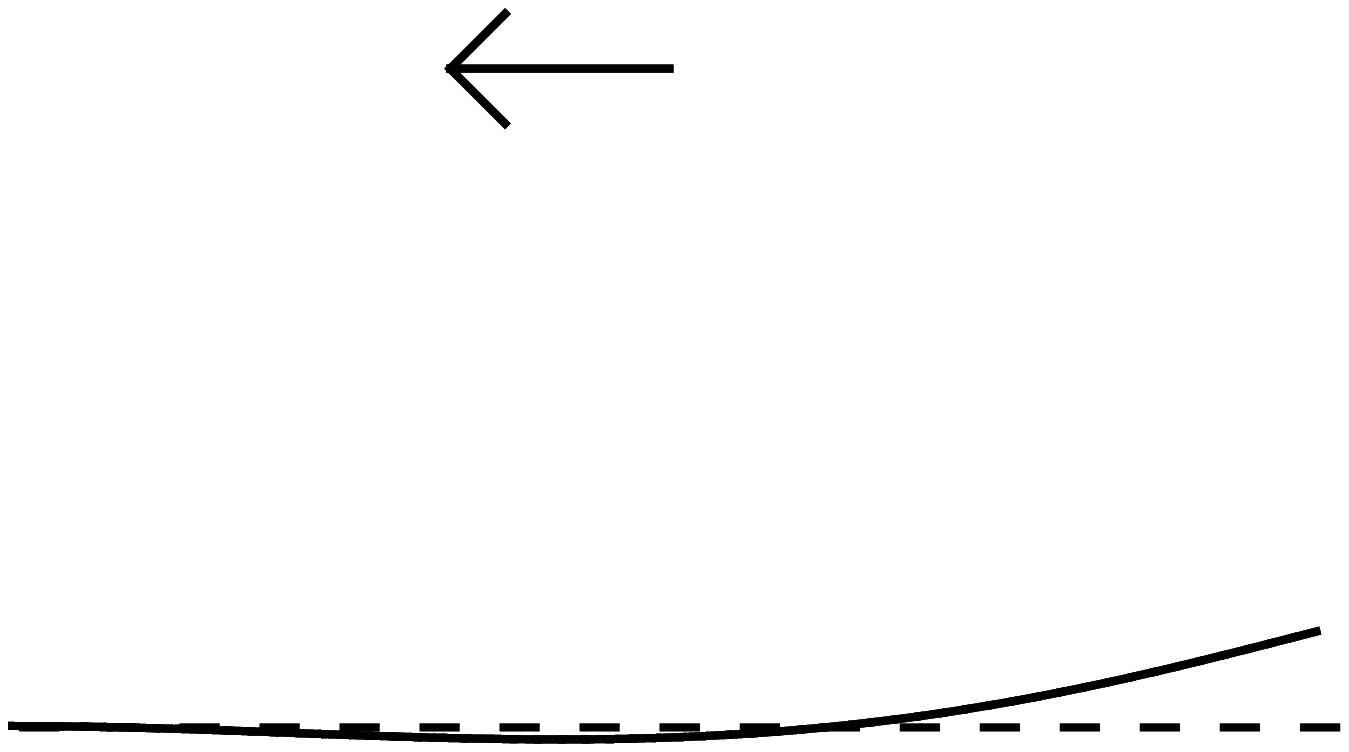}}
     \subfigure[$t=0.7$ms]{\includegraphics[scale=0.18]{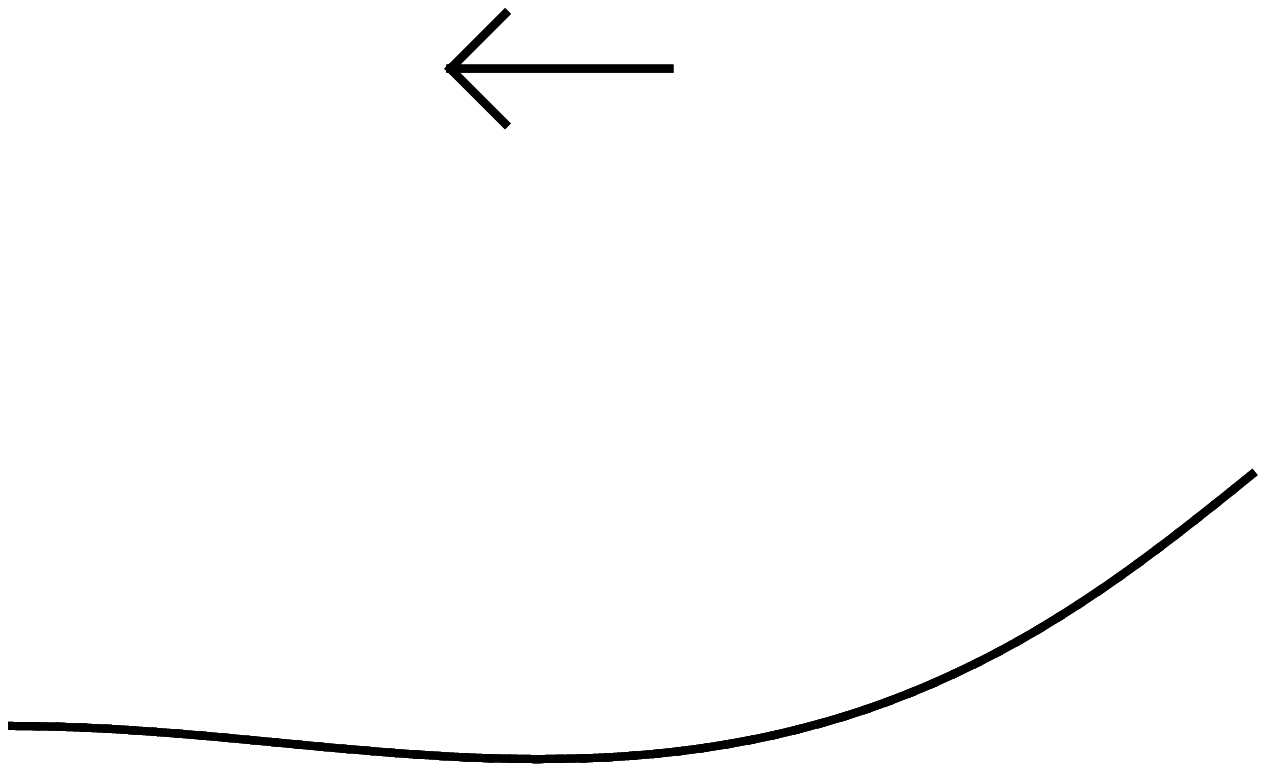}}
     \subfigure[$t=0.8$ms]{\includegraphics[scale=0.18]{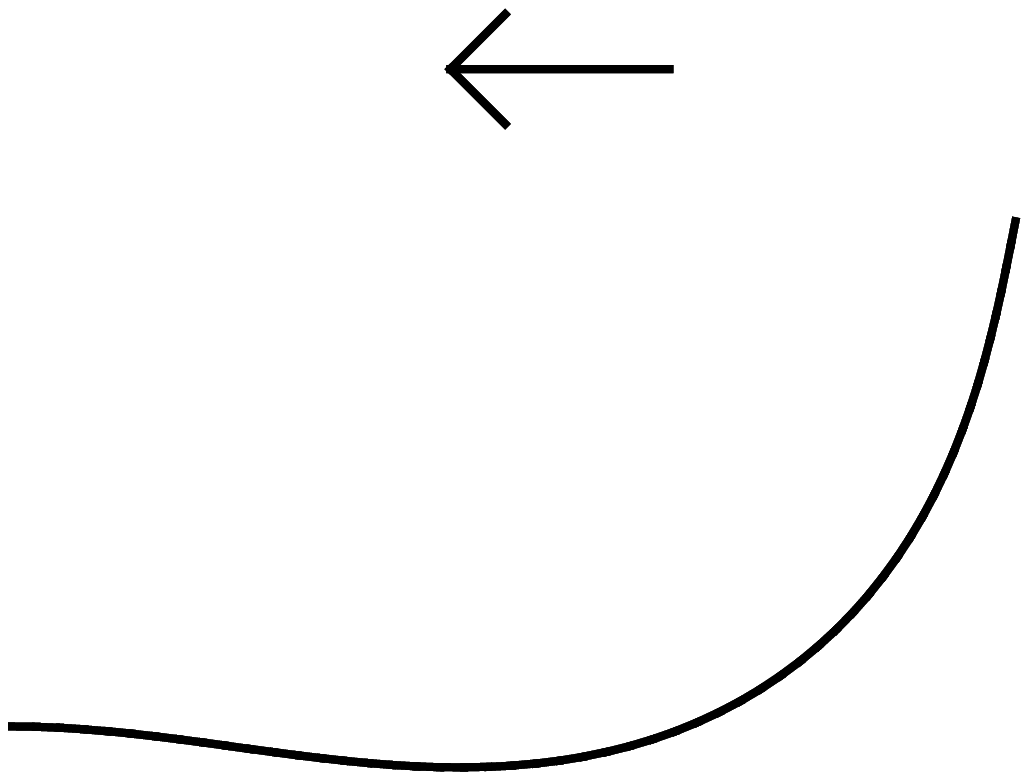}}
     \subfigure[$t=0.9$ms]{\label{case1_0p9}\includegraphics[scale=0.15]{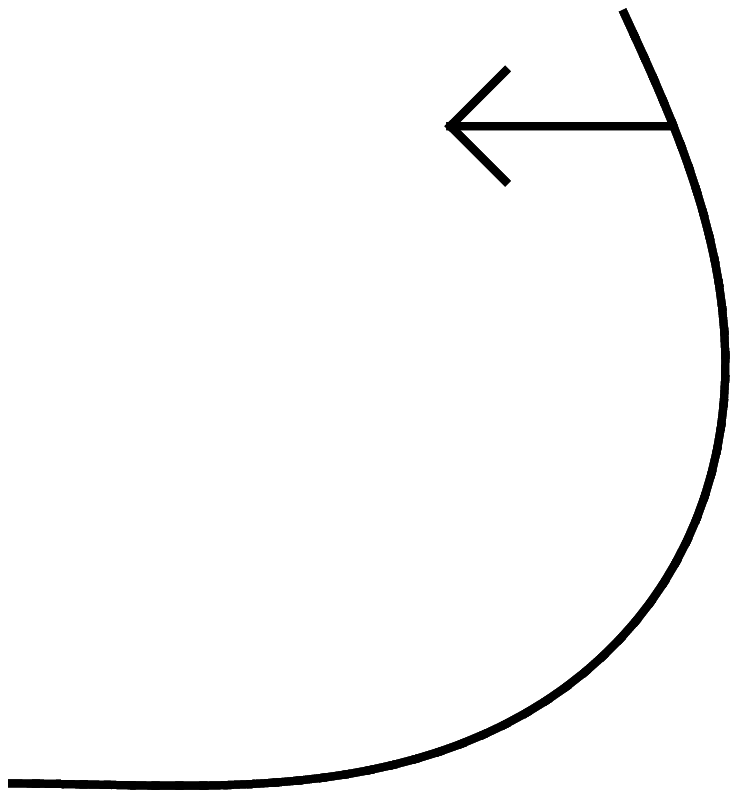}}
     \subfigure[$t=1.0$ms]{\includegraphics[scale=0.18]{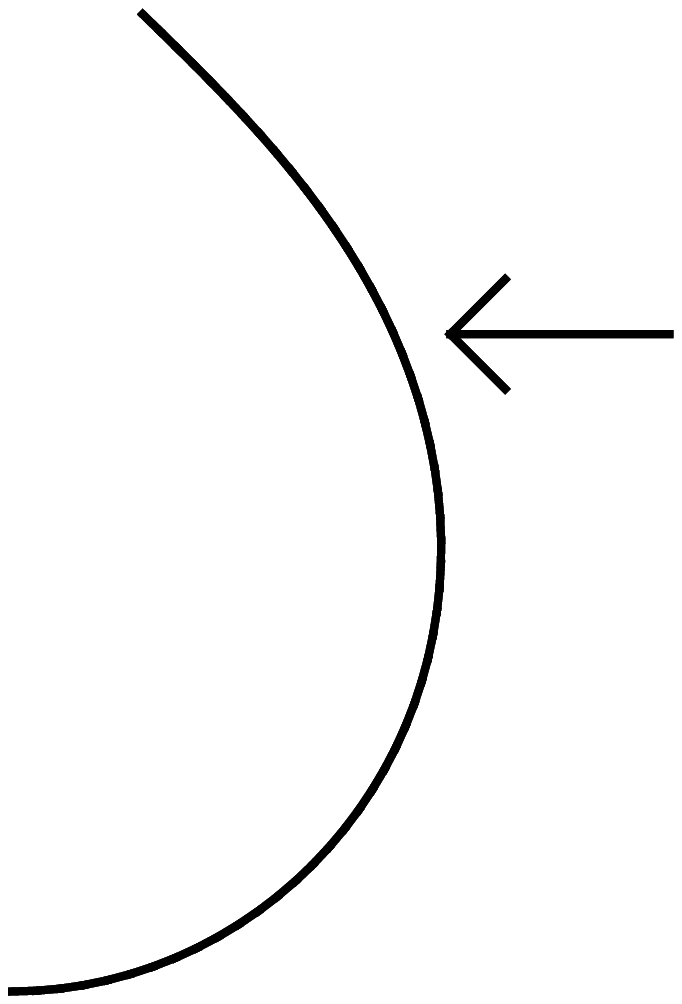}}
     \subfigure[$t=1.3$ms]{\includegraphics[scale=0.18]{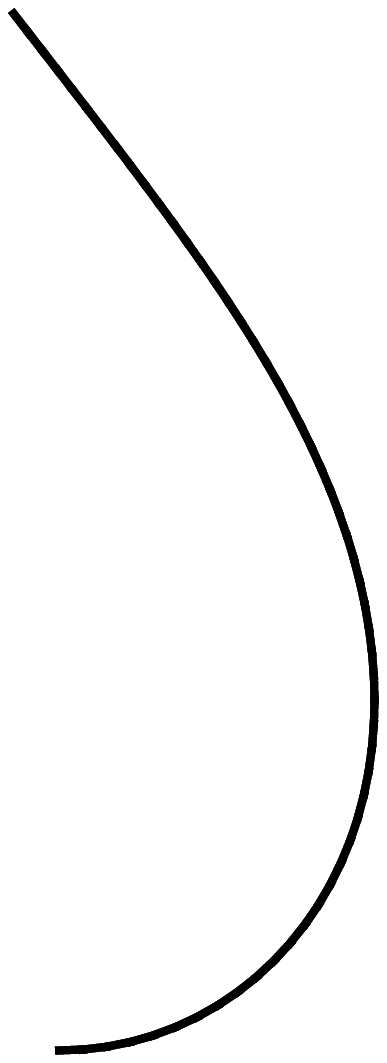}}
     \subfigure[$t=1.5$ms]{\includegraphics[scale=0.18]{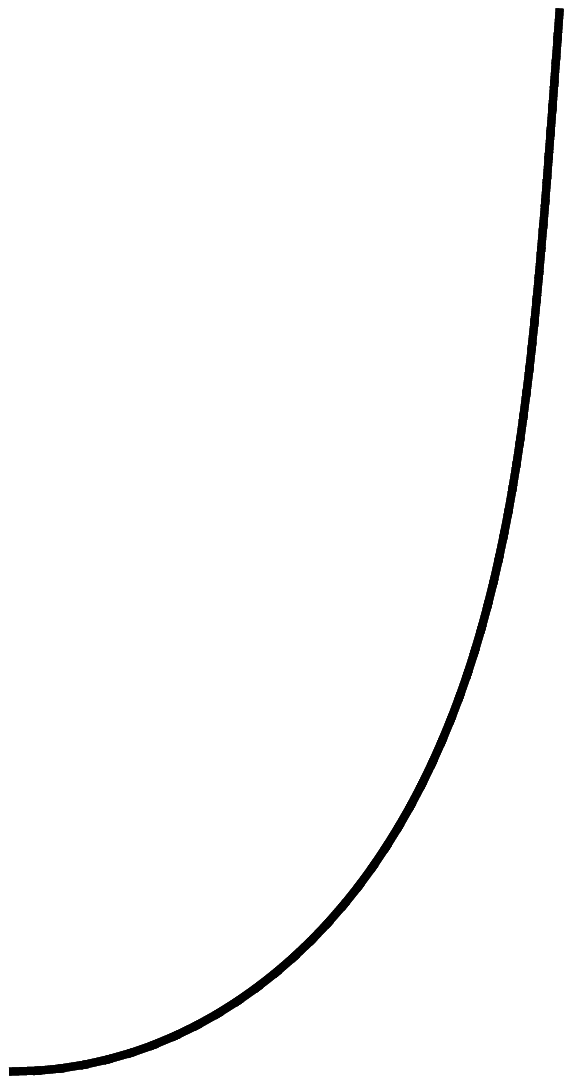}}
     \subfigure[$t=2.0$ms]{\includegraphics[scale=0.18]{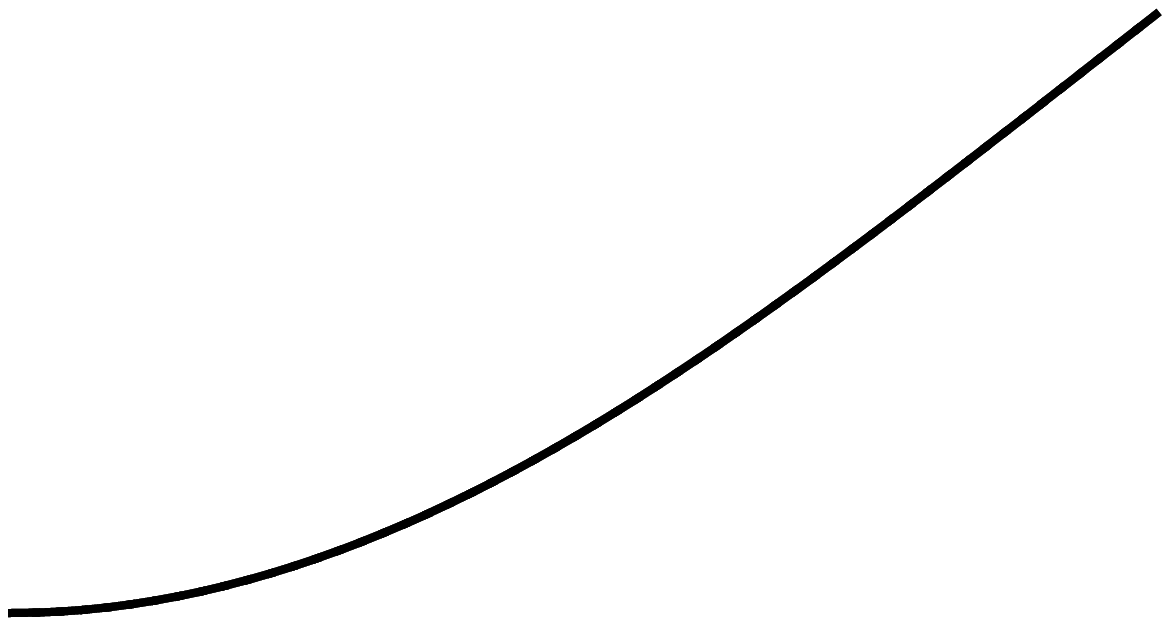}}
     \subfigure[Trajectory of the free end]{\label{fig:case1_trajectory}\includegraphics[scale=0.27]{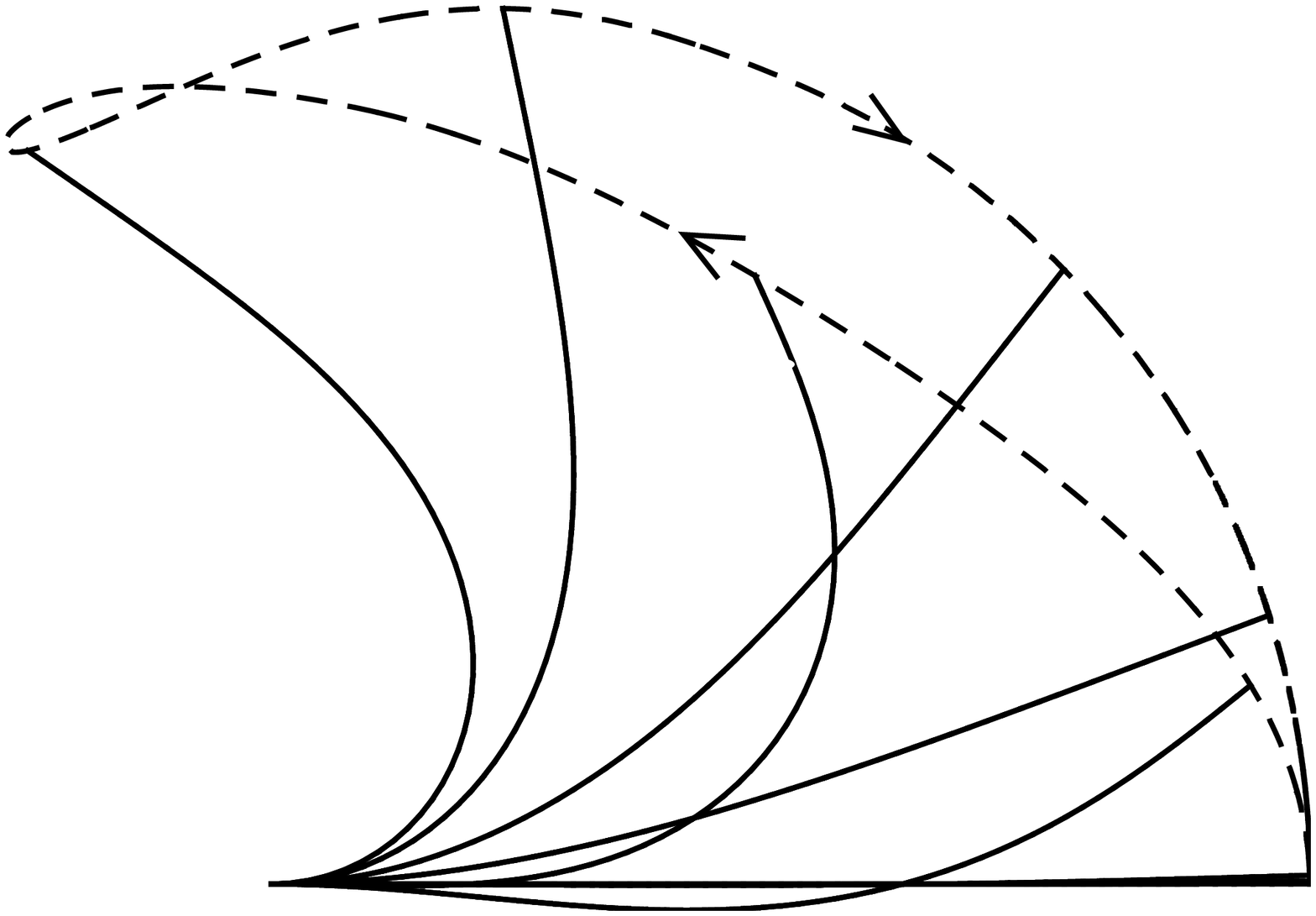}}

     \caption{Movement of a perturbed film with the applied magnetic field in opposite direction to the magnetization of the film.  The dashed line shows the initial position of the film. The arrow shows the direction of the applied field.}
\end{figure}
When it buckles the film tries to curl such that the magnetization in the film is aligned with the applied field (Fig.~\ref{case1_0p6}--\ref{case1_0p9}). When the applied field is removed, the film returns back to its initial position by elastic recovery. For this configuration,  the effective fluid propulsion will take place when the film recovers elastically.  Whether the film will buckle  up or  down depends critically on the sign of the initial imperfection (see Appendix \ref{sec:film_buckling}).
\\

\par
\begin{figure}[ht]\centering
     \subfigure[$t=0.0$ms]{\label{fig:case2_initial_geometry}\includegraphics[scale=0.15]{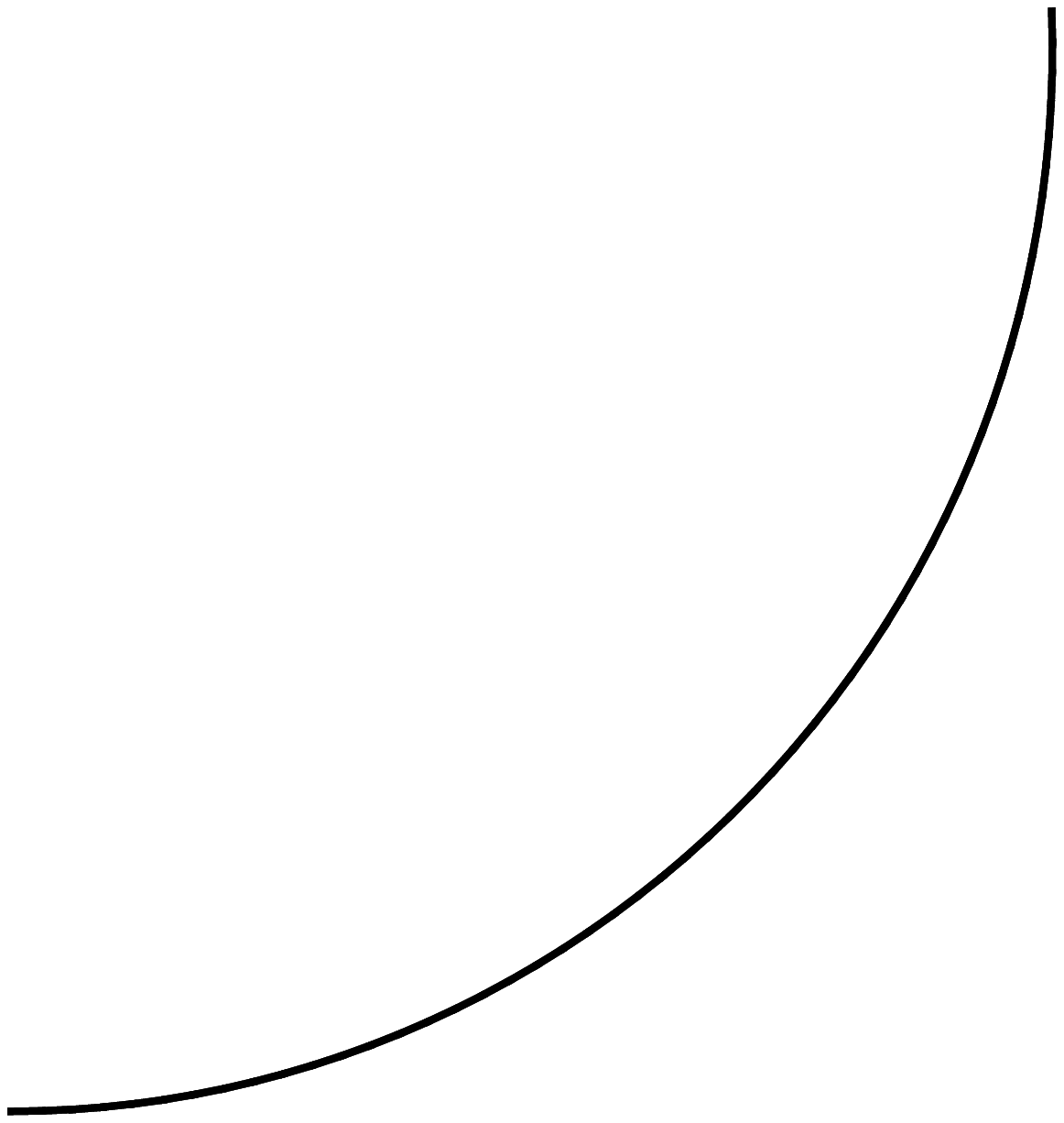}}
     \subfigure[$t=0.1$ms]{\includegraphics[scale=0.18]{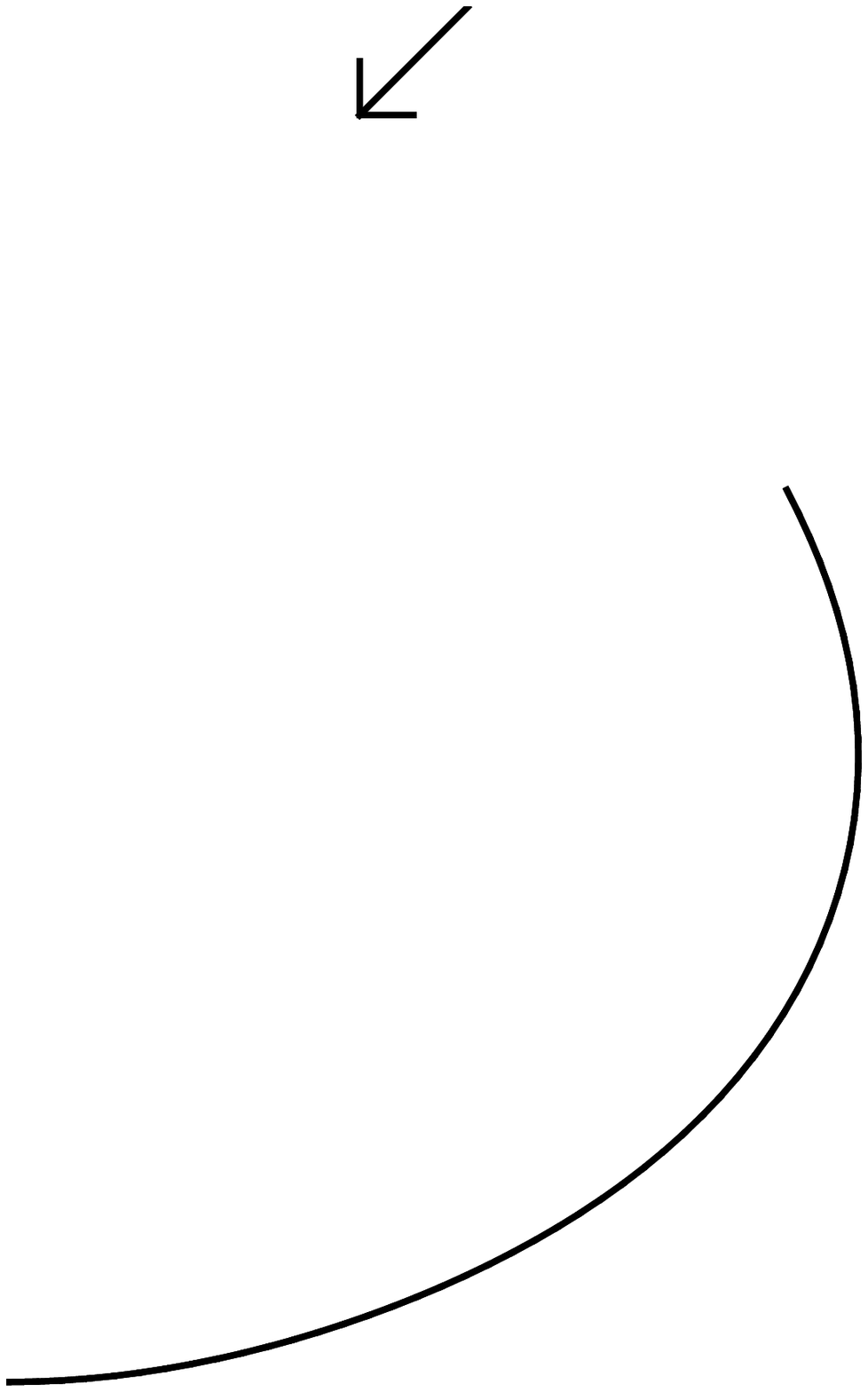}}
     \subfigure[$t=0.2$ms]{\includegraphics[scale=0.18]{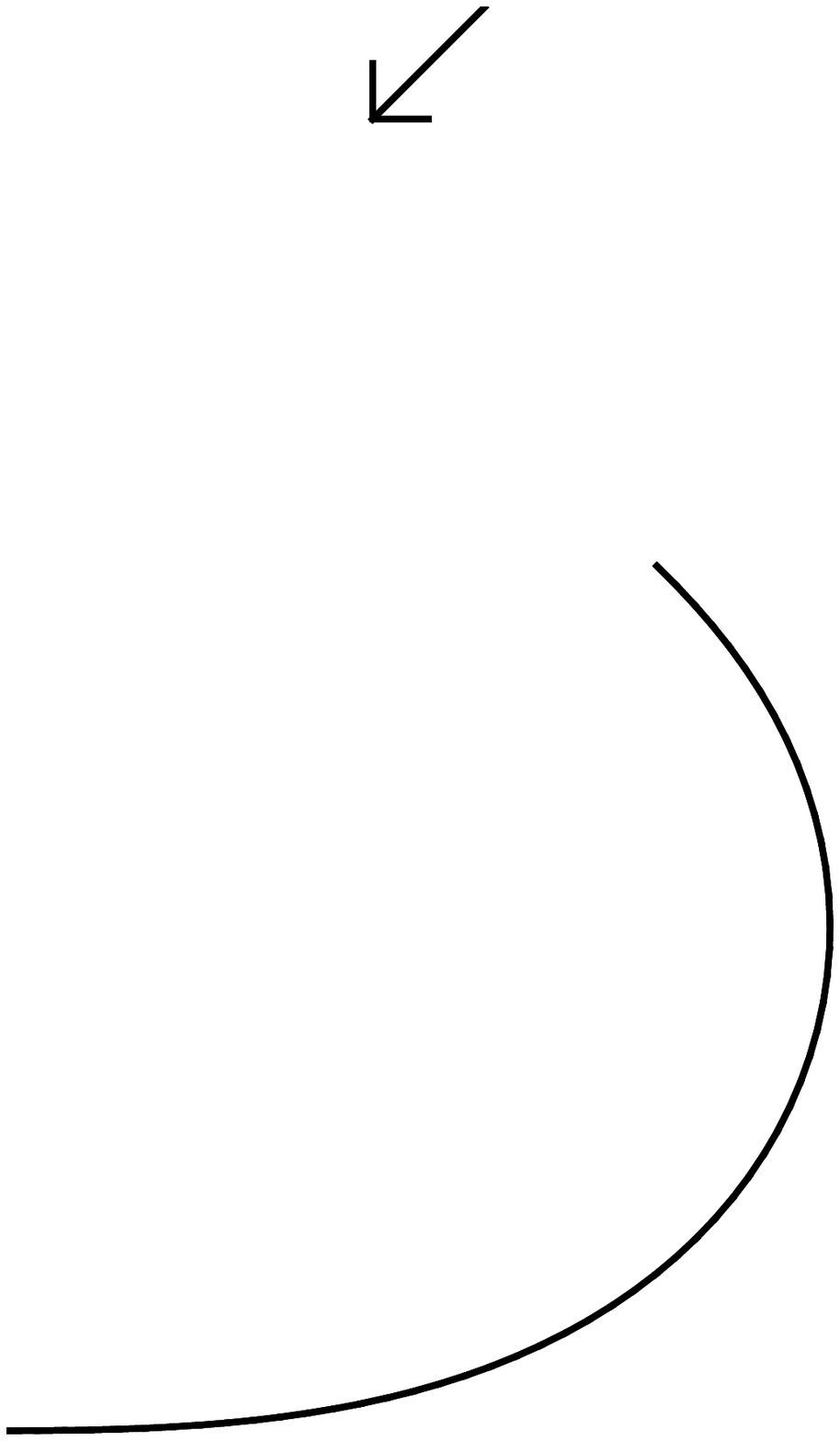}}
     \subfigure[$t=0.3$ms]{\includegraphics[scale=0.18]{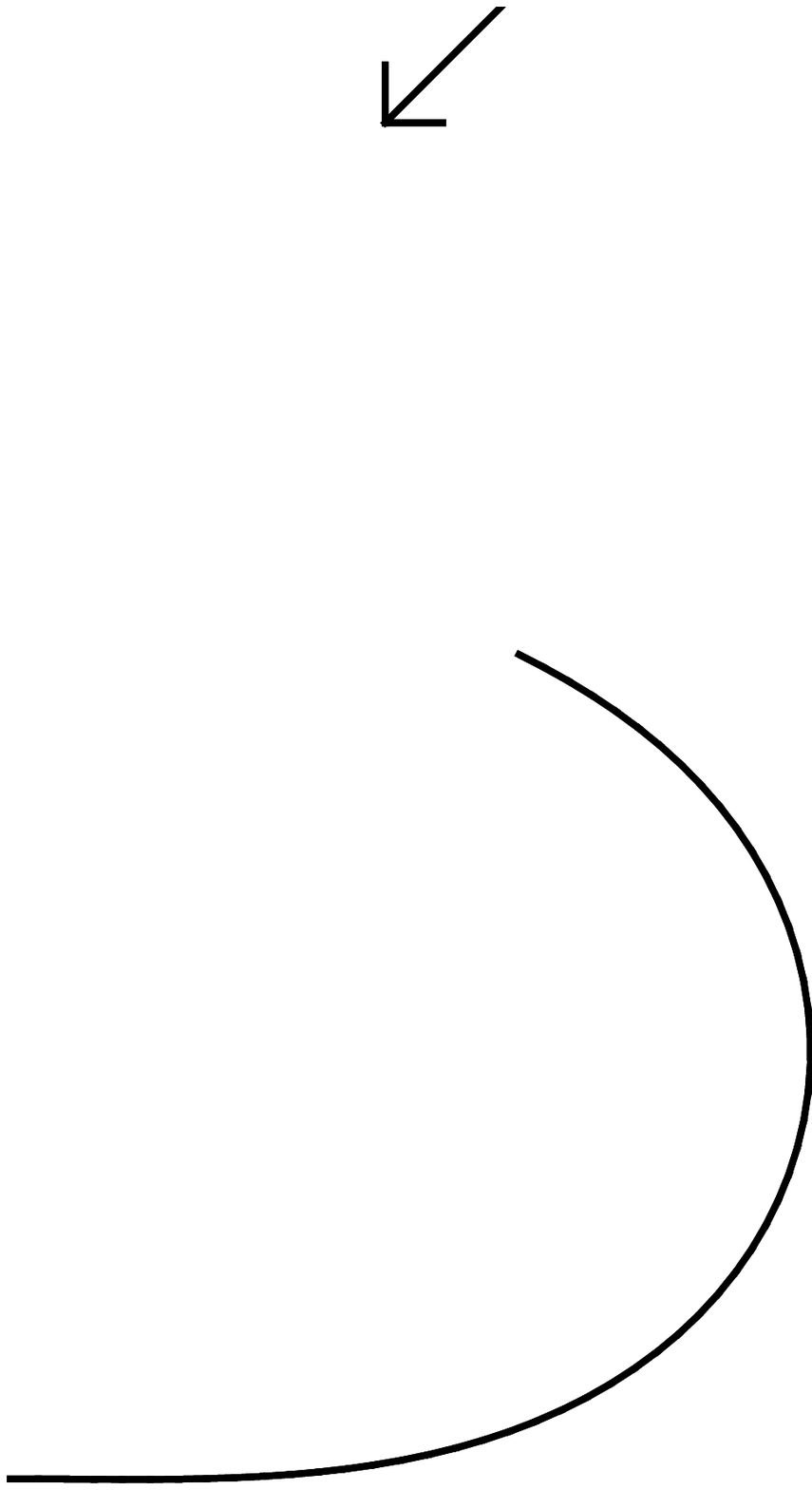}}
     \subfigure[$t=1.1$ms]{\includegraphics[scale=0.18]{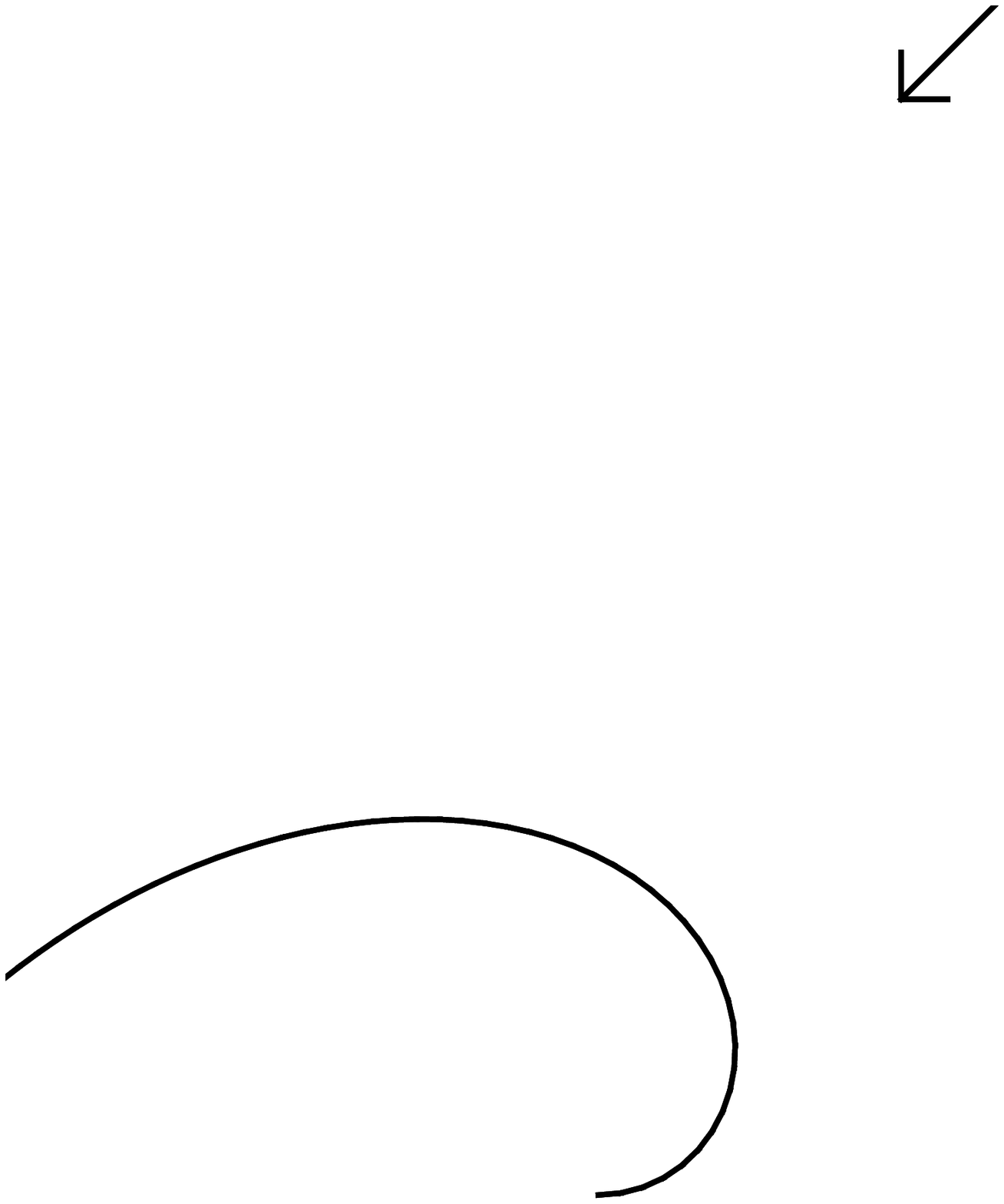}}
     \subfigure[$t=1.5$ms]{\includegraphics[scale=0.18]{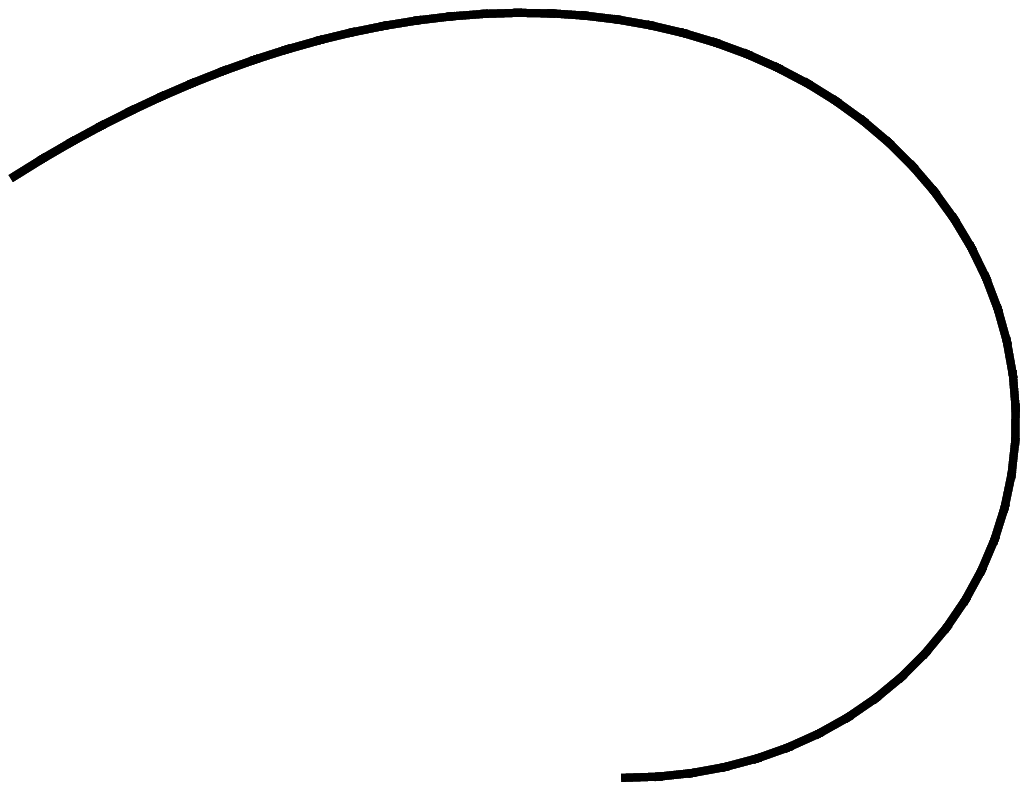}}
     \subfigure[$t=2.0$ms]{\includegraphics[scale=0.18]{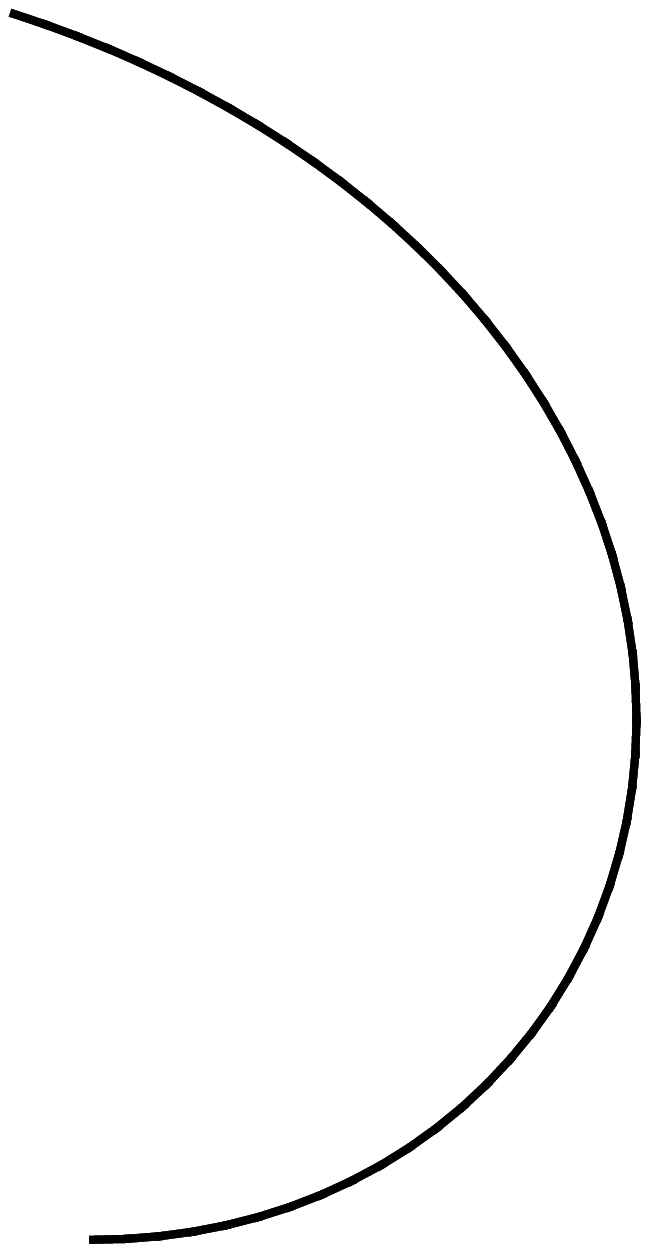}}
     \subfigure[$t=6.0$ms]{\includegraphics[scale=0.18]{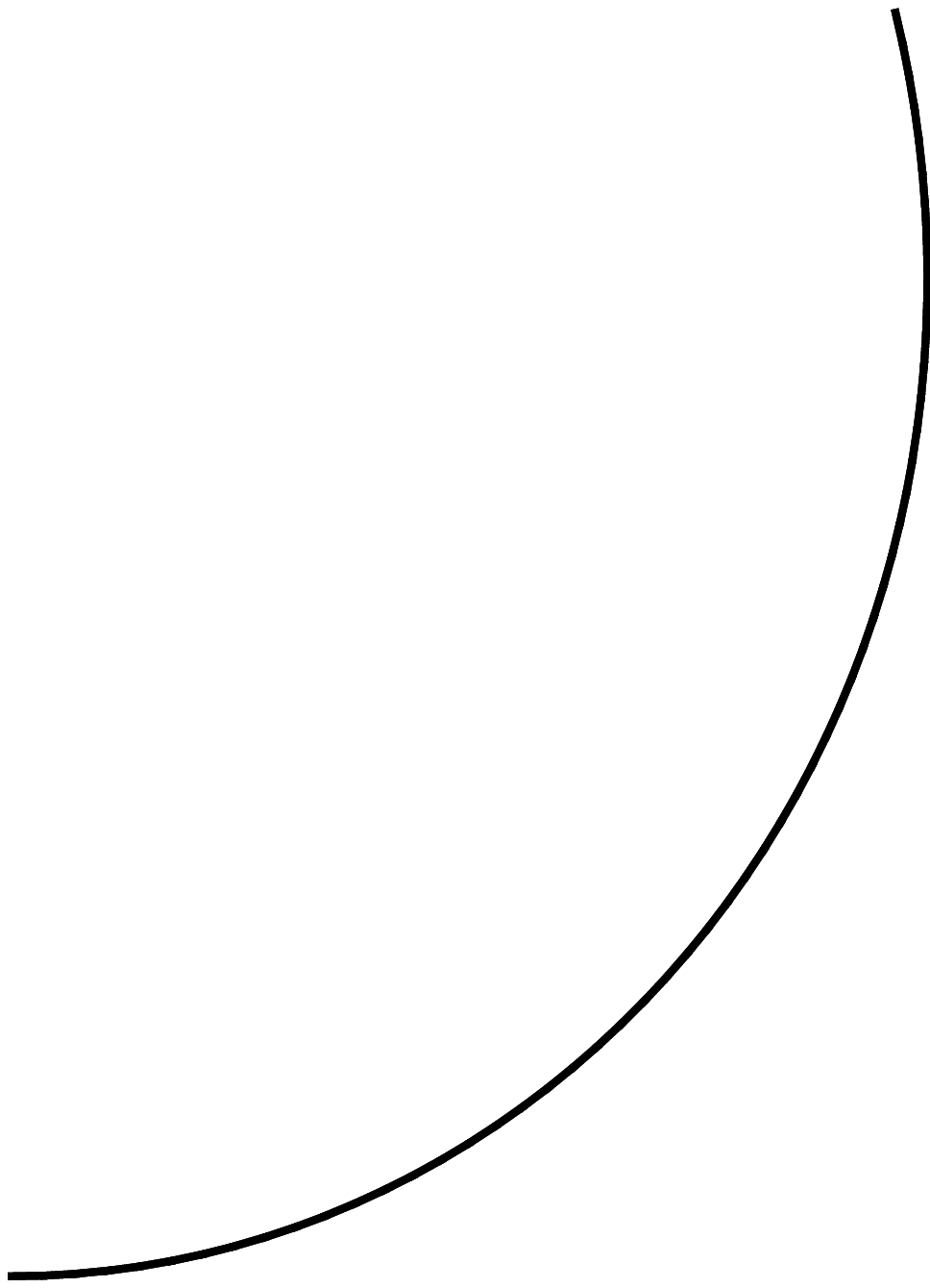}}
     \subfigure[Trajectory of the free end]{\label{fig:case2_trajectory}\includegraphics[scale=0.27]{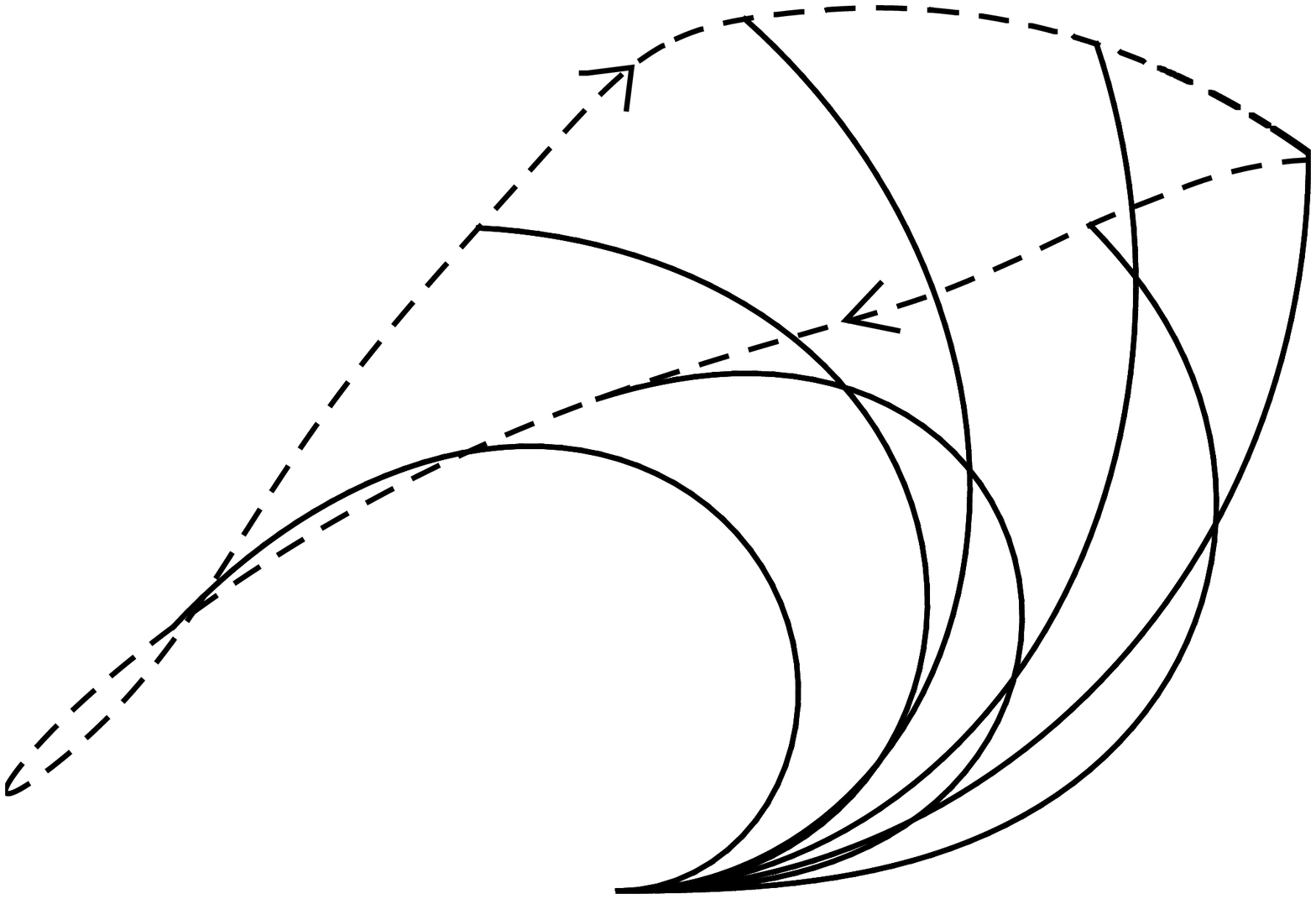}}
\caption{Movement of a curled permanently magnetic film. The arrow shows the direction of applied field.}\label{fig:case2_total_figure}
\end{figure}

\noindent
\textbf{3. A curled magnetic film.}  A curled film with remnant magnetization is subjected to a uniform magnetic field. The initial geometry of the film is shown in Fig.~\ref{fig:case2_initial_geometry}. The left edge is the clamped end. The direction of the magnetization is along the film with the magnetization vector pointing from the clamped end to the free end. The remnant magnetization of the film is taken to be 15 kA/m. An external  field of magnitude $9$ mT is applied at $225^\circ$ to the $x$ axis from $t = 0$ ms to $t = 1$ ms and then linearly reduced to zero in the next $0.2$ ms. The drag coefficients used are $C_x = 5$ $\text{Ns/m}^3$ and $C_y = 60$ $\text{Ns/m}^3$. The radius of curvature of the film is 100 $\mu$m. These system properties lead to a magneto-elastic number of $0.831$, a fluid number of $4.55$ and an inertia number of $0.24$. The physical mechanism of asymmetry is akin to the previous case, except that the film is made  to buckle in a  predetermined way by a curled geometry,  which turns out to  exhibit a large  asymmetry in motion. The propulsive action in the effective stroke takes place during elastic recovery. 
%

In the portion near the fixed end, the magnetic couple acts in a clockwise sense and in the portion near the free end,  in a counter-clockwise sense. As a result, these couples tend to bend the film, such that  the curvature of the film increases, bringing the ends  of the film closer together (this  behavior is similar to that of a  natural cilium). Here, the asymmetry is larger than the previous case as can be observed from Figs.~\ref{fig:case1_trajectory} and \ref{fig:case2_trajectory}.\\

\par
\begin{figure}[ht]\centering
     \subfigure[$t=1.0$ms]{\includegraphics[scale=0.17]{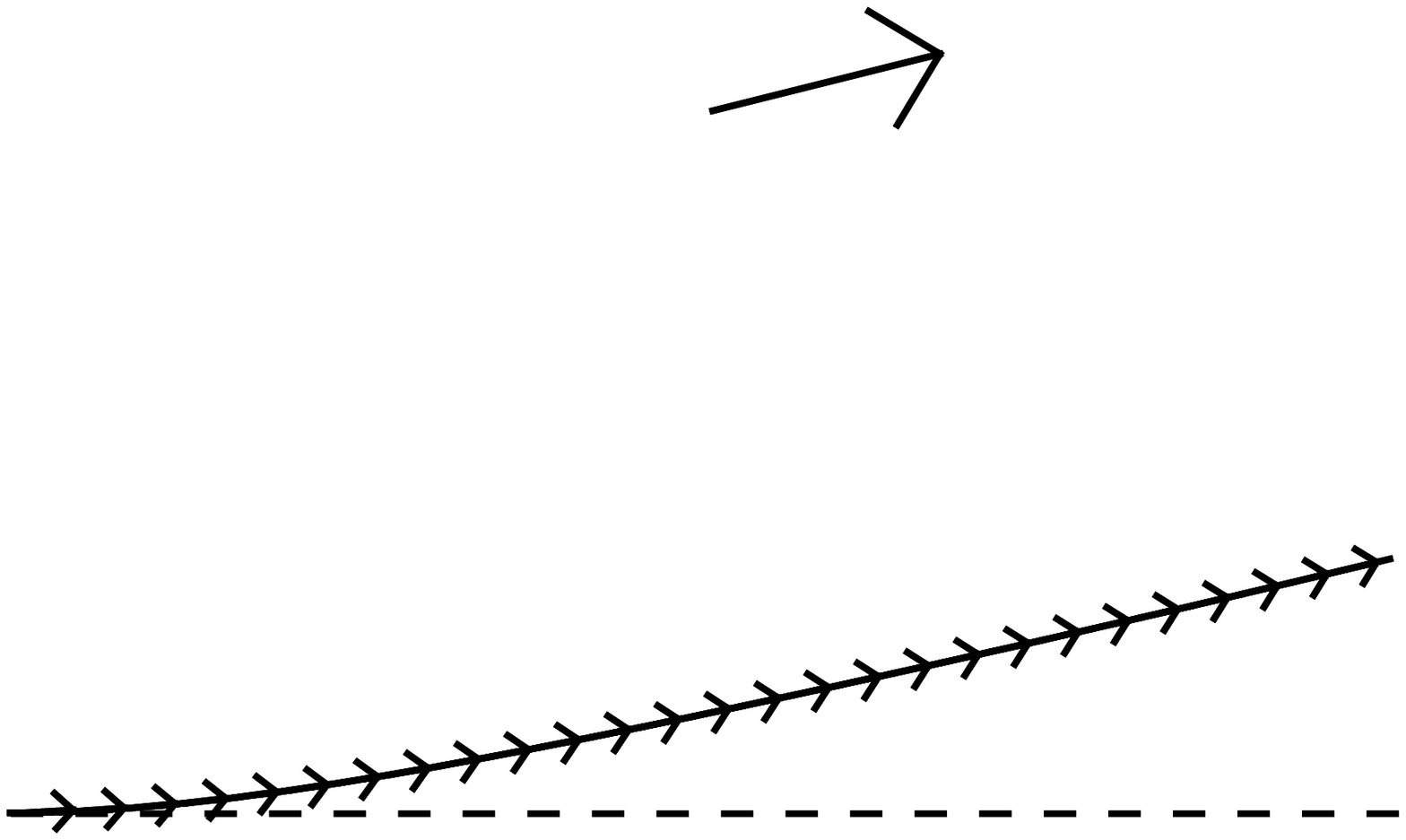}}
     \subfigure[$t=3.0$ms]{\includegraphics[scale=0.17]{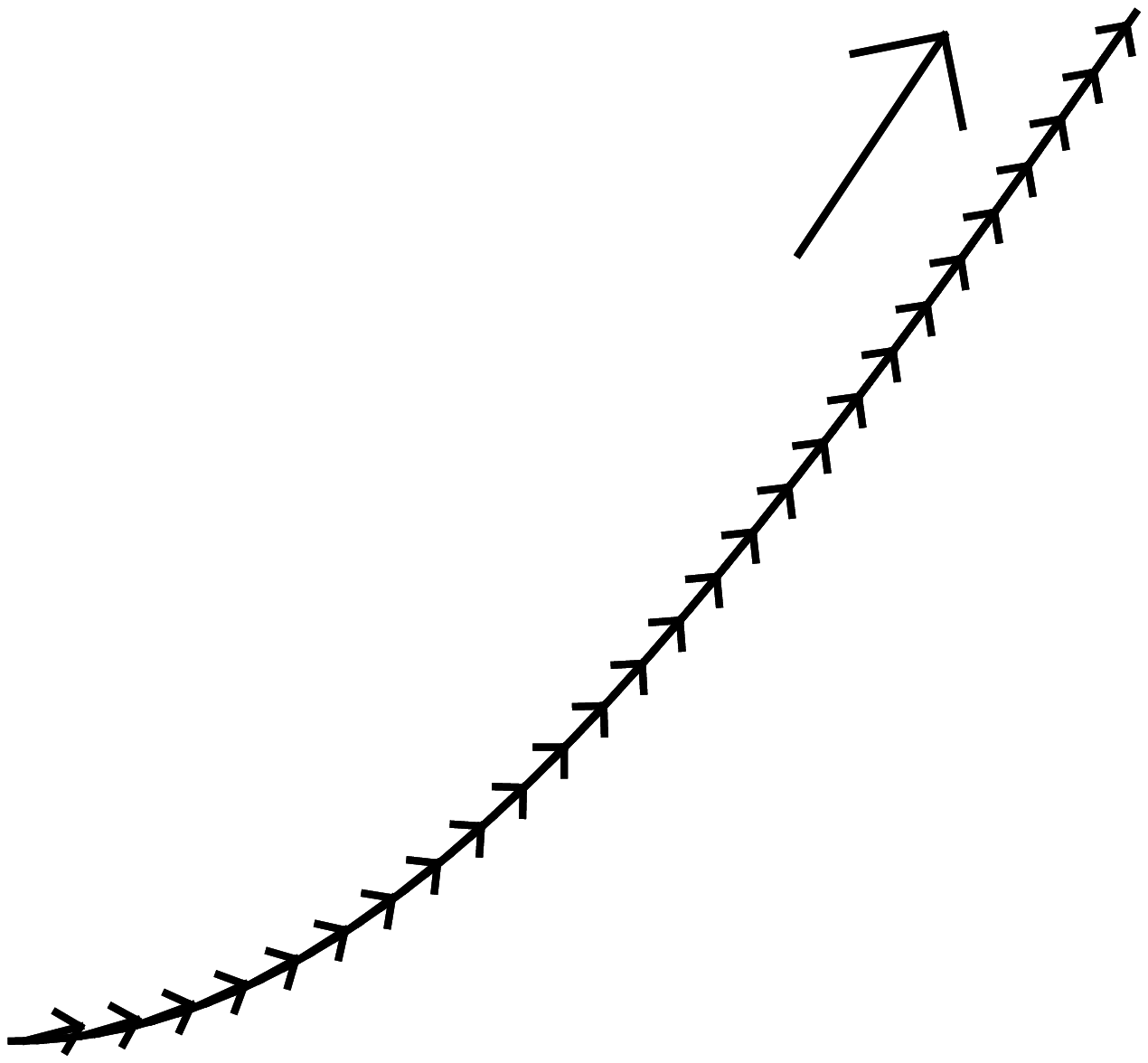}}
     \subfigure[$t=5.0$ms]{\includegraphics[scale=0.17]{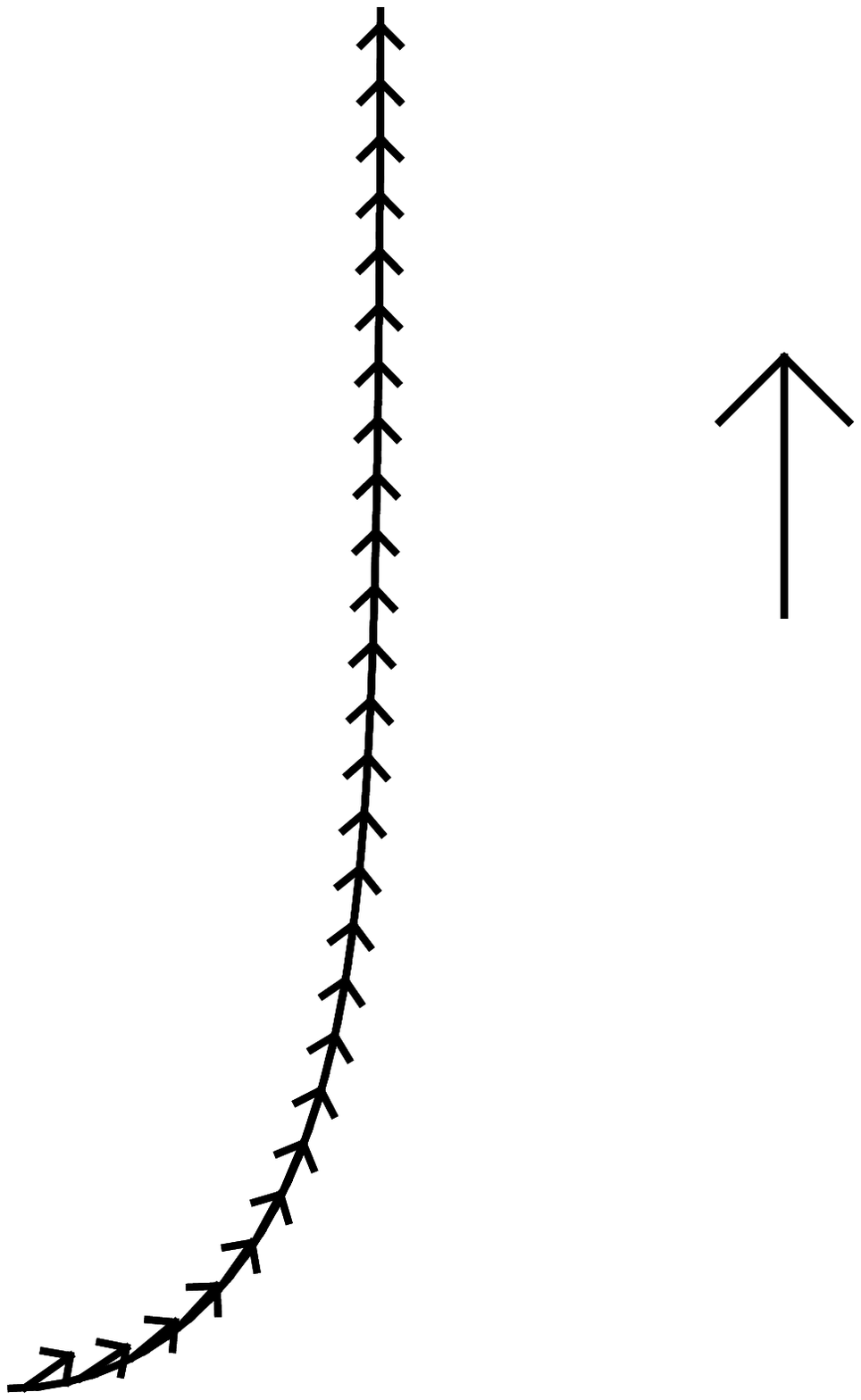}}
     \subfigure[$t=6.0$ms]{\includegraphics[scale=0.17]{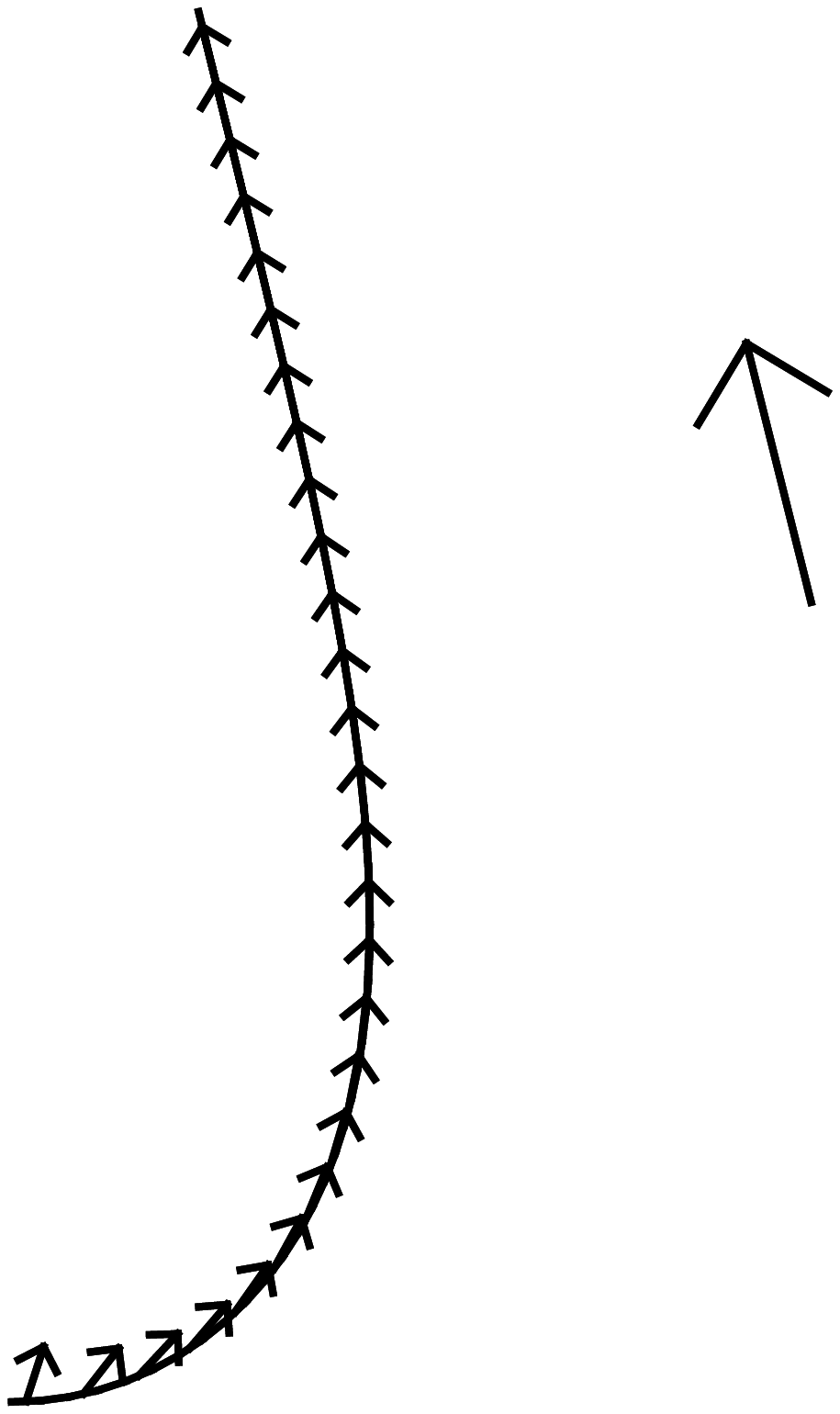}}
     \subfigure[$t=7.0$ms]{\label{fig:case3_7ms}\includegraphics[scale=0.17]{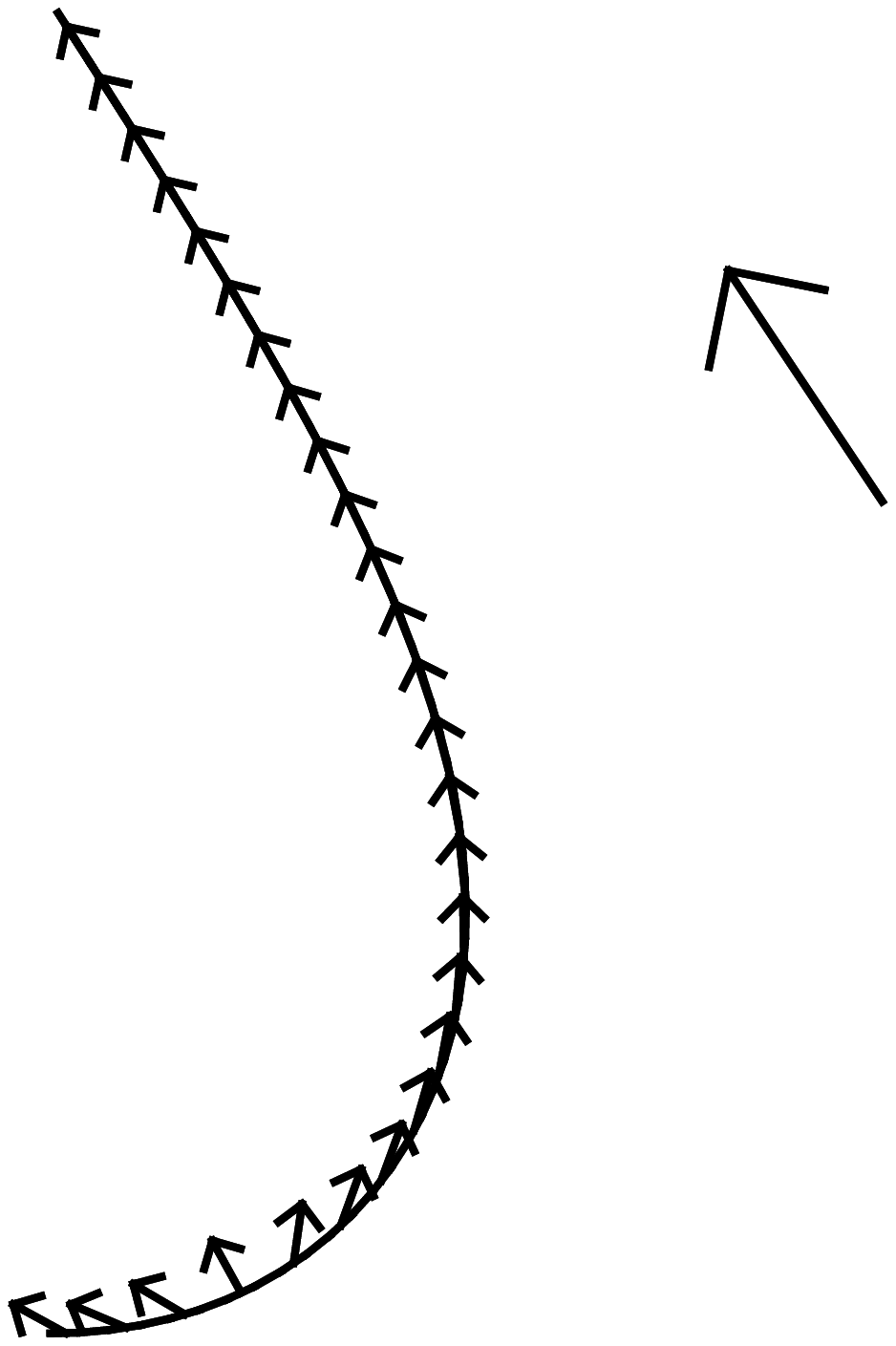}}
     \subfigure[$t=8.0$ms]{\label{fig:case3_8ms}\includegraphics[scale=0.17]{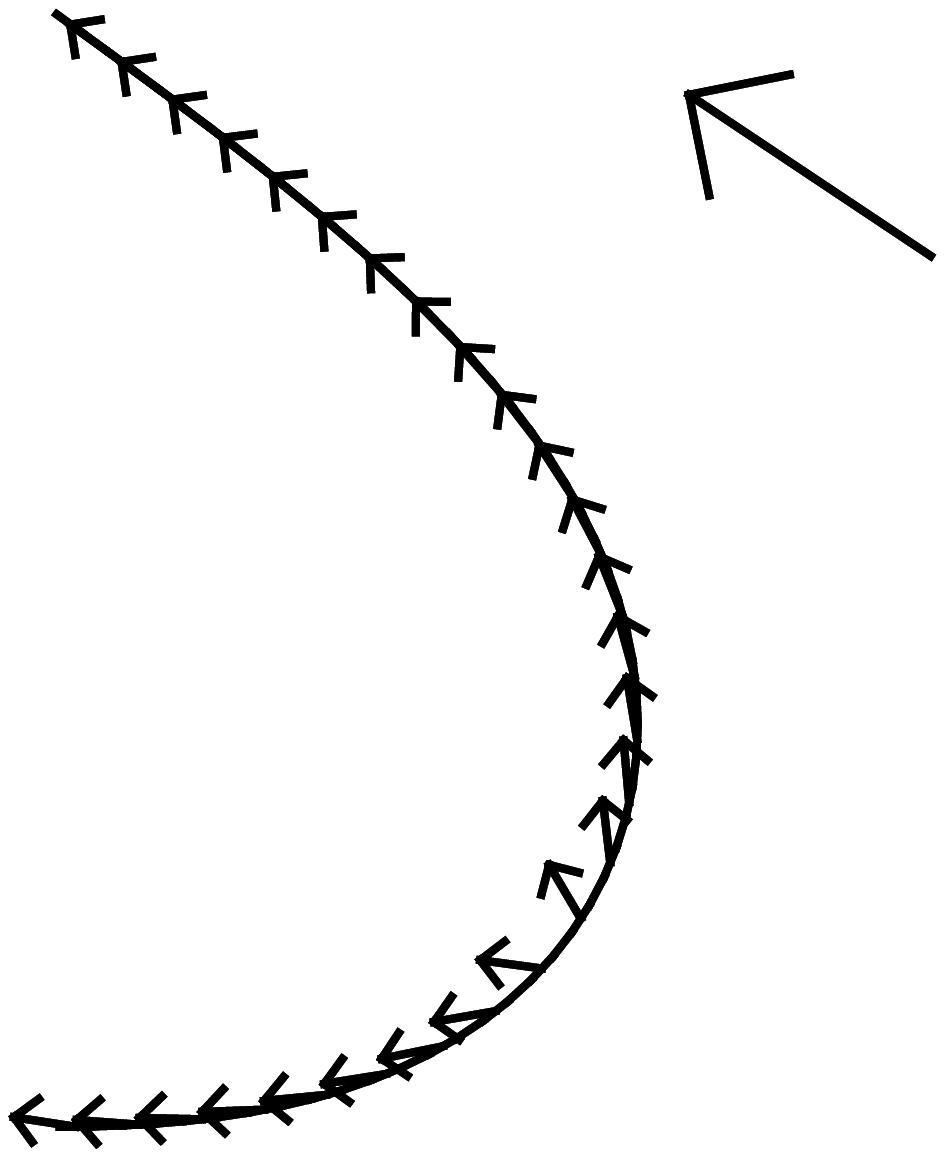}}
     \subfigure[$t=8.5$ms]{\label{fig:case3_8p5ms}\includegraphics[scale=0.17]{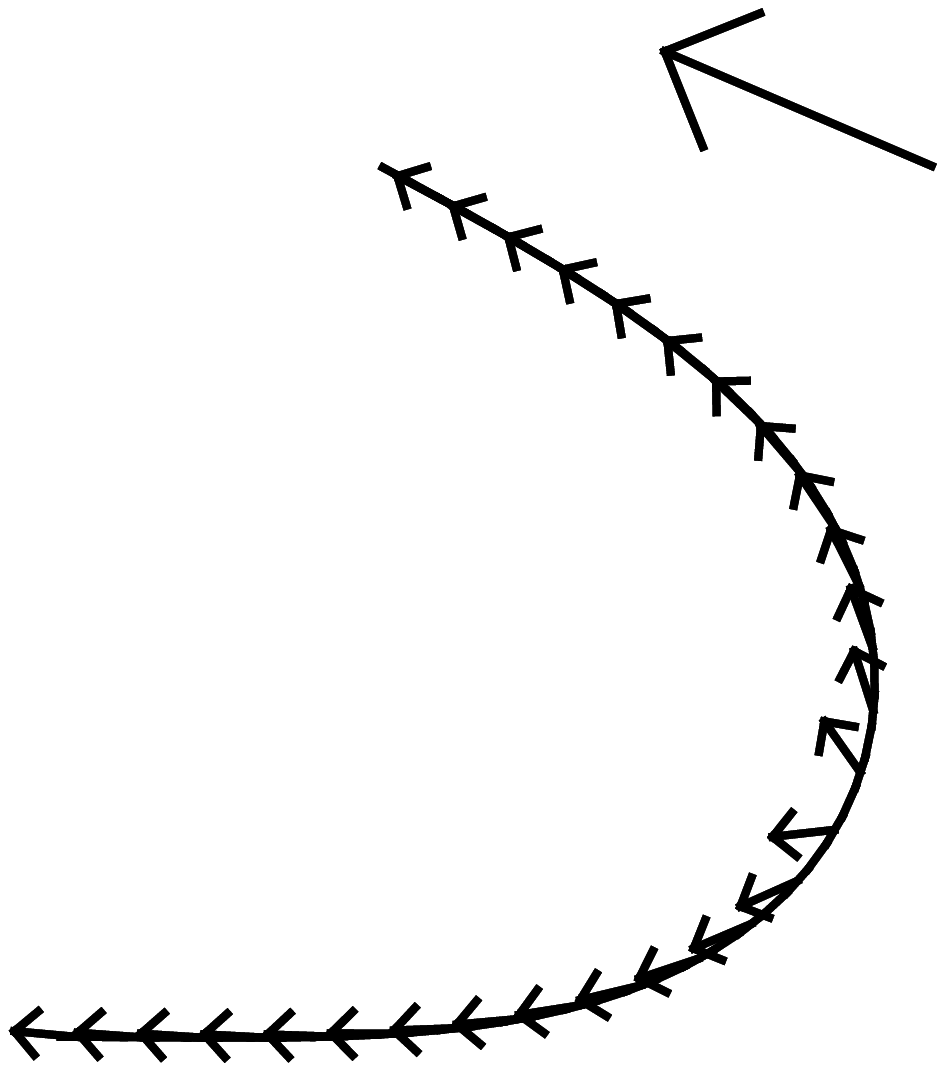}}
     \subfigure[$t=10.0$ms]{\includegraphics[scale=0.17]{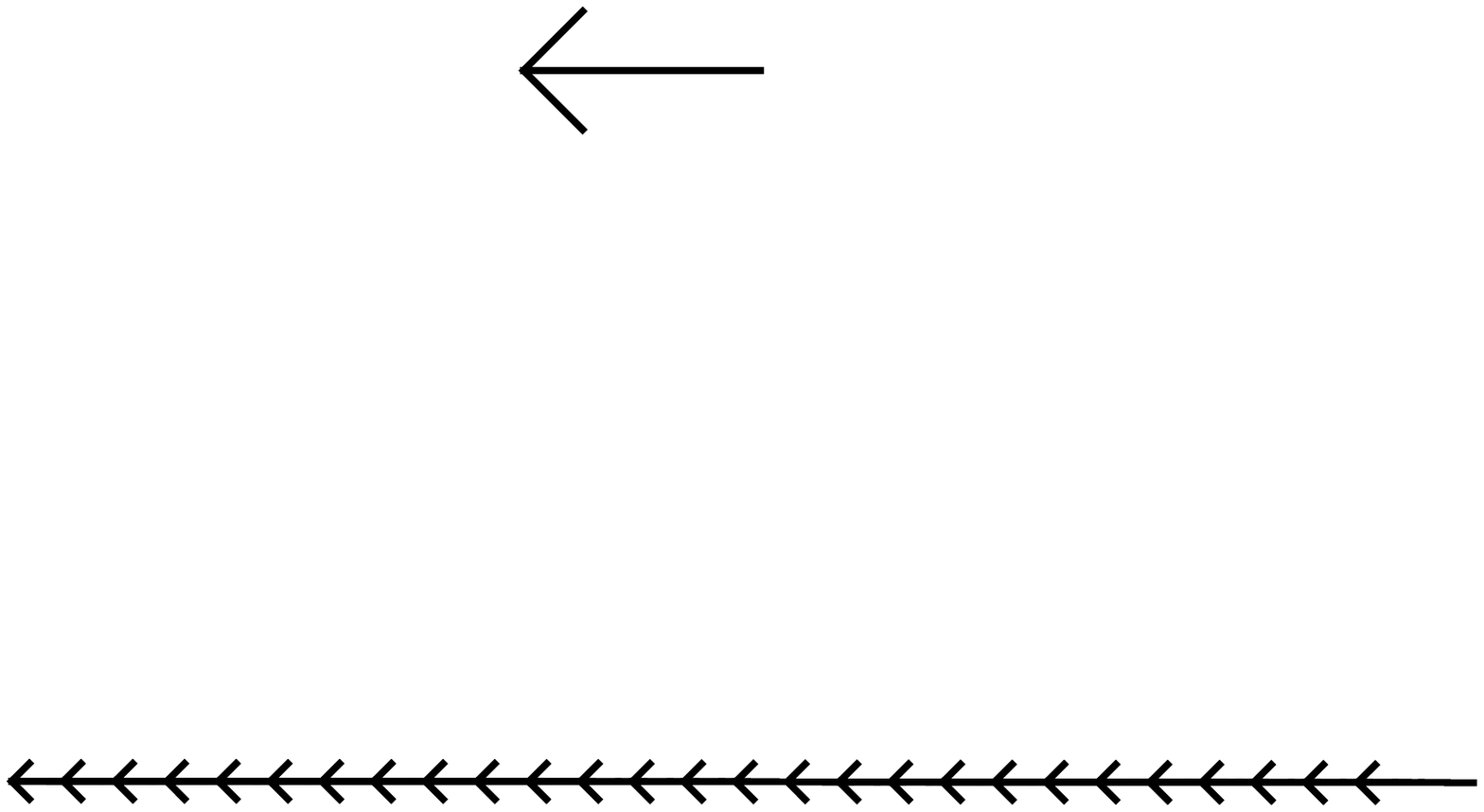}}
     \subfigure[Trajectory of the free end]{\includegraphics[scale=0.27]{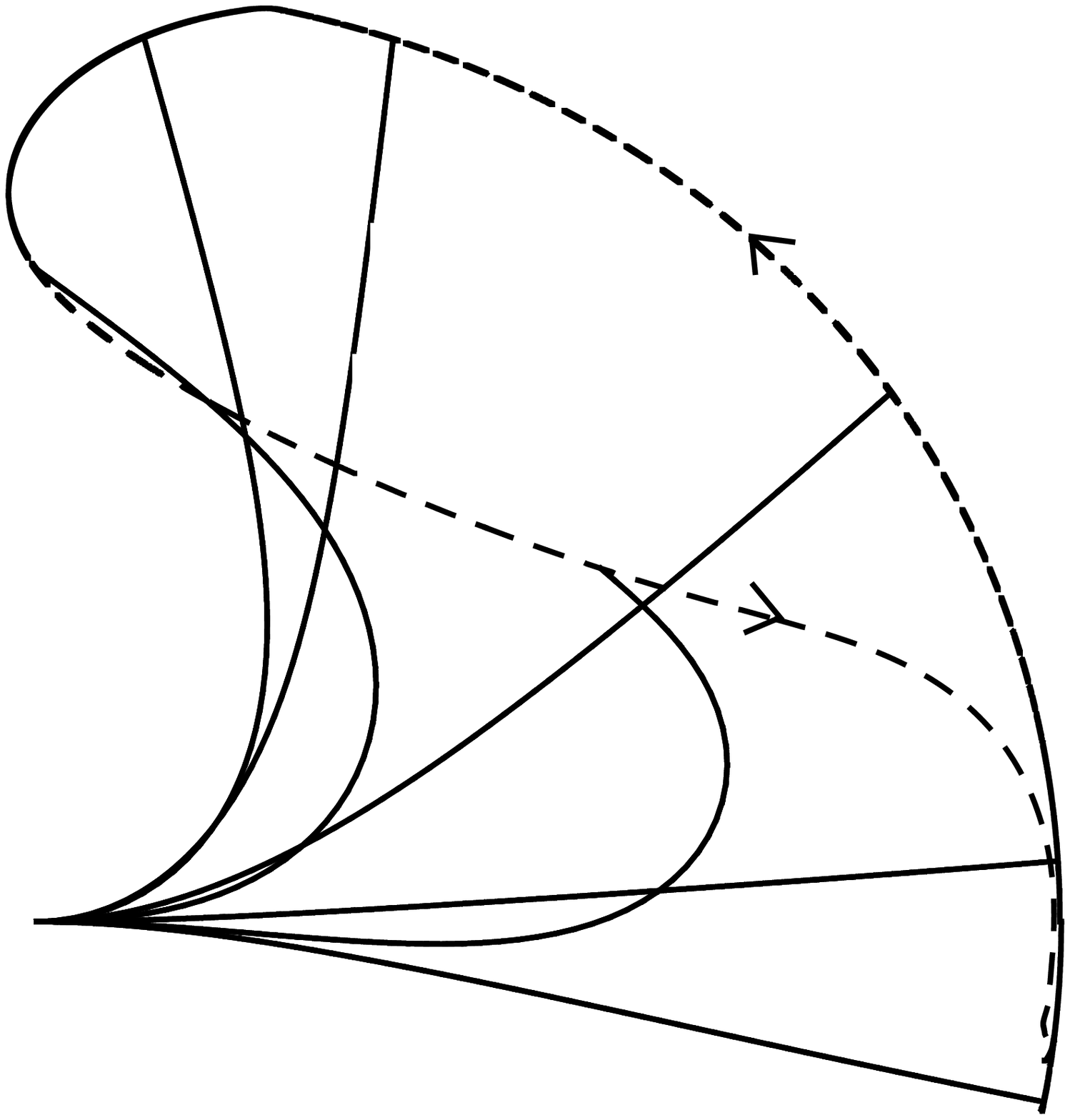}}

     \caption{Super-paramagnetic film in a rotating magnetic field. The dashed line shows the initial position of the film. The big arrow shows the direction of the applied magnetic field. The small arrows show the magnetization along the film.}\label{fig:case3_total_figure}
\end{figure}

\noindent\textbf{4. Super-paramagnetic film.} A super-paramagnetic film which is anisotropic in  its magnetic susceptibility and which is tapered along its length is subjected to a rotating magnetic field. The assumed susceptibilities are  4.6 and 0.8 in the tangential and normal directions, respectively. The  thickness of the film varies linearly along its length, being 2 $\mu$m at the left (attached) end and decreasing to $1 \mu$m at the right end. A magnetic field of $31.5$ mT is rotated from $0^\circ$ to $180^\circ$ in $t = 10$ ms and then kept constant for the rest of the time. The drag coefficients used are $C_x = 30$ $\text{Ns/m}^3$ and $C_y = 60$ $\text{Ns/m}^3$. These system properties lead to a magneto-elastic number of $1.96$, a fluid number of $0.075$ and an inertia number of $4.0\times 10^{-4}$.
When the rotating field is applied,  the free end portion of the beam is rotated through $135^\circ$ and a U bend is formed near the fixed end (Fig.~\ref{fig:case3_7ms} and \ref{fig:case3_8ms}). It is to be noted that the portion of the beam near the free end is nearly straight. This is because the magnetization of the film in this region is almost aligned to the applied field (which is evident from the magnetization vectors shown on the film in Fig.~\ref{fig:case3_total_figure}). As a result,  the  normal component of the field is low, and hence  the moment is less as well. In this situation it is in the part of the film near the U bend, where the magnetic couple distribution (which tends the film to bend) balances the elastic forces (which tends the film to become straight again), hence "freezing-in" the bent shape. When compared with other parts of the beam, at the U bend portion of  film the  magnetic couples are large.
On  the portion of the film  between the free end and the U bend,   anticlockwise moments are acting. On the portion between the fixed end and the U bend, clockwise moments are acting (Figs.~\ref{fig:case3_7ms}, \ref{fig:case3_8ms}, \ref{fig:case3_8p5ms}). Due to the tapered nature of the film, the clockwise moments are larger than the anticlockwise moments. Under the influence of such a system of moments, the film becomes more curved, the end to end distance decreases and  the film  recovers. As the beam recovers elastically the U bend  propagates to  the free end of the beam. It is to be noted that the film recovers in the presence of magnetic forces, i.e. the recovery is not an elastic one, but controlled by magnetic forces, keeping the film low. This phenomena can be exploited to provide a large asymmetry in motion of the film during the forward and return stroke. A larger field  will result in larger curvature of the U bend, forcing the film to stay lower, enhancing the efficiency of the return stroke.

   \newcommand{\cfdfluidpropulsionwidth}{70 mm}
\subsection{Fluid propulsion}\label{sec:drag_calib}
In the magneto-mechanical model, the presence of the fluid is incorporated through drag relations that account for the surface tractions as a function of the velocity. The specific values for the drag coefficients used in the previous Section have been calibrated through a comparison  with a fully-coupled solid-fluid model. This model and the calibration procedure will be discussed in this Section. Moreover, we will also investigate how efficient the four asymmetric motions are in actually propelling fluid. We will demonstrate that the swept area by the film tip is one-to-one related to the fluid propelled.

The Lagrangian solid dynamics model used to study the magneto-mechanical behavior of the films (Section \ref{sec:magnetomechanical_model}) is coupled to an Eulerian fluid dynamics code based on the method detailed in \cite{vanloon}. Fluid inertia is neglected, so the fluid model effectively solves the Stokes equations.
The explicit coupling between the two domains is established through Lagrange multipliers.
Input to the fluid dynamics model are the positions and velocities of the film at all times which result in a full velocity field  in the fluid.  We calculate the drag forces on the film as tractions via the stress tensor in the fluid. The traction distribution is subsequently imposed as surface tractions in the solid dynamics model through Eqn.~ \ref{eqn:ext_v_work}.

\begin{figure}[h!]\centering
     \subfigure[Partly magnetic film with cracks.]{\includegraphics[width =\cfdfluidpropulsionwidth ]{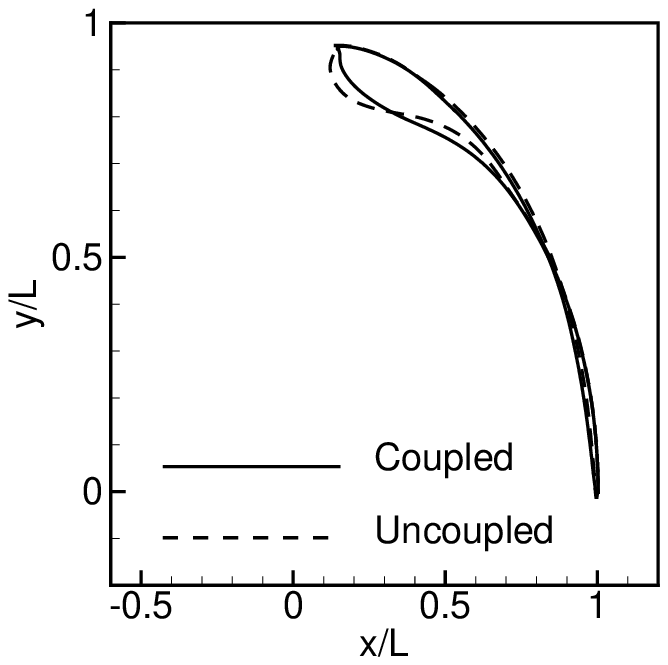}}
     \subfigure[Buckling of a straight magnetic film.]{\includegraphics[width =\cfdfluidpropulsionwidth ]{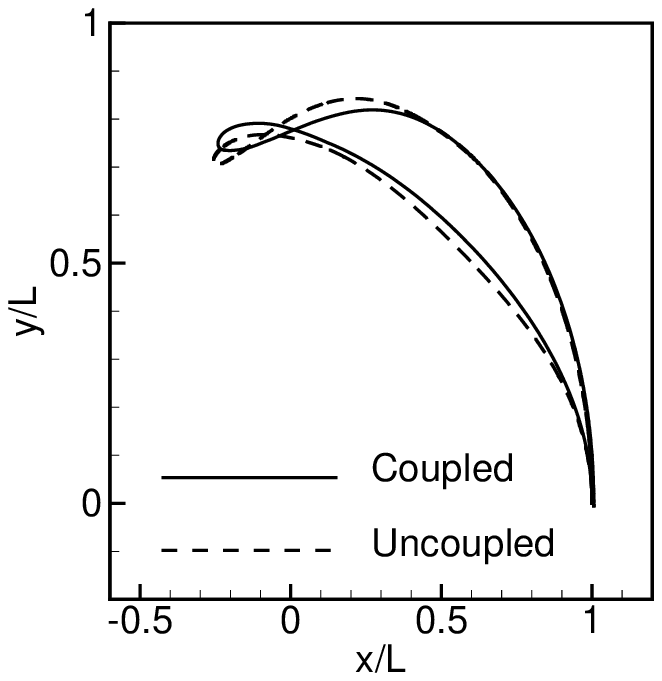}}
     \subfigure[Curled magnetic film.]{\includegraphics[width =\cfdfluidpropulsionwidth ]{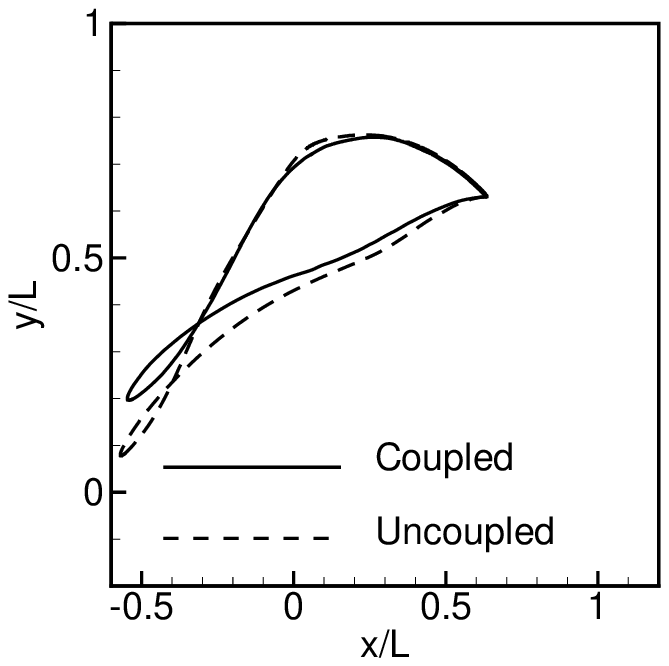}}
     \subfigure[Super-paramagnetic film]{\includegraphics[width =\cfdfluidpropulsionwidth ]{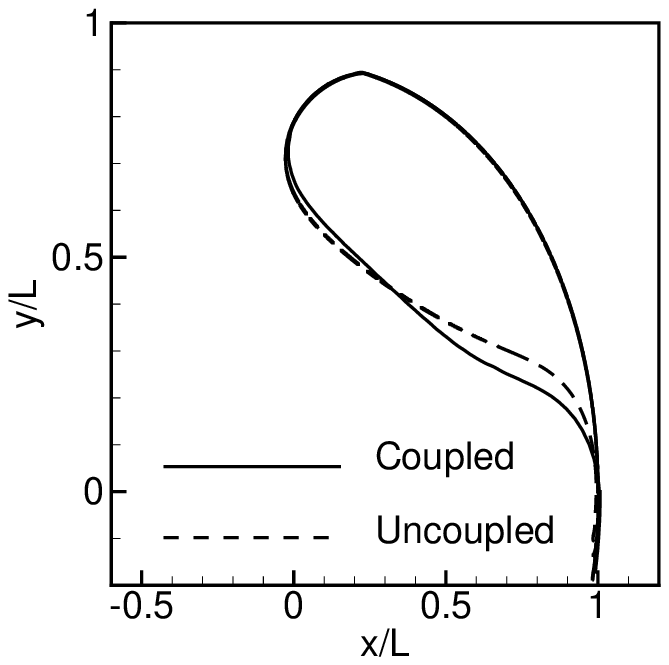}}
     \caption{Trajectory of the free end for various cases obtained when calibrating the drag coeffients used in the solid-drag simulations of Section \ref{sec:configuration} with the fully coupled solid-fluid simulations. For the magnetic film with cracks, the crack length is taken to be $0.1 \mu$m; the parameters of the three other configurations coincide with Section \ref{sec:configuration}}
     \label{fig:calib_drag}
\end{figure}

 To calibrate the drag coefficients, coupled solid-fluid simulations are performed using a periodic arrangement of cilia in a micro-fluidic channel.  The dimensions of the unit-cell analyzed are $400 \mu\text{m}$ in width (horizontal) and  $500 \mu\text{m}$ in height. No-slip boundary conditions are applied at the top and bottom boundaries of the channel and periodic boundary conditions at the left and right ends of the unit-cell.  The viscosity of the fluid is taken to be that of water ($1\times 10^{-3}$ Pas). The film is placed at the center of the channel to avoid any interaction with the boundary.  From the coupled solid-fluid simulations we get the trajectory of the free end. We then perform simulations using the magneto-mechanical model (Section \ref{sec:magnetomechanical_model}) with the assumption that the fluid exerts tractions that are proportional to the velocity, on the film as mentioned in Section \ref{sec:eom}. We  vary the drag coefficients   to match the trajectory of the free end obtained from the coupled solid-fluid simulations. The results are depicted in Fig.~\ref{fig:calib_drag}, clearly showing that the agreement  between the uncoupled and fluid-coupled trajectories is very good. The drag coefficients obtained here have been used for the simulations presented in Section \ref{sec:configuration}.  \\

{\par\noindent
At low Reynolds numbers, asymmetric motion is required to propel the fluid. To study how much fluid is propelled for a given asymmetric motion, we studied two configurations: the curled permanently magnetic film (Fig.~\ref{fig:case2_total_figure}) and the SPM film (Fig.~\ref{fig:case3_total_figure}). As a measure for the asymmetry, we compute the area swept by the free end of the film. To vary this area we vary the magnitude of the magnetic field applied and perform coupled solid-fluid simulations. The corresponding swept area and the total flow across the channel in one cycle is computed.    The domain for which the calculations are performed, is the same as that  used for the calibration of  the drag coefficients, discussed above.
The dependence of fluid propelled (the area flow per cycle) on the swept area is shown, for both the cases, in  Fig.~\ref{fig:linearity_of_flow_rate_with_area}. The cycle times for the two cases are 15 and 10 ms for the curled magnetic and the super-paramagnetic films, respectively. The volumetric flowrate can be calculated from Fig.~\ref{fig:linearity_of_flow_rate_with_area} by multiplying the area flow per cycle with the out-of-plane dimension of the channel divided by the cycle time. The flow rate is seen to vary linearly with the swept area.  This suggests that   the swept area can be used as a measure of effectiveness of the cilium, representing the fluid volume displaced. In the following Section, we will  use this to investigate the efficiency of  fluid propulsion as a function of the system parameters.

\begin{figure}[h!]\centering
     \subfigure[Curled  magnetic film.]{\includegraphics[width =\cfdfluidpropulsionwidth]{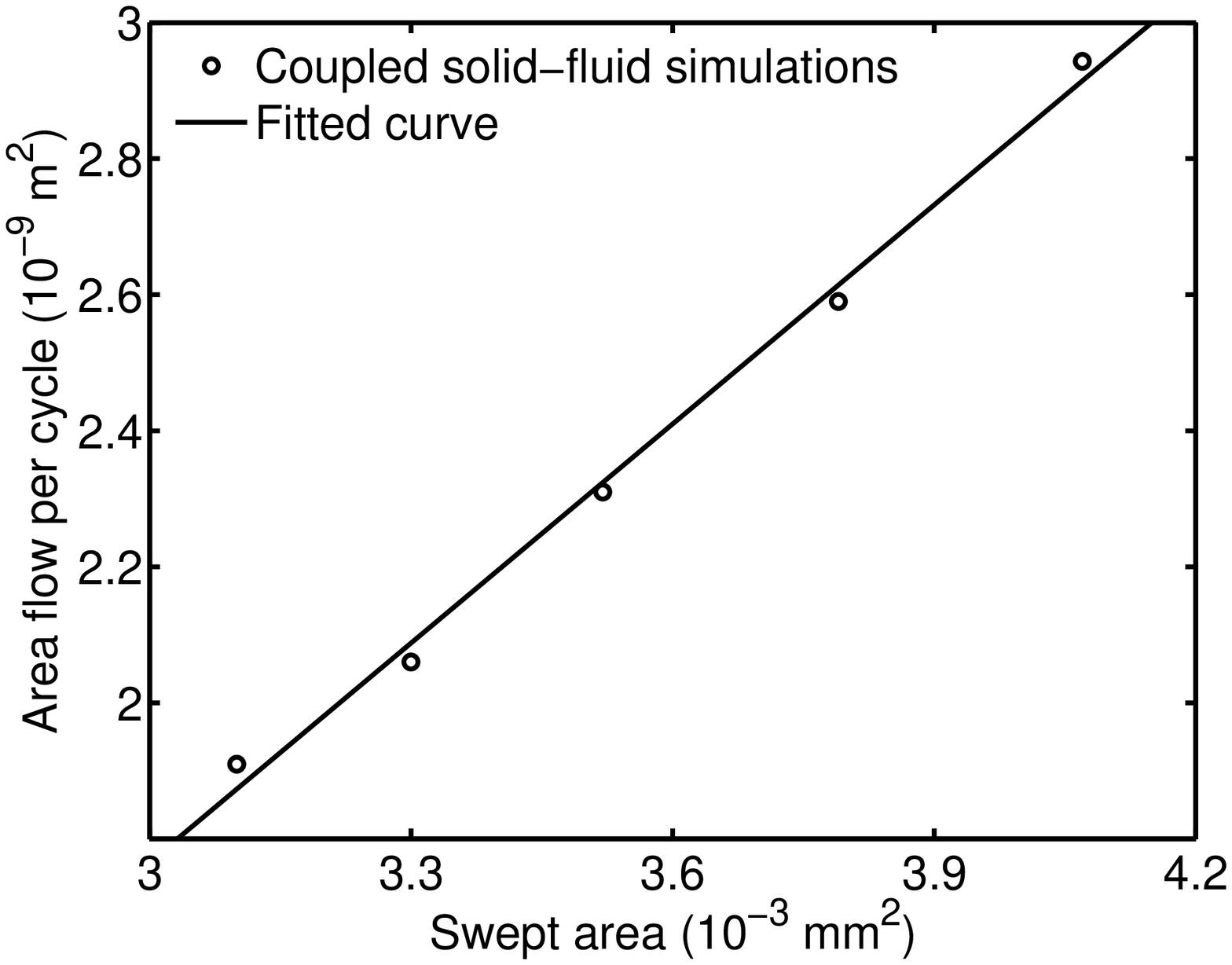}}
     \subfigure[Super-paramagnetic film.]{\includegraphics[width =\cfdfluidpropulsionwidth]{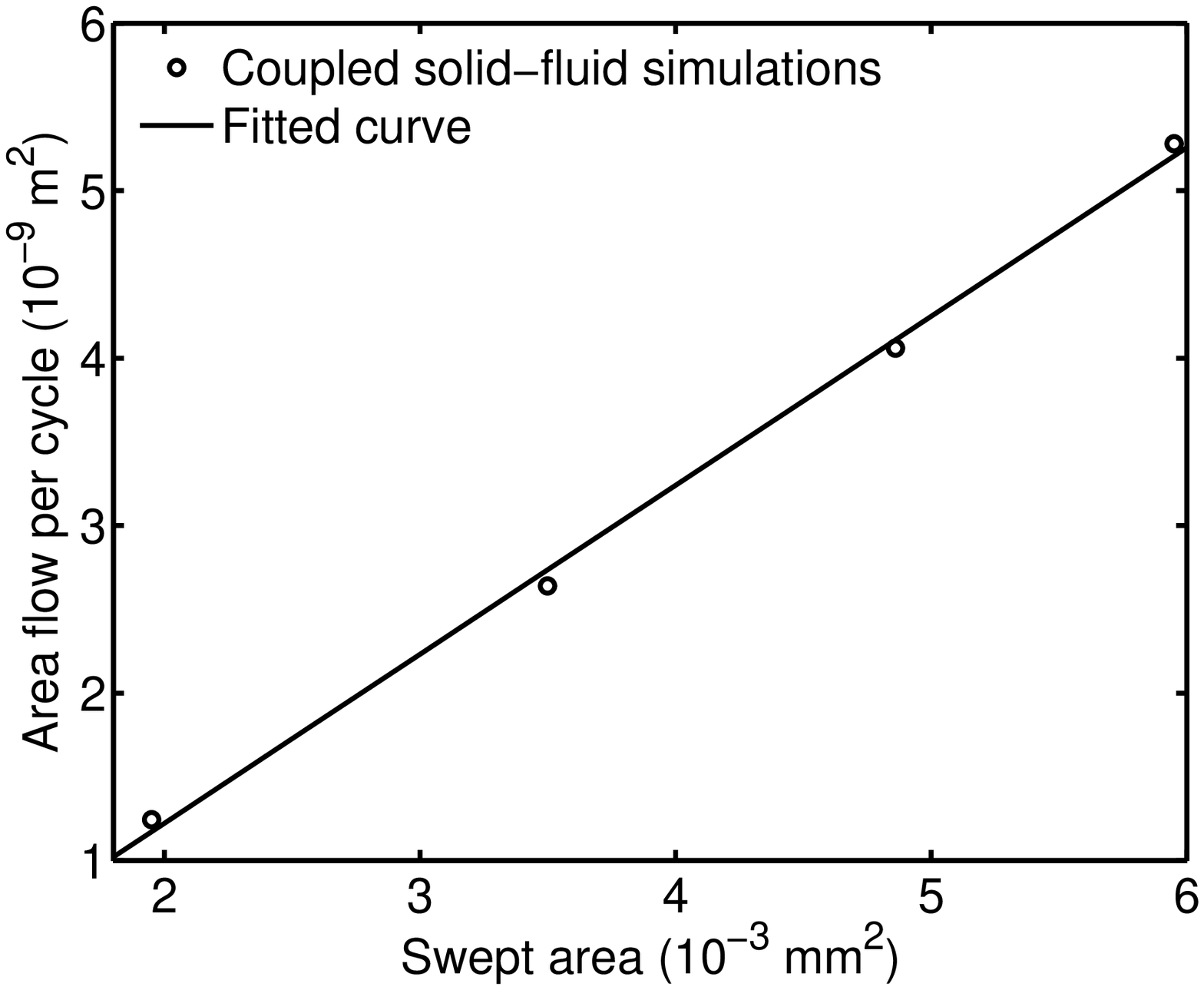}}
     \caption{Relation between swept area and fluid propelled.}
     \label{fig:linearity_of_flow_rate_with_area}
\end{figure}

%

\section{Efficiency survey and discussion}

\newcommand{\dimanaconfwidth}{70 mm}

%
A parametric study of the curled magnetic film and the super-paramagnetic film is performed next. The parameters that govern the response of the actuator system are the inertia number, the fluid number and the magneto-elastic number. The  performance of the system is quantified in terms of the swept area (as a measure of the asymmetry and associated fluid flow) and  the time taken by the film to reach its initial position.

The actuator system based on its material properties, its geometry and the fluid to be propelled can have different fluid and inertia numbers. Fixing them establishes the system. Different set of these numbers refer to different systems. For a set of these systems the magneto-elastic number is varied and the swept area and the time to reach the initial position is noted.

The ingredients in the dimensionless parameters are as follows: $L$ is taken to be the length of the film, $h$ is the thickness of the film, the length is varied from 100 to 1000 $\mu\text{m}$ and the aspect ratio ($L/h$) is varied from 10 to 200. The density $\rho$ of the film is varied from 600 to 8000 $\text{kg/m}^3$, the drag coefficient $C_y$ is varied from 6 to 600 $\text{Ns/m}^3$, while the ratio $C_y/C_x$ is kept the same, and the time scale $\tau$ is taken to be the time in which the load is applied. For a magnetic film it  is the time during which the magnetic field is applied and for the SPM it is the time during which the magnetic field is rotated.

The  area swept by the free end of the film  for different systems (varying inertia number and fluid number)  keeping the magneto-elastic number fixed is shown in Figs.~\ref{fig:case2_area_mn_30} and \ref{fig:case3_area_mn_30}, for the magnetic film and SPM film respectively. The area is normalized with the maximum area that can be swept by a given system (maximum area that can be swept is $\pi L^2 /2$). This plot shows that for a given magneto-elastic number, the swept area decreases with an increase of fluid number. As the fluid number increases, the drag force opposing the motion of the film increases, hence the deformation of the film is less, thus sweeping lesser area. 


The magnetic number needed to sweep  a given area for different systems is shown in Figs.~\ref{fig:case2_mn_area_p2} and \ref{fig:case3_mn_area_0p45}. It can be seen that the same area can be swept by all the systems. As the fluid number increases, we need a large magneto-elastic number to sweep a given area , for a given inertia number. When the fluid number is kept constant, we need a larger magneto-elastic number to sweep the same area at low inertia numbers.
 For the curled magnetic film,  dependency of the swept area of the film is large at small fluid numbers and this dependency  decreases at high fluid numbers. This is due to the fact that the viscous forces dominate over the inertial forces at high fluid numbers.

When sweeping an area of Fig.~ \ref{fig:case2_mn_area_p2} and \ref{fig:case3_mn_area_0p45}, time taken by different systems to return to their initial position  is shown in Figs.~\ref{fig:case2_mn_time_p2} and \ref{fig:case3_time_area_0p45}, respectively. It can be seen that it takes longer for the film to return to its initial position when the fluid number is increased and is independent of the inertia number. But in the case of SPM film, because of the whip like return stroke, some dynamic effects are observable at larger inertia numbers, hence these systems take longer time to return to their initial position.

The lines of constant magneto-elastic number for different systems which sweep the same area are shown schematically in Fig.~\ref{fig:case2_schematic} and \ref{fig:case3_schematic} for the magnetic film and the super-paramagnetic film, respectively. These figures nicely summarize the dependency of swept area on the non-dimensional parameters. The arrow on broken line points in the direction of increasing magneto-elastic number, which means that to get the same swept area we need higher magneto-elastic numbers at high fluid numbers. The lines of constant magneto-elastic number have a slope, this shows the dependency  on the inertia number, i. e. as the inertia number increases, to sweep the same area we need relatively less magneto-elastic number. For the curled magnetic film (Fig.~\ref{fig:case2_schematic}), the change in slope of the constant magneto-elastic number implies the decrease in dependency on the inertia number at high fluid numbers.   For the curled magnetic film, the upper bound for the Inertial number, shown in Fig.~\ref{fig:case2_schematic} has a slope of 2.0  The lines of constant shape ($L/h$ constant) and constant size ($L$ constant) are  also shown in the schematic pictures. It can be seen  that reducing either the size or the aspect ratio decreases the fluid number. Hence to sweep a given area the film needs  lower magnetic number when either the aspect ratio of the film or its size is decreased.

\begin{figure}
     \centering
     \subfigure[]{\includegraphics[width = \dimanaconfwidth]{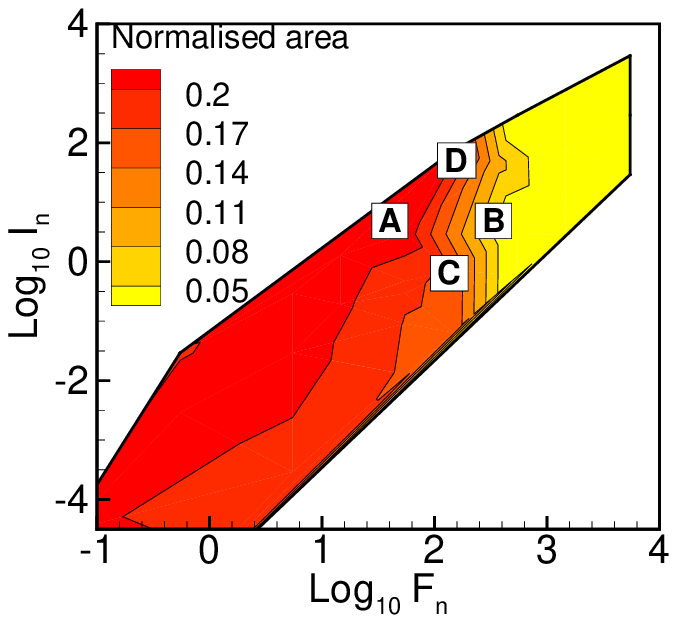}\label{fig:case2_area_mn_30}}
     \subfigure[]{\includegraphics[width = \dimanaconfwidth]{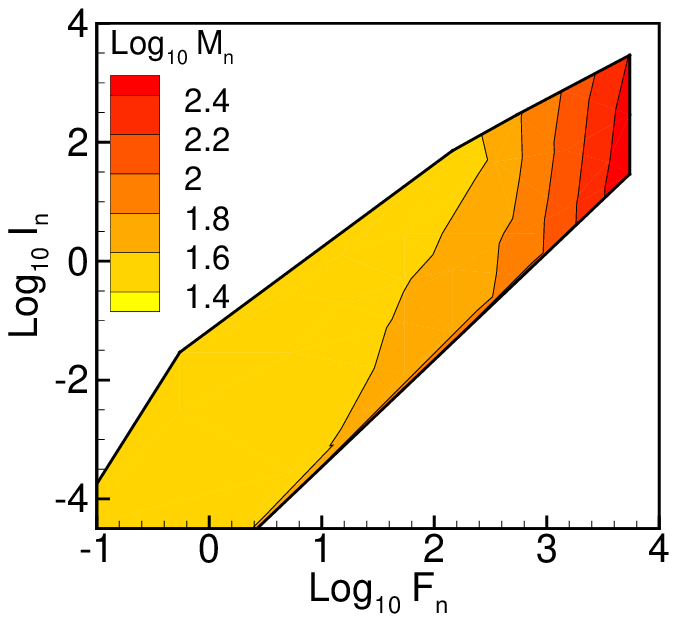}\label{fig:case2_mn_area_p2}}
     \subfigure[]{\includegraphics[width = \dimanaconfwidth]{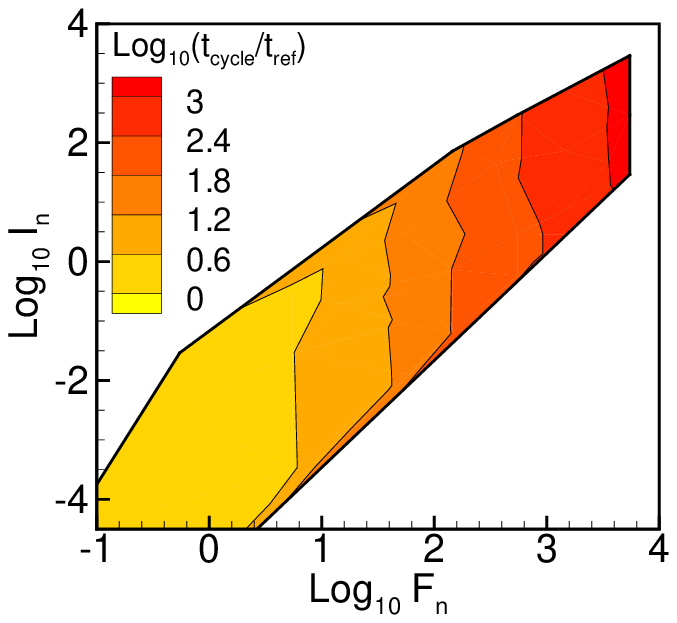}\label{fig:case2_mn_time_p2}}
     \subfigure[]{\includegraphics[scale = 0.5 ]{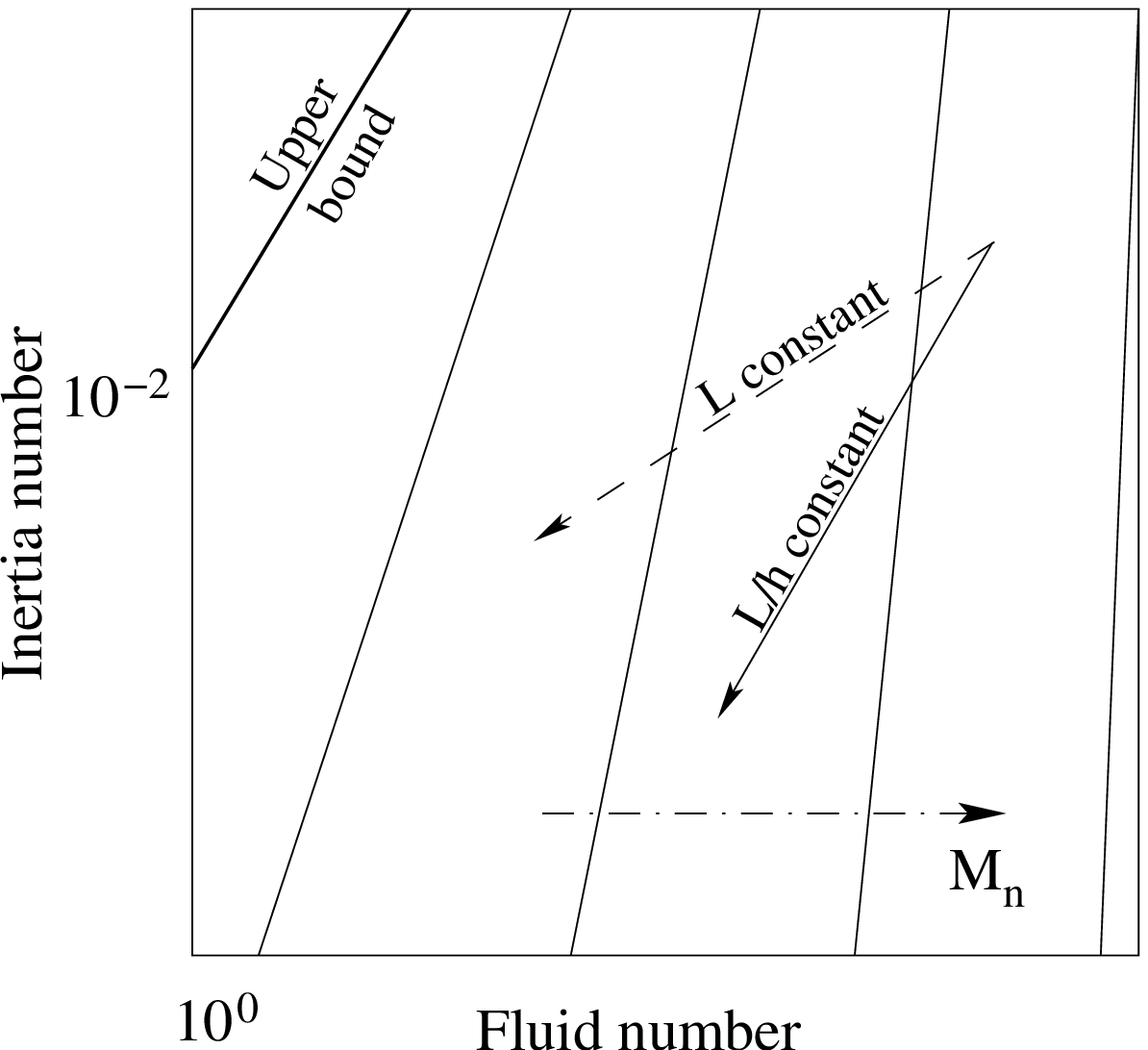}\label{fig:case2_schematic}}

      \caption{Parametric study of a curled permanently magnetic film.
       \subref{fig:case2_area_mn_30} Contours of swept area for  $M_n=30.0$, 
       \subref{fig:case2_mn_area_p2} Contours of $M_n$ needed to sweep an area of $0.2$, 
       \subref{fig:case2_mn_time_p2} Time taken by the film to return to its initial position when the tip sweeps an area of $0.2$, 
       and
       \subref{fig:case2_schematic} Schematic representation of the lines of constant magneto-elastic number.  The arrow on the broken, dashed and solid lines represent the direction of increasing magneto-elastic number, decreasing aspect ratio and decreasing size, respectively.
}
     \label{fig:case2_contours}
\end{figure}
\begin{figure}
     \centering 
     \subfigure[]{\includegraphics[width = \dimanaconfwidth]{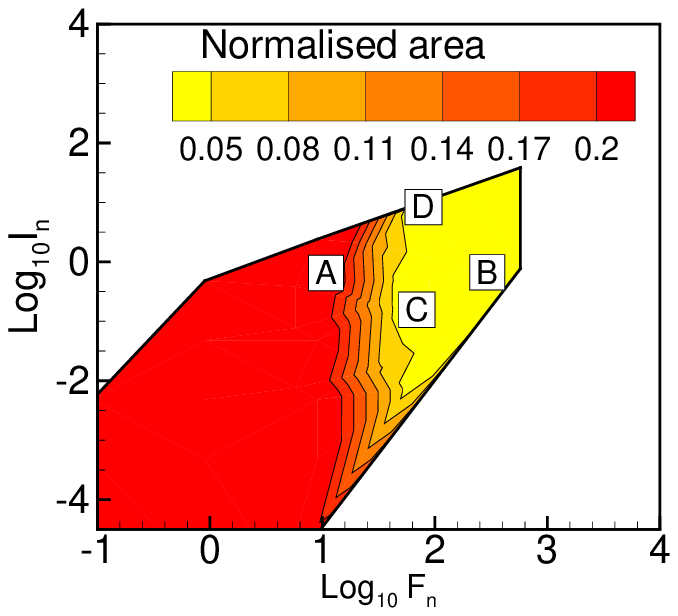}\label{fig:case3_area_mn_30}}
     \subfigure[]{\includegraphics[width = \dimanaconfwidth]{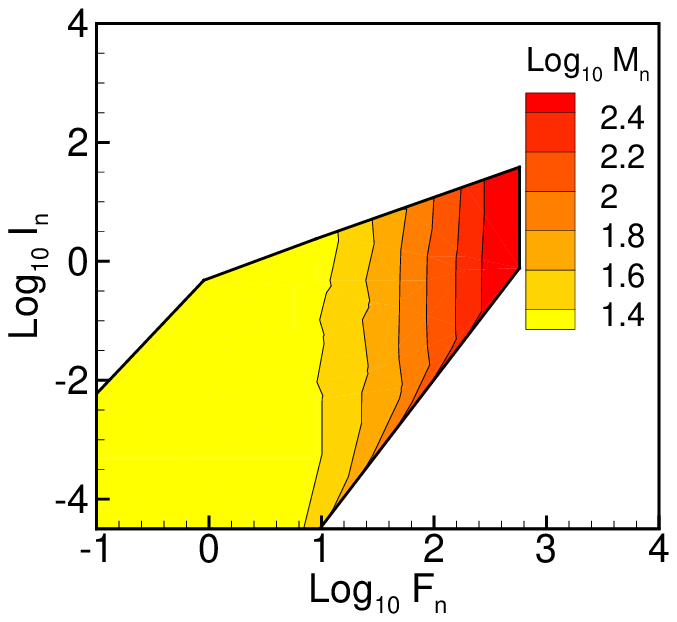}\label{fig:case3_mn_area_0p45}}
     \subfigure[]{\includegraphics[width = \dimanaconfwidth]{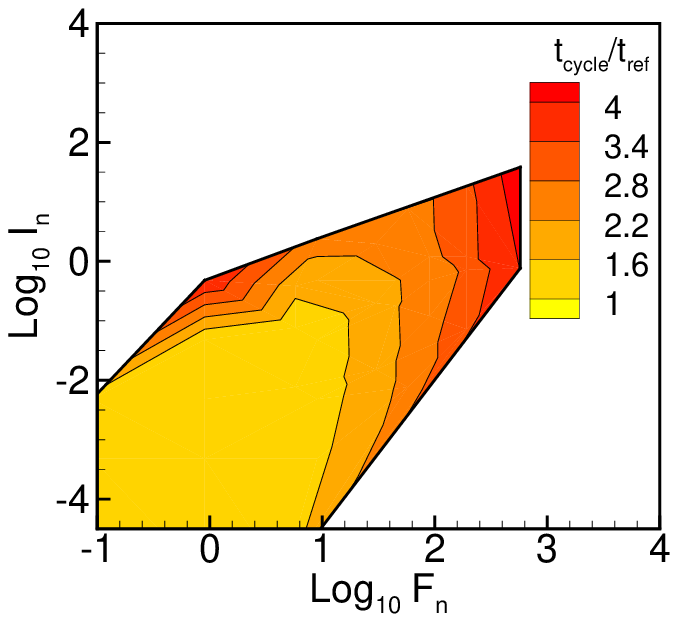}\label{fig:case3_time_area_0p45}}
     \subfigure[]{\includegraphics[scale = 0.5 ]{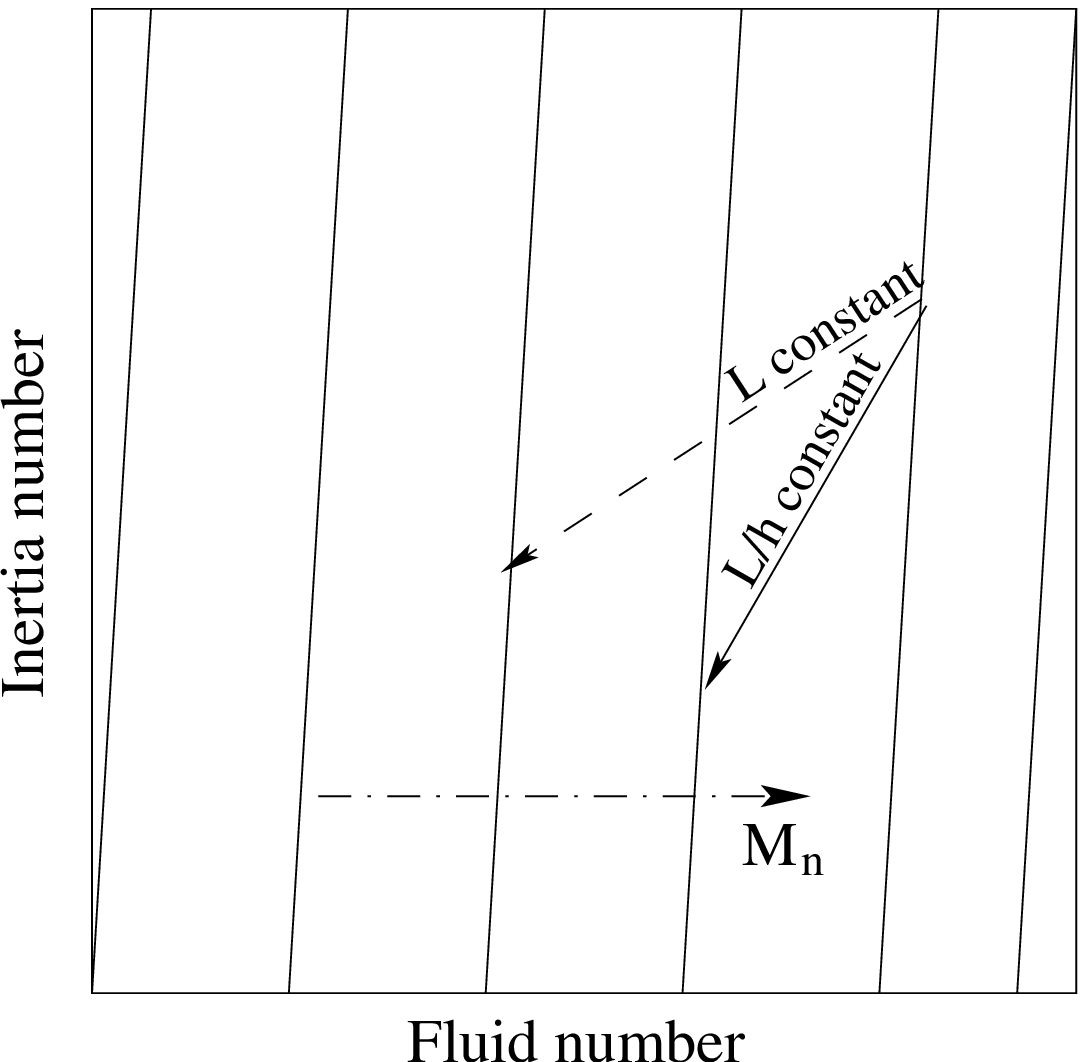}\label{fig:case3_schematic}}
     \caption{Parametric study of a curled permanently magnetic film.
       \subref{fig:case3_area_mn_30} Contours of swept area for  $M_n=30.0$, 
       \subref{fig:case3_mn_area_0p45} Contours of $M_n$ needed to sweep an area of $0.45$, 
       \subref{fig:case3_time_area_0p45} Time taken by the film to return to its initial position when the tip sweeps an area of $0.45$ and
       \subref{fig:case3_schematic} Schematic representation of the lines of constant magneto-elastic number. The arrow on the broken, dashed and solid lines represent the direction of increasing magneto-elastic number, decreasing aspect ratio and decreasing size, respectively.
    }
     \label{fig:case3_contours}
\end{figure}


\section{Conclusion}
This work has addressed the numerical  design of magnetically driven actuators for  fluid propulsion in micro-fluidic channels.  The  key characteristics of such actuators,  is that they have to move in an asymmetric manner, due to the low Reynolds number. Such actuators are proposed to be made of polymer films with embedded magnetic particles which can be manipulated by an externally applied field. Four possible configurations have been identified which result in an asymmetric motion of the actuators. In the configuration based on permanently magnetic films the effective stroke is slower   compared to the recovery stroke. This is due to the fact that the effective stroke is driven by the elastic forces, while the recovery is due to the applied magnetic field. This nature of slow effective stroke and fast recovery stroke has also been reported by Kim and Netz \cite{kim_netz}, in their study of the propulsion efficiency of a  periodically beating elastic filaments anchored to a substrate. An important result from our work is that the flow across the channel varies linearly with the area swept by the free end of the film, promoting the swept area  as an efficiency parameter of the actuator system. The permanently magnetic film is very sensitive to the applied field at large values of magnetization. The SPM film is equally sensitive to the applied field at all values of susceptibility.

\section*{Acknowledgements}\nonumber
This work is a part of the $\text{6}^\text{th}$ Framework European project 'Artic', under contract STRP 033274.
\appendixpage
\begin{appendices}

\section{Interpolation Functions}\label{int_func}
\begin{equation}\begin{split}
\mb{N_u}&=\begin{array}{c c c c c c c}[N_1 & 0& 0& N_2& 0& 0]\end{array}\\
\mb{N_v}&=\begin{array}{c c c c c c c}[0 & H_1& H_2& 0& H_3& H_4 ]\end{array},
\end{split}\end{equation}
where
\begin{equation}
\begin{split}
	N_1=&1-\frac{x}{l},\\
	N_2=&\frac{x}{l},\\
	H_1=&\frac{1}{l^3}(2x^3-3x^2l+l^3),\\
	H_2=&\frac{1}{l_0l^3}(x^3l-2x^2l^2+xl^3),\\
	H_3=&\frac{1}{l^3}(-2x^3+3x^2l),\\
	H_4=&\frac{1}{l_0l^3}(x^3l-x^2l^2)
\end{split}
\end{equation}
and $l$ is the length of the element and $l_0$. is a reference length.

\section{Magnetic Buckling Analysis}\label{sec:film_buckling}
In this Section we study the buckling behavior of a straight permanently magnetic film. When the applied field is  opposite to the magnetization and the film is straight, no couple is induced for the motion of the film (see Fig.~\ref{fig:film_buckling_schematic}).     But, if the film is slightly perturbed, couples will be acting on the film. These couples will increase when the film is deflected further from the initial, straight configuration. Clearly, the straight configuration is an unstable equilibrium state, so that buckling will occur above a critical value of the magnetic field.  This situation is similar to a cantilever with an end compressive load. The cantilever buckles because of the moment of compressive load and the film buckles because of the magnetic couple.
\begin{figure}[htpb]
   \centering
   \includegraphics[scale=0.5]{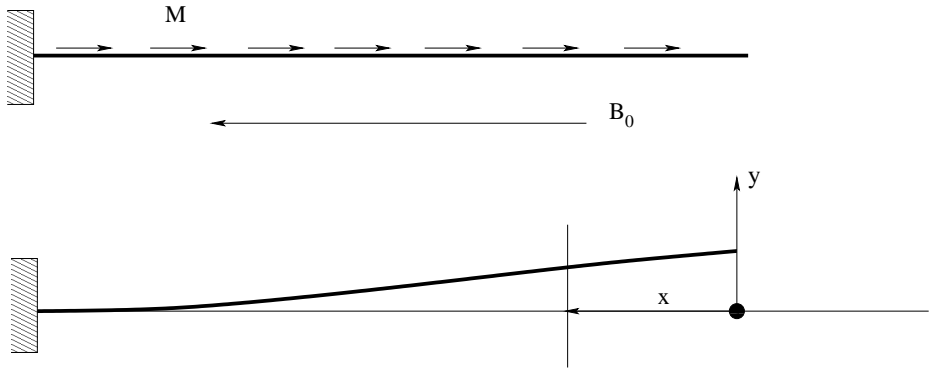}
   \caption{Buckling of a film under a magnetic field}\label{fig:film_buckling_schematic} 
\end{figure}
The total moment of the body couple on an infinitesimal element of length $dx$ at  $x$  is
\begin{equation}
   dN=MAB\sin\theta dx,
\end{equation}where $\theta$ is the angle between  the beam axis and the horizontal.
The total moment at $x$ is 
\begin{eqnarray}
     N(x)&=&\int^x_0 dN_x dx=\int_0^x MAB\sin\theta dx =\int_0^x MAB\theta dx=\int_0^x MAB\frac{dv}{dx}dx\nonumber\\ 
         &=&\int_0^x MAB dv=MAB\left(v(x)-v_0\right).
\end{eqnarray}
The equilibrium equation of the beam is
\begin{eqnarray}
 EI\frac{d^2 v(x)}{dx^2}&=&-MAB(v(x)-v_0) \nonumber\\
\frac{d^2v(x)}{dx^2}&=&-k^2(v(x)-v_0)\nonumber\\
 \frac{d^2v(x)}{dx^2}+k^2v(x)&=&k^2v_0,
\end{eqnarray}
where, $k^2=\frac{MAB}{EI}$ and $EI$ is the bending stiffness of the beam.  The solution is
\[v(x)=A\cos k x+B\sin k x+v_0.\]
Using the boundary condition at the clamped end
\[ v(l)=0;\ \ \ \frac{d v(l)}{d x}=0, \]
we get,
\begin{equation}v(x)=v_0(1-\cos k(l-x)).\end{equation}
The displacement at $x=0$ is $v_0$, which yields
\begin{eqnarray}
      v_0&=&v_0(1-\cos kl)\nonumber,
\end{eqnarray}
so that  the  buckling criterion can be written as
\begin{equation}\label{eqn:buckling_criteria}
MB =\frac{EI n^2\pi^2}{4Al^2}.
\end{equation}
For a given $M$, the critical $B$ can be found from Eqn.~\ref{eqn:buckling_criteria}.
Note that this  kind of behavior cannot be observed in a super-paramagnetic film.

\end{appendices}

\end{document}